\documentclass[a4paper]{article}
\usepackage[utf8]{inputenc}
\usepackage[english]{babel}
\usepackage[T1]{fontenc}
\usepackage{geometry}
\usepackage{amsthm,amsmath,amssymb,amsfonts,stmaryrd,dsfont, makecell}
\usepackage{mathpartir,etoolbox,xstring,ifthen}
\usepackage{thm-restate}
\usepackage[hidelinks]{hyperref}
\usepackage[capitalise]{cleveref}
\usepackage{tikz,tikz-cd,mytikz}
\usepackage{mathtools}
\usepackage{macros}

\usetikzlibrary{decorations.pathmorphing, tikzmark}

\title{Globular weak $\omega$-categories\\as models of a type theory}

\author{Thibaut Benjamin\footnote{Ecole Polytechnique}, Eric
  Finster\footnote{University of Cambridge} and Samuel Mimram\footnote{Ecole
    Polytechnique}} \date{}

\makeatletter \newcommand{\neutralize}[1]{\expandafter\let\csname
  c@#1\endcsname\count@} \makeatother

\newtheorem{thm}{Theorem} \newtheorem*{thm*}{Theorem}
\newtheorem{prop}[thm]{Proposition} \newtheorem*{prop*}{Proposition}
\newtheorem{lemma}[thm]{Lemma} \newtheorem*{lemma*}{Lemma}
 \newtheorem{coro}[thm]{Corollary}

\theoremstyle{definition} \newtheorem{defi}[thm]{Definition}

\newtheorem{lemma-defi}[thm]{Definition/Lemma} \newenvironment{lemma-defibis}[1]
{\renewcommand{\thethm}{\ref*{#1}$'$}%
  \addtocounter{thm}{-1}%
  \begin{lemma-defi}} {\end{lemma-defi}}

\theoremstyle{remark} \newtheorem{rem}[thm]{Remark} \newtheorem{note}[thm]{Note}
\newtheorem{ex}[thm]{Example}

\begin{document}
\maketitle

\begin{abstract}
  We study the dependent type theory $\CaTT$, introduced by Finster
and Mimram, which presents the theory of weak $\omega$-categories, following the
idea that type theories can be considered as presentations of generalized
algebraic theories. Our main contribution is a formal proof that the models of
this type theory correspond precisely to weak $\omega$-categories, as defined by
Maltsiniotis, by generalizing a definition proposed by Grothendieck for weak
$\omega$-groupoids: those are defined as suitable presheaves over a
cat-coherator, which is a category encoding structure expected to be found in an
$\omega$-category. This comparison is established by proving the initiality
conjecture for the type theory $\CaTT$, in a way which suggests the possible
generalization to a nerve theorem for a certain class of dependent type theories.

\end{abstract}

\newpage
\setcounter{tocdepth} {2}
\tableofcontents
\newpage

\section*{Introduction}
The notion of weak $\omega$-category has emerged quite naturally by generalizing
situations encountered in algebraic topology: it consists in an algebraic
structure which comprises cells of various dimensions, which can be composed in
various dimensions, and satisfy the expected laws. We are however interested in
\emph{weak} such structures here, which means that we want to encompass
situations where the laws do not hold up to equality, but only up to
higher-dimensional cells, which thus play the role of witnesses that those laws
are satisfied. Those cells should themselves satisfy coherence laws, which
should only hold up to higher cells, which should themselves satisfy coherence
laws, and so on. Because of these towers of coherence cells, coming up with a
suitable definition of weak $\omega$-category is quite difficult. Historically,
definitions of weak $\omega$-groupoids (also called $\infty$-groupoids) were
first proposed, such as Kan complexes~\cite{joyal2002quasi}: those are weak
$\omega$-categories in which every cell is reversible, and are thus closer to
spaces encountered in algebraic topology. Then, around the beginning of the
century, various definitions for weak $\omega$-categories have been proposed: we
refer the reader to the surveys on the topic~\cite{leinster2002survey,
  cheng2004higher} for a general presentation of those. The comparison between
the proposals is still an ongoing research topic, and seems to be technically
out of reach for now for some of them.

While originating from topology, unexpected connections were found with type
theory: a series of works around 2010 revealed that the iterated identity types
in Martin-Löf type theory endow each type with the structure of a weak
$\omega$-groupoid~\cite{lumsdaine2009weak, van2011types,
  altenkirch2012syntactical}. This key observation is in fact one of the
motivations that lead to the development of homotopy type
theory~\cite{hottbook}. Based on this, and following Cartmell's insight that
type theory could be used to formulate generalized algebraic
theories~\cite{cartmell1986generalised}, Brunerie managed to extract from the
rules generating identity types, a definition of weak
$\omega$-groupoids~\cite{brunerie2016homotopy}, that he could show to be
equivalent to a definition proposed by Grothendieck~\cite{pursuing-stacks}. The
novelty of this definition lies in the fact that it is itself formulated as a
type theory.


Following Brunerie's approach, Finster and Mimram~\cite{catt} gave a
definition of weak $\omega$-categories in the form of a type theory
called $\CaTT$. Their definition follows the lines of a generalization
of Grothendieck's weak $\omega$-groupoids to weak $\omega$-categories,
proposed by Maltsiniotis~\cite{maltsiniotis}. The goal of this article
is to show that the type theory $\CaTT$ is equivalent to one of the
definitions proposed by Maltsiniotis. Moreover
Ara~\cite{ara2010groupoides} has proved, up to a conjecture later
proved by Bourke~\cite{bourke2020iterated} that this definition is
equivalent to a definition proposed by
Leinster~\cite{leinster2004higher} following a method introduced by
Batanin~\cite{batanin1998monoidal}. Our result completes this circle
of ideas, establishing that these three definitions are three sides of
the same story, expressed in different languages.

After brief general reminders about semantics of type theory in
\cref{sec:sem-tt}, we introduce a type theory for globular sets in
\cref{sec:glob}, which serves both as a basis and as a baby version of
our main proof. We then briefly present the Grothendieck-Maltsiniotis
definition of weak $\omega$\nbd-categories in \cref{sec:gm-def}, in
order to recall the motivations for the introduction of the type
theory \CaTT{}, which is introduced in \cref{sec:catt} along with some
examples of derivations in this theory, and some of its properties. We
then study in \cref{sec:syn-catt} the syntactic category of this
theory and begin relating it to the Grothendieck-Maltsiniotis
definition of weak $\omega$-categories. Finally, in
\cref{sec:models-catt}, we study the models of this type theory, and
show that they are equivalent to the aforementioned definition of weak
$\omega$-categories. The reader who wishes to familiarize themselves
with the type theory along the way may also experiment with the
implementation~\cite{gitcatt}.




\section{Categorical semantics of type theory}
\label{sec:sem-tt}
We begin by recalling the categorical framework we use here to study
type theory, together with an associated notion of model. We do not
introduce any type theory just yet, but only the categorical
description of type theories. We refer the reader
to~\cref{sec:globular-tt} for a presentation of the notion of type
theory considered here. We denote $\Cat$ the category of categories,
and $\Cath$ the $2$-category of categories. In general, we underline
the $2$-categories to distinguish them visually.

\subsection{Categories with families.}
The categorical models of type theory considered here are categories with
families, which were introduced by Dybjer~\cite{dybjer}. This particular choice
has little impact on the developments performed here since most other notions of
model, such as Cartmell's categories with
attributes~\cite{cartmell1986generalised}, are known to be equivalent to this
one.

We write $\Fam$ for the category of \emph{families}, where an object is a family
$(A_i)_{i\in I}$ consisting of sets $A_i$ indexed by a set~$I$, and a morphism
$f:(A_i)_{i\in I}\to (B_j)_{j\in J}$ is a pair consisting of a function
$f:I\to J$ and a family of functions $(f_i:A_i\to B_{f(i)})_{i\in I}$.
%

Suppose given a category~$\C$ equipped with a functor $T:\op{\C}\to\Fam$. Given
an object $\Gamma$ of~$\C$, its image under~$T$ is a family denoted
\[
  T\Gamma \qeq \pa{\Tm^\Gamma_A}_{A\in\Ty^\Gamma}
\]
\ie we write $\Ty^\Gamma$ for the indexing set and $\Tm^\Gamma_A$ for the elements
of the family. Given a morphism $\Gg:\GD\to\GG$ in~$\C$ and an element
$A\in\Ty^\GG$, we write $A[\Gg]$ for the object $\Ty^\Gg(A)$ of
$\Ty^\GD$. Similarly, given an element $t\in\Tm^\GG_A$, we write
$t[\Gg]$ for the element $\Tm^\Gg_A(t)$ of $\Tm^\GD_{A[\Gg]}$. With
these notations, the functoriality of~$T$ is equivalent to the following equations
\begin{align*}
  A[\gamma\circ\delta]
  &=
    A[\gamma][\delta]
  &
    t[\gamma\circ\delta]
  &=
    t[\gamma][\delta]
  \\
  A[\id{}]&=A
  &
    t[\id{}]&=t
\end{align*}
for composable morphisms of $\C$.

A \emph{category with families}  consists of a category $\C$
together with a functor $T:\op\C\to\Fam$ as above, such that $\C$ has a terminal
object, denoted $\emptycontext$, and such that there is a \emph{context
  comprehension} operation: given an object $\Gamma$ and type $A\in\Ty^\Gamma$,
there is an object $\pa{\Gamma,A}$, together with a projection morphism
$\pi:\pa{\Gamma,A} \to \Gamma$ and a term $p\in\Tm^{\pa{\Gamma,A}}_{A[\pi]}$,
such that for every morphism $\Gg : \Delta\to\Gamma$ in $\C$ together with a
term $t \in\Tm^{\Delta}_{A[\Gg]}$, there exists a unique morphism
$\sub{\Gg,t} : \Delta\to\pa{\Gamma,A}$ such that $p[\sub{\Gg,t}] = t$
and the following diagram commutes:
\[
  \begin{tikzcd}
    &(\Gamma,A)\ar[d,"\pi"]\\
    \Delta\ar[ur,dotted,"\sub{\gamma,t}"]\ar[r,"\gamma"']&\Gamma
  \end{tikzcd}
\]
In a category with families, the class of \emph{display maps} is the smallest
class of morphisms containing the projection morphisms $\pi:(\Gamma,A)\to\Gamma$
and closed under composition and identities.

A \emph{morphism} between two categories with families $(\C,T)$ and
$(\C',T')$, is a functor $F : \C\to\C'$ together with a natural
transformation $\phi : T\to T'\circ \op{F}$, such that $F$ preserves the
terminal object and the context comprehension operation on the nose.
In this article we consider the category of categories with families
that we denote $\CwF$, as well as the $2$-categories obtained by
considering natural transformations between the morphisms of
categories with families as functors, we denote it $\CwFh$.

\paragraph{Models of a category with families.}
We define a large category with families in a similar way, as a large
category equipped with a functor into families of large sets indexed
by a large set, and satisfying the exact same properties. Note that a
category with families can be seen as a large category with families.
There is a structure of a category with large families on the
category $\Set$, where, given a set $X$, $\Ty^X$ is the (large) set of
all $X$-indexed families of sets $(Y_{x})_{x\in X}$. Given such a
family $Y = (Y_{x})$, the set $\Tm^{X}_{Y}$ is the set of $X$-indexed
families of elements $(y_{x})_{x\in X}$ with $y_{x}\in Y_{x}$. For a
map $f : X' \to X$ the action of $f$ is given by
$(Y[f])_{x} = Y_{f(x)}$ and $(y[f])_{x} = y_{f[x]}$. We define the
category of \emph{models} of a category with families $\C$, denoted
$\Mod\C$, to be the category whose objects are the morphisms of
categories with families from~$\C$ to $\Set$. Explicitly we have
\[
  \Mod\C = \CwFh(\C,\Set)
\]



\paragraph{Pullbacks in a category with families.}
The structure of category with families enforces a compatibility condition
between context comprehension and the action of morphisms on types, expressed by
the following lemma: it states that all pullbacks along display maps exist and
that they can be explicitly computed from the given structure.

\begin{lemma}\label{lemma:pullback-display-map}
  In a category with families~$\C$, for every morphism $f:\Delta\to\Gamma$
  in~$\C$ and $A\in\Ty^\Gamma$, the square
  \[
    \begin{tikzcd}
      (\Delta,A[f])\ar[d,"\pi'"'] \ar[r,"\sub{f\circ\pi',p'}"]& (\Gamma,A)\ar[d,"\pi"]\\
      \Delta \ar[r,"f"'] & \Gamma
    \end{tikzcd}
  \]
  is a pullback, where $\pi':(\Delta,A[f])\to\Delta$ and
  $p'\in\Tm^{(\Delta,A[f])}_{A[f][\pi']}$ are obtained by context comprehension.
\end{lemma}
%
%
\begin{proof}
  Consider a diagram of the following form in~$\C$, without the dotted arrow:
  \[
    \begin{tikzcd}
      \Theta\ar[ddr, "\delta"', bend right = 20]\ar[drr, "\gamma", bend left = 20]\ar[dr,dotted,"\sub{\delta,p[\gamma]}"description]& & \\
      & (\Delta,A[f])\ar[d,"\pi'"] \ar[r,"\sub{f\circ\pi',p'}"]& (\Gamma,A)\ar[d,"\pi"]\\
      & \Delta \ar[r,"f"] & \Gamma
    \end{tikzcd}
  \]
  Given a term $p\in\Tm^{(\Gamma,A)}_{A[\pi]}$, we have
  $p[\gamma]\in\Tm^\Theta_{A[\pi][\gamma]} =
  \Tm^\Theta_{A[f][\delta]}$. By context extension, we obtain a map
  $\sub{\delta,p[\gamma]} : \Theta\to(\Delta,A[f])$ such that
  $\pi'\circ\sub{\delta,p[\gamma]} = \delta$ and
  $p'[\sub{\delta,p[\gamma]}] = p[\gamma]$. Since moreover
  $p' = p[\sub{f\circ\pi',p'}]$, the previous equality amounts to
  $p[\gamma] = p[\sub{f\circ\pi',p'}\circ\sub{\delta,p[\gamma]}]$.
  This condition is necessary for the upper triangle to commute. Hence
  the uniqueness of the map. We just have to show that this map makes
  the upper triangle commute. Notice that
  $\pi\circ\sub{f\circ\pi',p'}\circ\sub{\delta,p[\gamma]} =
  \pi\circ\gamma$, and
  $p[\gamma] = p[\sub{f\circ\pi',p'}\circ\sub{\delta,p[\gamma]}]$, by
  universal property of the extension for morphisms, this implies the
  commutativity of upper triangle.
\end{proof}

\noindent
The structure of a category with families can be thought of as a way
of ensuring that the pullbacks of the form of the above lemma exist,
while also enforcing that they are split. This means that the choice
of the pullbacks is such that taking a pullback along a composite
morphism $g\circ f$ gives the same result as taking the pullback along
$f$ and then along $g$. In the formalism of categories with families,
this means that we have $(\GG,A[\Gd\circ\Gg]) = (\GG,A[\Gd][\Gg])$.
Since the structure of category with families provides these
pullbacks, and since the morphisms of categories with families preserve
this structure, these morphisms also preserve these pullbacks, as
witnessed by the following result.

\begin{lemma}\label{lemma:cwf-morphism}
  Let $\C$ and $\catD$ be two categories with families, together with a morphism
  $(F,\phi) : \C\to\catD$. Then, for any object $\Gamma$ in $\C$ together with an
  element $A\in \Ty^\GG$, and for any morphism $\Gg:\GD \to \GG$ in $\C$, the
  following equation is satisfied:
  \[
    F(\Delta,A[\Gg]) = (F\Delta,(\phi_\Gamma A)[F\Gg])
  \]
\end{lemma}
\begin{proof}
  By definition of a morphism of categories with families, we have
  \[
    F(\Delta,A[\Gg]) = (F(\Delta),(\phi_{\Delta}(A[\Gg])))
  \]
  and, by naturality of $\phi$, the following square commutes
  \[
    \begin{tikzcd}
      Ty^{\Gamma} \ar[r,"\phi_\Gamma"]\ar[d,"\_\bra{\gamma}"'] & Ty^{F(\Gamma)}\ar[d,"\_\bra{F\Gg}"] \\
      Ty^{\Delta} \ar[r,"\phi_\Delta"] & Ty^{F(\Delta)}
    \end{tikzcd}
  \]
  This proves in particular the $\phi_{\Delta}(A[\Gg]) = (\phi_\Gamma A)[F\Gg]$,
  from which follows the desired equality.
\end{proof}

\noindent
Lemma~\ref{lemma:pullback-display-map} allows us to understand this
result as the fact that $F$ preserves pullbacks along display maps. In
fact, the following result shows that preserving these pullbacks is
precisely the condition that a functor from a category with families
to sets has to satisfy in order to be a model of the category with
families.

\begin{lemma}\label{lemma:models-cwf}
  The models of a category with families $\C$ are in bijective
  correspondence with the functors $\C\to\Set$ that preserve the
  terminal object and the pullbacks along display maps.
\end{lemma}
\begin{proof}
  By Lemma~\ref{lemma:cwf-morphism}, the underlying functor of a
  morphism of categories with families preserves the pullbacks along
  display maps and, by definition, such a functor has to preserve the
  terminal object as well. So it suffices to prove that a functor
  $F:\C\to\Set$ preserving the initial object and pullbacks along
  display maps gives rise to a unique model. Consider such a functor
  $F$, together with an object $\GG$ in $\C$ and a type $A\in\Ty^\GG$.
  Suppose defined $\phi$ such that $(F,\phi)$ is a model of~$\C$, then
  necessarily
  $F(\GG,A) = (F\GG,\phi_{\GG}A) = \bigsqcup\limits_{x\in
    F\Gamma}(\phi_\GG (A))_{x}$ by definition of the context
  comprehension in $\Set$. Thus $F(\Gamma,A)$ is uniquely determined
  by $F(\GG,A) = \bigsqcup\limits_{x\in F\Gamma}(\phi_\GG (A))_{x}$.
  Similarly, for a term $t\in\Tm^\GG_A$, there is a morphism
  $\sub{\id\GG,t} : \GG\to(\GG,A)$, so we have
  $F(\sub{\id\GG,t}) = \sub{\id{F\GG},\phi_{\GG,A}(t)}$. By definition
  of the structure of category with families on $\Set$, this
  chompletely characterize the map $F(\sub{\id\GG,t})$ as the one
  sending every element $x$ of $F(\Gamma)$ onto
  $(\phi_{\GG,A}(t))_{x}$. Together with the equation
  $\phi_{\GG,A}(t) = p[F(\sub{\id\GG,t})]$, this completely
  characterizes the map $\phi_{\GG,A}$. Conversely, these assignments
  define a natural transformation $\phi$, which make $(F,\phi)$ into a
  model of $F$.
\end{proof}
\noindent This condition relies on the specific structure of category
with families of $\Set$, and previous lemma does not extend as a
characterization of morphisms between arbitrary categories with
families. It also justifies retrospectively why it is not that
important to be precise about size issues with $\Set$, as one may as
well ignore the structure of category with families on $\Set$
altogether, and define a model as a functor $\C\to\Set$ that preserves
the terminal object and the pullback along the display maps.

\subsection{Contextual categories.}
In order to carry on some inductive constructions on the syntax of a theory, and
handle them in full generality, we also need introduce the notion of
\emph{contextual categories}, due to Cartmell~\cite{cartmell1986generalised},
and studied by Streicher~\cite{streicher1991contextual} and
Voevodsky~\cite{voevodsky2015c} under the name of \emph{C-systems}. These equip
categories with families with the extra structure required in order to perform
those inductive constructions.

\begin{defi}
  A \emph{contextual category} consist in a category with families $\C$ together
  with a function~$\ell$ associating to each object $\Gamma$ of $\C$ a natural
  number $\ell(\Gamma)$, called its \emph{length}, such that
  \begin{itemize}
  \item the terminal object $\emptycontext$ is the unique object such that
    $\ell(\emptycontext) = 0$,
  \item for every object $\Gamma$ and type $A \in\Ty^\Gamma$, $\ell(\Gamma,A) =
    \ell(\Gamma) + 1$,
  \item for every object $\Gamma$ such that $\ell(\Gamma)>0$, there is
    a unique object $\Gamma'$ together with a unique type
    $A \in\Ty^{\Gamma'}$ such that $\Gamma = (\Gamma',A)$.
  \end{itemize}
\end{defi}

Note that a contextual category is usually defined to be a category
with attributes satisfying such properties. However, since categories
with families and categories with attributes are equivalent, we will
also refer to these as contextual categories. The notion of contextual
category is not invariant under equivalences of categories and relies
on a particular presentation of a category. Its use is justified by
the fact that the syntax of a type theory gives a particular
presentation of a category with families, which happens to be a
contextual category.

Given a contextual category $\C$, an object $\Gamma$ whose length is strictly positive
decomposes in a unique way as $\Gamma',A$, and we simply write $\pi_\Gamma: \Gamma
\to \Gamma'$ (or even~$\pi$) instead of $\pi_{\Gamma',A}$. We also write
$x_{\Gamma}$ for the term $p_{\Gamma',A}$ in $\Tm^{\Gamma}_{A[\pi]}$, thought of
as a variable. More generally, we declare that a term is a \emph{variable} when
it is of the form $x_\Gamma[\pi]$ where $\pi$ is a display map. Note that in a
contextual category, if $\pi : \Delta\to\Gamma$ is a display map, then necessarily
$l(\Delta) > l(\Gamma)$. This implies that the variables of a non-empty context
$(\Gamma,A)$ are either $x_{(\Gamma,A)}$, or of the form $x[\pi_{(\Gamma,A)}]$
where $x$ is a variable of $\GG$.

The following lemma shows that a map in a contextual category is entirely characterized by
its action on the variables in its target context.

\begin{lemma}
  \label{lemma:funext-c-system}
  Consider two maps $\gamma,\delta : \Delta \to \Gamma$, in a contextual category, such
  that, for every variable $x$ in $\Gamma$, $x[\Gg] = x[\Gd]$. Then we have $\Gg =
  \Gd$.
\end{lemma}
\begin{proof} 
  We prove this result by induction on the length of the object $\GG$.
  \begin{itemize}
  \item Suppose that $\GG$ is of length $0$. Then necessarily, $\GG = \emptycontext$ is
    the terminal object, and thus~$\Gg = \Gd$.
  \item Suppose that $\GG$ is of length $l+1$. Then it is of the form
    $(\GG',A)$, and there is a map $\pi:\GG\to \GG'$. Suppose moreover
    that there are two maps $\Gg,\Gd : \GD \to \GG$, such that for
    every variable $x$ of $\GG$, we have $x[\Gg] = x[\Gd]$. Note that
    we necessarily have $\Gg = \sub{\pi\circ\Gg,x_\GG[\Gg]}$ and
    $\Gd = \sub{\pi\circ\Gd,x_\GG[\Gd]}$, as it is the case for every
    maps. Then for the variable $x_\GG$, we have
    $x_\GG[\Gg] = x_\GG[\Gd]$. Moreover, for every variable $x$ of
    $\GG'$, $x[\pi]$ is a variable of $\GG$, and thus
    $x[\pi][\Gg] = x[\pi][\Gd]$, which proves
    $x[\pi\circ\Gg] = x[\pi\circ\Gd]$, and, by induction hypothesis,
    $\pi\circ\Gg = \pi\circ\Gd$. We thus have proved that
    $\sub{\pi\circ\Gg,x_\GG[\Gg]} = \sub{\pi\circ\Gd,x_\GG[\Gd]}$,
    \ie{} $\Gg = \Gd$.\qedhere
  \end{itemize}
\end{proof}

\section{A type theory for globular sets}\label{sec:glob}
We first describe a type theory whose models are globular sets, on which we rely
in order to introduce the type theory~$\CaTT$. It was previously considered by
Brunerie~\cite{brunerie2016homotopy} and Finster and Mimram~\cite{catt}, and we
expand here on their work. This type theory is quite poor, as it has no term
constructors (the only terms are variables): it will also serve as a simple
setting in order to present the techniques and properties which will be
generalized later on to the more complex type theory~$\CaTT$.

\subsection{The category of globular sets.}
Globular sets are a generalization of graphs, which comprise not only points and
arrows, but also higher dimensional cells. Similarly to graphs, the category of
globular sets can be defined as a presheaf category.

\paragraph{The category of globes.}
The \emph{category of globes} $\G$ is the category whose objects are the natural
numbers and morphisms are generated by
\[
  {\sigma}_i,{\tau}_i:{i}\to{i+1}
\]
subject to following \emph{coglobular relations}:
\begin{equation}
  \label{eq:coglob-rel}
  {\sigma}_{i+1}\circ{\sigma}_i={\tau}_{i+1}\circ{\sigma}_i \qquad\qquad
  {\sigma}_{i+1}\circ{\tau}_i={\tau}_{i+1}\circ{\tau}_i
\end{equation}
The category of \emph{globular sets} $\GSet = \widehat{\G}$ is the presheaf
category over the category $\G$. Given a globular set $G$, we write $G_n$
instead of $G\,n$. Equivalently, a globular set is a family of sets
$\pa{G_n}_{n\in\N}$ equipped with maps $s_i,t_i : G_{i+1} \to G_i$ satisfying
the \emph{globular relations}, dual to \eqref{eq:coglob-rel}:
\begin{equation}
  \label{eq:glob-rel}
  s_i\circ s_{i+1} \qeq s_i\circ t_{i+1} \qquad\qquad t_i\circ s_{i+1}
  \qeq t_i\circ t_{i+1}
\end{equation}
An element of $G_{n}$ is called a $n$-cell, and the maps $s_{i},t_{i}$
are caleld respectively the source and target. Often, the indices of
the source and target maps can be inferred from the context and we
therefore omit them and write $s$ and $t$ to simplify notations.

Given an object~$n$, the associated representable $\Yoneda({n})$ is called the
\emph{$n$-disk} and is usually written $D^n$. It can be explicitly described by
\[
  (D^n)_i = \left\{
    \begin{array}{l}
      \text{$\set{\ast_0,\ast_1}$ if $i < n$}\\
      \text{$\set{\ast}$ if $i = n$} \\
      \text{$\emptyset$ if $i > n$}
    \end{array}
  \right.
\]
with $s(\_) = \ast_0$ and $t(\_) = \ast_1$. Throughout this paper, we use the
Greek lower cases $\sigma$ and $\tau$ to denote the morphisms in the category
$\G$, or to denote the image of the morphisms in $\G$ via a functor
$F : \G\to\C$, and we use their equivalent Latin lower cases $s,t$ to denote the
image of the morphisms in $\G$ via a functor $F : \op\G\to\C$.

\paragraph{The $n$-sphere globular set.}
Given $n\in\N$, the \emph{$n$-sphere} $S^n$ is the globular set,
equipped with an inclusion $\iota^n:S^n\hookrightarrow D^n$, defined
by
\begin{itemize}
\item $S^{-1} = \emptyset$ is the initial object, and
  $\emptyset \hookrightarrow D^1$ is the unique arrow,
\item $S^{n+1}$ and $\iota^{n+1}$ are obtained by the pushout
  \[
    \begin{tikzcd}
      S^n \ar[r,"\iota_n", hookrightarrow]\ar[d,"\iota_n"', hookrightarrow] \ar[rd,phantom,"\ulcorner",very near end] & D^n \ar[d] \ar[ddr, bend left, "\sigma_n"] & \\
      D^n \ar[r] \ar[rrd, bend right, "\tau_n"'] & S^{n+1} \ar[rd, dotted, "\iota_{n+1}" description] & \\
                                                                                                                      & & D^{n+1}
    \end{tikzcd}
  \]
\end{itemize}
This definition is well defined since, as a presheaf category, the
category of globular sets is cocomplete (and the colimits are computed
pointwise).

\paragraph{Finite globular sets.}
A globular set~$G$ is \emph{finite} if it can be obtained as a finite colimit of
representable objects. It can be shown that this is the case precisely when the set
$\bigsqcup\limits_{i\in\N} G_i$ is finite, because all representables themselves
satisfy this property. We write $\FinGSet$ for the full subcategory of $\GSet$
whose objects are the finite presheaves. We sometimes call a finite globular set
a \emph{diagram}, and describe it using a diagrammatic notation. For instance,
the diagram
\[
  \begin{tikzcd}
    x \ar[r, bend left = 30, "f"] \ar[r,bend right = 30, "g"'] \ar[r, phantom,
    "\phantom{\scriptstyle\alpha}\!\Downarrow\!{\scriptstyle\alpha}"]& y \ar[r, "h"] & z
  \end{tikzcd}
\]
denotes the finite globular set $G$, whose only non-empty cell sets are
\begin{align*}
  G_0&= \set{x,y,z}
  &
    G_1&= \set{f,g,h}
  &
    G_2&= \set{\alpha}
\end{align*}
and whose the sources and targets are defined by
\begin{align*}
  s(f)&=x
  &
  s(g) &= x
  &
  s(h) &= y
  &
  s(\alpha) &= f
  \\
  t(f) &= y
  &
  t(g) &= y
  &
  t(h) &= z
  &
  t(\alpha) &= g
\end{align*}
Disks and spheres are finite globular sets. In small dimensions, they can be
depicted as
\[
  \begin{array}{l@{\ =\ }c}
    D^0 &
          \begin{tikzcd}
            \cdot
          \end{tikzcd}\\[1.5ex]
    D^1 &
          \begin{tikzcd}[ampersand replacement = \&]
            \cdot \ar[r] \& \cdot
          \end{tikzcd}\\[1.5ex]
    D^2 &
          \begin{tikzcd}[ampersand replacement = \&]
            \cdot \ar[r,bend left = 30] \ar[r,bend right = 30] \ar[r, phantom,
            "\Downarrow"] \& \cdot
          \end{tikzcd} \\[1.5ex]
    D^3 &
          \begin{tikzcd}[ampersand replacement = \&]
            \cdot \ar[r,bend left = 30] \ar[r,bend right = 30] \ar[r, phantom,
            "\Downarrow \Rrightarrow \Downarrow"] \& \cdot
          \end{tikzcd}
  \end{array}
  \qquad\qquad
  \begin{array}{l@{\ =\ }c}
    S^0 &
          \begin{tikzcd}[ampersand replacement = \&]
            \cdot \& \cdot
          \end{tikzcd}\\[1.5ex]
    S^1 &
          \begin{tikzcd}[ampersand replacement = \&]
            \cdot \ar[r,bend left = 30] \ar[r,bend right = 30] \& \cdot
          \end{tikzcd} \\[1.5ex]
    S^2 &
          \begin{tikzcd}[ampersand replacement = \&]
            \cdot \ar[r,bend left = 30] \ar[r,bend right = 30] \ar[r, phantom,
            "\Downarrow \phantom{a} \Downarrow"] \& \cdot
          \end{tikzcd} \\[1.5ex]
    \multicolumn{2}{c}{}
  \end{array}
\]
By definition, $\FinGSet$ is the free cocompletion of $G$ under all finite
colimits (see \cref{sec:models-glob}).

\subsection{The theory $\Glob$.}
\label{sec:globular-tt}
In this section, we introduce our notation for the type theories we consider,
and describe a particular type theory describing globular sets. The precise
relation between this type theory and the category of globular sets is detailed in
\secr{syn-glob} and \secr{models-glob}.

\paragraph{Signature.}
We consider a countably infinite set whose elements are called
\emph{variables}. A \emph{term} is this theory is simply a variable
(later on, we consider theories where terms are more general than just
variables). A \emph{type} is defined inductively to be either
\[
  \Obj \qquad\text{or}\qquad \Hom Atu
\]
where $A$ is a type and $t,u$ are terms. A \emph{context} is a list
\[
  (x_1:A_1,\ldots,x_n:A_n)
\]
of variables $x_1,\ldots,x_n$ together with types $A_1,\ldots,A_n$, the empty
context is denoted $\emptycontext$. A \emph{substitution} is a list
\[
  \sub{x_1\mapsto t_1,\ldots,x_n\mapsto t_n}
\]
of variables $x_1,\ldots,x_n$ together with terms $t_1,\ldots,t_n$. From now on,
we use the following naming conventions
\begin{align*}
  \text{variables} &: x,y,\ldots &
  \text{terms} &: t,u,\ldots &
  \text{types} &: A,B,\ldots\\
  \text{contexts} & : \Gamma,\Delta,\ldots &
  \text{substitutions} & : \Gg,\Gd,\ldots
\end{align*}

\paragraph{Judgments.}
The theory consists in four different kinds of \emph{judgments}, for which we
give the notations, along with the intuitive meaning:

\smallskip
\begin{tabular}{l@{\ :\ }l}
  $\Gamma\vdash$ & the context $\Gamma$ is well-formed \\
  $\Gamma\vdash A$ & the type $A$ is well-formed in the context $\Gamma$ \\
  $ \Gamma\vdash t:A$ & the term $t$ has type $A$ in context $\Gamma$ \\
  $ \GD\vdash \Gg : \GG$  & the substitution $\Gg$ goes from the context $\GD$ to the context $\GG$
\end{tabular}
\smallskip

\noindent
Most of the time, when we refer to a context, a type, a term or a substitution,
we implicitly mean such an object satisfying the adequate judgment. To emphasize
this convention we add the adjective \emph{raw} to designate an object as given
by the signature, without supposing that a corresponding judgment is derivable,
and we state that a property is \emph{syntactic} when it holds for raw
expressions.

\paragraph{Syntactic properties.}
Given a raw term $t$ (\resp a raw type $A$, a raw context $\Gamma$, a raw substitution
$\Gg$), we define the set of its \emph{free variables} $\Var t$ (\resp $\Var
A$, $\Var \Gamma$, $\Var \Gg$) by induction as follows
\begin{align*}
  \text{on terms:} & & \Var x & = \set{x} \\
  \text{on types:} & & \Var\Obj &= \emptyset & \Var{\Hom Atu} & = \Var A\cup\Var t \cup\Var u\\
  \text{on contexts:} & & \Var\emptycontext &= \emptyset  & \Var{\Gamma,x:A} & = \set{x}\cup\Var\Gamma\\
  \text{on substitutions:}& & \Var{\sub{}} & = \emptyset & \Var{\sub{\Gg,x\mapsto t}} &= \Var t \cup \Var\Gg
\end{align*}
\noindent Given a raw type $A$ in this theory, we define its \emph{dimension} $\dim(A)$ by induction by
\begin{align*}
  \dim(\Obj)&=-1
  &
  \dim(\Hom Atu)&= \dim(A)+1
\end{align*}
The choice of starting at $-1$ is dictated here by the correspondence
established in Lemma~\ref{lemma:up-disk-glob}.
We define the dimension of a context $\Gamma = (x_i:A_i)_{1\leq i\leq n}$ to be
\[
  \dim(\Gamma) = \max_{1\leq i\leq n}{\dim(A_i)}
\]
and the dimension of a term $t$ in the context \(\Gamma\), when the judgment
$\Gamma\vdash t:A$ holds to be
\[
  \dim(t)=\dim(A)+1
\]

\paragraph{Action of substitutions, composition, identity.}
Given a raw substitution $\gamma$, we define its action $t[\gamma]$ on a raw
term~$t$, its action $A[\gamma]$ on a raw type~$A$, and is composition
$\delta\circ\gamma$ with another raw substitution~$\delta$ by
\begin{align*}
  t[\sub{}] &= t & t[\sub{\Gg,x\mapsto u}] &= \left\{
  \begin{array}{l@{\quad}l}
    u & \text{if $t = x$} \\
    t[\Gg] & \text{otherwise}
  \end{array}\right.\\
  \Obj[\Gg] &= \Obj & (\Hom Atu)[\Gg] &= \Hom{(A[\Gg])}{(t[\Gg])}{(u[\Gg])}\\
  \sub{}\circ\Gg &=\sub{} & \sub{\Gd,x\mapsto t}\circ\Gg &= \sub{\Gd\circ\Gg, x\mapsto t[\Gg]}
  \intertext{We also define a special raw substitution associated to a raw context $\GG$, that we call the \emph{identity substitution} $\id\GG$, by induction by}
  \id\emptycontext &= \sub{} & \id{\GG,x:A} &=\sub{\id\GG,x\mapsto x}
\end{align*}

\paragraph{Typing rules.}
The inference rules for the theory $\Glob$ are given in figure~\ref{fig:rules-glob}. We say that a context (\resp a type, term substitution) is derivable if there is a derivation tree leading to its well-formedness judgment.
\begin{figure}[h]
  \centering
  \begin{tabular}{|cc|}
    \hline
    \multicolumn{2}{|l|}{\emph{For contexts:}}\\
    $\inferrule{\null}{\emptycontext\vdash}{\regle{ec}}$ & $\inferrule{\GG\vdash A}{\GG,x:A\vdash}{\regle{ce}}$\quad when $x\notin\Var\GG$\\
    \multicolumn{2}{|l|}{\emph{For types:}}\\
    $\inferrule{\GG\vdash}{\GG\vdash\Obj}{\regle{$\Obj$-intro}}$ &
    $\inferrule{\GG\vdash A \\ \GG\vdash t:A \\ \GG\vdash u:A}{\GG\vdash \Hom Atu}{\regle{$\Hom{}{}{}$-intro}}$\\
    \multicolumn{2}{|l|}{\emph{For terms:}}\\
    $\inferrule{\GG\vdash\\(x:A)\in\GG}{\GG\vdash x:A}{\regle{var}}$ & \\
    \multicolumn{2}{|l|}{\emph{For substitutions:}}\\
    $\inferrule{\GD\vdash}{\GD\vdash\sub{}:\emptycontext}{\regle{es}}$ & $\inferrule{\GD\vdash\Gg:\GG \\ \GG,x:A\vdash \\ \GD\vdash t:A[\Gg]}{\GD\vdash\sub{\Gg,x\mapsto t}:(\GG,x:A)}{\regle{se}}$\\
    \hline
  \end{tabular}
  \caption{Derivation rules of the theory $\Glob$}
  \label{fig:rules-glob}
\end{figure}

\noindent
We have defined the sets of free variables as a syntactic function on terms and types, thus
independent of the judgments, but we are often rather interested in the variables
of a typed term together. To express this, we write $\Var{t:A}$ for the
union $\Var{t}\cup\Var{A}$, with the implicit convention that in the current
context $\Gamma$, the judgment $\Gamma\vdash t:A$ is derivable.

The first few results that we mention about the theory $\Glob$ are proved by
induction on the rules of the theory. These induction are typically tedious and
uninformative, and we refer the reader to our \agda{}
formalisation~\cite{benjamin21formalization} for the details.

\begin{lemma}\label{lemma:glob-derivation}
  The following properties can be shown and are useful for later proofs
  \begin{itemize}
  \item if $\Gamma\vdash A$ then $\Gamma\vdash$,
  \item if $\Gamma\vdash t:A$ then $\Gamma\vdash A$,
  \item if $\Delta\vdash\Gg:\Gamma$ then $\Delta\vdash$ and $\Gamma\vdash$,
  \item if $\Gamma\vdash \Hom Axy$ then $\GG\vdash x:A$ and $\GG\vdash y:A$,
  \item if $\GG\vdash A$, then $\Var A\subseteq\Var\GG$,
  \item if $\GG\vdash t:A$ then $\Var{t:A}\subseteq \Var\GG$.
  \end{itemize}
\end{lemma}

\begin{lemma}
  \label{lem:gset-type-unique}
  A term admits at most one type in a given context: if both $\Gamma\vdash t:A$
  and $\Gamma\vdash t:B$ are derivable then $A=B$.
\end{lemma}

\begin{lemma}
  \label{lem:gset-der-unique}
  A given judgment admits at most one derivation.
\end{lemma}

\paragraph{Notational conventions.}
In a type $\Hom Atu$, the type $A$ is the common type of both $t$ and $u$ and is
sometimes left implicit. Similarly, when a substitution
$\gamma = \sub{x_i \mapsto t_i}_{1\leq i\leq n}$ is such that the judgment
$\Delta\vdash\gamma:\Gamma$ holds for some context
$\Gamma = (y_i:A_i)_{1\leq i\leq m}$, then necessarily $m = n$ and $x_i = y_i$
for $1 \leq i\leq n$. For this reason, when the context $\Gamma$ is fixed, we
may leave the variables $x_1,\ldots,x_n$ implicit and simply write
\[
  \gamma = \sub{t_1,\ldots,t_n} = \sub{t_i}_{1\leq i\leq n}
\]
Finally, following \cref{lem:gset-der-unique}, we sometimes abusively assimilate
a derivation with the judgment it derives.

\subsection{The syntactic category of $\Glob$.}
\label{sec:syn-glob}
Our main tool to study the semantics of a type theory is a category we
associate to it, called its syntactic category. We define it in the
special case of the theory $\Glob$, and state some results which
ensure that it is well-defined. We then study the structures present
in this category, and illustrate how one can use those in order to
study the semantics of the theory.
The construction will be analogous later on in the case of the type theory
$\CaTT$, albeit more technically involved.

\paragraph{Admissibility of the action of the substitutions.}
When introducing the type theory, we have defined the actions of substitution,
their compositions and the identity substitution syntactically, by induction on
the signature. By induction over the rules of the theory, we can check that
these operations preserves the derivability of the judgments. Again, this
properties have been formally verified in \agda~\cite{benjamin21formalization}.
\begin{prop}\label{prop:cwf}
  The following rules are admissible
  \[
  \begin{array}{c@{\qquad\qquad}c}
    \inferrule{\GG\vdash A \\ \GD\vdash \Gg:\GG}{\GD\vdash A[\Gg]} &
    \inferrule{\GG\vdash t:A \\ \GD\vdash \Gg:\GG}{\GD\vdash t[\Gg]:A[\Gg]} \\[1.2em]
    \inferrule{\GG\vdash \Gth : \GTH \\ \GD\vdash \Gg : \GG}{\GD\vdash \Gth\circ\Gg : \GTH} &
    \inferrule{\GG\vdash}{\GG\vdash \id\GG:\GG}
  \end{array}
  \]
\end{prop}

\paragraph{The syntactic category.}
The last two statements of Proposition~\ref{prop:cwf} ensure that the
composition of substitution and the identity substitution preserve derivability
and thus can be lifted as operations on derivable objects. We keep the same
notation for these operations.

\begin{prop}
  \label{prop:category}
  The following equalities hold:
  \begin{align*}
    \id\Gamma \circ \Gg &= \Gg
    &
    \Gg \circ \id\GD &= \Gg
    &
    \Gg \circ (\Gd \circ \Gth) &= (\Gg \circ \Gd) \circ \Gth
  \end{align*}
\end{prop}

\noindent
Note that we assume here that all the objects we manipulate are derivable, even
if we leave their derivation implicit, in particular, although the second
equation holds syntactically, it is not the case for the first equation which
only holds for a derivable substitution $\GD\vdash \Gg:\GG$, nor for the last
equation which only holds for three a derivable substitutions $\Gg,\Gd$ and
$\Gth$ which are composable.

The last two results of Proposition~\ref{prop:cwf} as well as
Proposition~\ref{prop:category} ensure that we can build a category~$\Syn\Glob$,
called the \emph{syntactic category} of the theory $\Glob$, whose objects are
the contexts $\GG$ such that $\Gamma\vdash$ and morphisms $\GD\to\GG$ are the
substitutions $\GD\vdash\Gg:\GG$. The first two statements of
Proposition~\ref{prop:cwf} can then be read as the fact that it acts on
derivable types and terms:

\begin{prop}
  The composition of substitutions and the identity substitution are compatible
  with the action of the substitution on types and terms. More precisely, the
  following equations hold, for derivable objects:
  \begin{align*}
    A[\id\GG] &= A & A[\Gg\circ\Gd] &= A[\Gg][\Gd] \\
    t[\id\GG] &= t & t[\Gg\circ\Gd] &= t[\Gg][\Gd]
  \end{align*}
\end{prop}

\noindent
Propositions~\ref{prop:cwf},~\ref{prop:category}
and~\ref{prop:with-families} can be summarized into the following
proposition, which is crucial for studying the semantics of type
theories:

\begin{prop}
  \label{prop:with-families}
  The category $\Syn{\Glob}$ carries a structure of category with families, such
  that, for an object $\Gamma$ of $\Syn\Glob$, the set $\Ty^\Gamma$ consists in
  the types derivable in $\Gamma$ and, for $A$ such a type, the set
  $\Tm^\Gamma_A$ consists in terms of type $A$ in $\Gamma$.
\end{prop}

\begin{note}
  Here we have given a presentation with named variables, but one could also
  give a presentation of the same type theory using unnamed variables, such as
  de Bruijn indices. This would lead to a slightly different notion of the
  syntactic category, which is essentially the previously defined syntactic
  category quotiented under renamings (or $\alpha$-conversion) of contexts. From
  now on, we suppose given such a presentation with unnamed variables, so that
  the renamings are not explicitly taken in account in the syntactic
  category. Since there is no variable binders, this operation of quotienting is
  straightforward.
\end{note}

\paragraph{Disks and spheres contexts.}
In the category $\Syn\Glob$, there are two classes of contexts which play an
important role, the \emph{$n$-disk context}~$D^n$ and the \emph{$n$-sphere
  context}~$S^n$. Their precise role in our theory are made clear by the
Lemma~\ref{lemma:up-disk-glob} and by the understanding of the syntactic
category provided by the Theorem~\ref{thm:syn-glob}. These contexts are defined
inductively by
\[
  \begin{array}{l@{\qeq}l@{\qquad\qquad}l@{\qeq}l}
    S^{-1} & \emptycontext & D^0 & (d_0 : U_0) \\
    S^n & (D^n, d_{2n+1} : U_n) & D^{n+1} & (S^n,d_{2(n+1)} : U_{n+1})
  \end{array}
\]
where the types $U_n$ are inductively defined by
\[
  \begin{array}{l@{\qeq}l}
    U_0 & \Obj\\
    U_{n+1} & \Hom{U_n}{d_{2n-2}}{d_{2n-1}}
  \end{array}
\]
and where the $d_i$ are a family of distinct variables. We reserve the notation
$d_i$ for these specific variables throughout this paper. This is simply a
convenient writing convention, since ultimately we consider everything up to
renaming.
\begin{prop}
  For any integer $n$, the contexts $D^n$ and $S^n$ are well-formed, \ie the
  following rules are admissible.
  \[
    \inferrule{\null}{D^n\vdash} \qquad\qquad\qquad \inferrule{\null}{S^n\vdash}
  \]
\end{prop}
\begin{proof}
  We prove the validity of these contexts by induction. First notice that
  $S^{-1} = \emptycontext$ is well defined by the rule \regle{ec}, and that by
  applying successively the rules \regle{ce} and \regle{obj}, $D^0$ is also well
  defined. Then, suppose that $S^{k-1}$ and $D^k$ are valid contexts. The rule
  \regle{ax} ensures that $D^k\vdash d_{2k}:U_k$, and by
  Lemma~\ref{lemma:glob-derivation}, this proves that $D^k\vdash U_k$, since
  moreover $d_{2k+1}\notin \Var{D^k}$, the rule \regle{ce} applies and shows
  $S^k\vdash$. Moreover, the rule \regle{ax} applies twice to show both
  $S^k\vdash d_{2k} : U_k$ and $S^k\vdash d_{2k+1}:U_k$, hence by the rule
  \regle{hom}, this proves $S^k\vdash U_{k+1}$ and since $d_{2(k+1)}\notin S^k$,
  the rule \regle{ce} applies and proves $D^{k+1}\vdash$.
\end{proof}

We can also define two substitutions $D^{n+1}\vdash s_n:D^n$ and $D^{n+1}\vdash
t_n:D^n$, with the following formulas
\begin{align*}
  s_n &= \sub{d_0\mapsto d_0,d_1\mapsto d_1,\ldots,d_{2n-1} \mapsto d_{2n-1},d_{2n}\mapsto d_{2n}}
  \\
  t_n &= \sub{d_0\mapsto d_0,d_1\mapsto d_1,\ldots,d_{2n-1} \mapsto d_{2n-1},d_{2n}\mapsto d_{2n+1}}
\end{align*}
One can check that the morphisms define this way satisfy the globular
relations~\eqref{eq:glob-rel}, hence the disks objects are coglobular objects
in the category $\Syn{\Glob}$. We reformulate this fact by the following
definition.
\begin{defi}
  We define the functor $D^\bullet:\op\G\to\Syn{\Glob}$, sending every object
  $n$ on the disk context $D^n$ and the morphisms $\Gs_n$ (\resp{} $\Gt_n$) on
  the substitution $s_n$ (\resp{} $t_n$) in $\Syn\Glob$.
\end{defi}

\paragraph{Familial representability of types.}
The following lemma is central in our study of the type theory $\Glob$. It
allows to understand both types and terms as special cases of substitutions, and
the action of substitution then becomes pre-composition.
\begin{lemma}\label{lemma:up-disk-glob}
  For any natural number $n$, the map
  \[
    \begin{array}{rcl}
      \Syn{\Glob}(\GG,S^{n-1}) & \to & \setof{A \in \Ty^\GG}{\dim(A) = n-1} \\
      \Gg & \mapsto & U_{n}[\Gg]
    \end{array}
  \]
  is an isomorphism natural in $\GG$. Given a type $A$ of dimension $n-1$, we
  denote the associated substitution
  \[
    \Gc_A : \GG \to S^{n-1}
  \]
  Moreover, the maps
  \[
    \begin{array}{rcl}
      (\Syn{\Glob}/S^{n-1})(\GG\xrightarrow{\Gc_A}S^{n-1},D^n\xrightarrow{\pi} S^{n-1}) & \to & \Tm^\GG_A \\
      \Gg & \mapsto & d_{2n}[\Gg]
    \end{array}
  \]
  are also isomorphisms, natural in~$\Gamma$ (the source is a hom-set in the
  slice category of $\Syn{\Glob}$ over~$S^{n-1}$). Given a term $t\in\Tm^\GG_A$
  of type $A$, we denote the associated substitution over $\Gc_A$ by
  $\Gc_t:\GG\to D^{n}$, in such a way that the following diagram commutes
  \[
    \begin{tikzcd}
      \GG \ar[dr,"\Gc_A"']\ar[r,"\Gc_t"] & D^n\ar[d,"\pi"] \\
      & S^{n-1}
    \end{tikzcd}
  \]
\end{lemma}
\begin{proof}
  We first prove that the first part of the statement implies the second one,
  for a given natural number $n$. This is a consequence of the fact that the
  context $D^n$ is defined to be $(S^{n-1},d_{2n}:U_n)$. Indeed, an object in
  $\Syn{\Glob}/S^{n-1}$ is a context that comes equipped with a substitution
  $\Gg : \GG \to S^{n-1}$, and the universal property of the context
  comprehension operation states exactly that there is a natural isomorphism
  $\Syn{\Glob}(\GG,D^n) \isoto \Tm^\GG_{U_n[\Gg]}$. Using the previous natural
  isomorphism, one can write $\Gg$ as $\Gc_A$ and $U_n[\Gg]$ then simplifies to
  $A$, which proves the natural isomorphism $\Syn{\Glob}(\GG,D^n) \isoto
  \Tm^\GG_A$. We now prove by induction over the dimension $n$ that the first
  part of the judgment holds.
  \begin{itemize}
  \item Case $n = 0$. The context $S^{-1} = \emptycontext$ is terminal: there is
    always exactly one substitution $\Gamma\vdash \sub{} : \emptycontext$.
    Similarly there is always exactly one type of dimension $-1$ derivable in
    $\Gamma$, which is the type $\Obj$, and which is the type $U_0$ by
    definition.
  \item Suppose that the result holds for the sphere $S^{n-1}$. Then, by the
    second part of the result that we have already proven, we get a natural
    isomorphism $(\Syn{\Glob}/S^{n-1})(\GG,D^n) \isoto \Tm^\GG_A$. Substitutions
    $\Gamma\vdash\gamma : S^{n}$ are exactly the ones of the form
    $\sub{\gamma',u}$ and are derived by the following application of the
    rule~\regle{se}
    \[
      \inferrule{\Gamma\vdash\gamma' : D^n \\ D^n\vdash U_n \\
        \Gamma\vdash u : U_n[\gamma']}{\Gamma\vdash \sub{\gamma',u} : S^n}
    \]
    The substitutions $\GG\vdash \Gg:S^n$ are thus naturally isomorphic to pairs
    $\Gg',u$, with $\GG\vdash \Gg' : D^n$ and $\GG\vdash u : U_n[\Gg']$. By
    induction, the substitutions $\GG\vdash \Gg':D^n$ are of the form $\Gc_t$,
    for $\GG\vdash t:A$ a term in $\GG$. Then the type $U_n[\Gc_t]$ rewrites as
    $A$ by naturality of the previous transformation. So these substitutions are
    naturally isomorphic to pairs of terms of dimension $n$ and of the same type
    in $\GG$, which are exactly the types in $\GG$ of dimension $n$.
  \end{itemize}
\end{proof}
\begin{note}
  This proof does not rely on how the terms are constructed, so no matter what
  the term constructors are, this result will always hold.
\end{note}

\noindent
We can reformulate this result in several ways. First, we can collect together
all the isomorphisms $\setof{A\in\Ty^\GG}{\dim A = n-1} \isoto
\Syn{\Glob}(\GG,S^{n-1})$ into a single natural isomorphism
\[
  \Ty^\GG \isoto \coprod_{n\in\N} \Syn{\GG,S^{n-1}}
\]
In other words, we have proven that the family $S^{\bullet}$ \emph{familially
  represents} the functor $\Ty$. We can also unravel a bit this proposition,
showing that any type $\GG\vdash A$ corresponds uniquely to a substitution
$\GG\vdash \Gc_A:S^{n-1}$, and that any term $\GG\vdash t:A$ corresponds
uniquely to a substitution $\GG\vdash \Gc_t : D^{n}$, in such a way that the
following diagram commutes
\[
  \begin{tikzcd}
    \GG \ar[r,"\Gc_t"]\ar[dr,"\Gc_A"'] & D^n \ar[d,"\pi"] \\
    & S^{n-1}
  \end{tikzcd}
\]
To simplify things further, we write $\ty : D^n \to S^{n-1}$ for the projection
substitution, so that we have $\ty\circ\Gc_t = \Gc_A$. In other words, $\ty$
acts on terms by giving their associated types. The definition of the morphism
$\ty$ along with \cref{lemma:pullback-display-map} shows the following:
\begin{lemma}\label{lemma:ext-glob}
  In the category $\Syn{\Glob}$, a context of the form $(\GG,x:A)$ is obtained
  as the pullback
  \[
    \begin{tikzcd}
      (\GG,x:A) \ar[r]\ar[d,"\pi"']\ar[dr,phantom,"\lrcorner"very near start] & D^n \ar[d,"\ty"]\\
      \GG \ar[r,"\Gc_A"'] & S^{n-1}
    \end{tikzcd}
  \]
\end{lemma}

It is straightforward using \cref{lemma:ext-glob} to check that the sphere
contexts can be obtained as iterated pullbacks of the disks contexts, dually to
the way topological spheres can be obtained as pushout of topological disks.
\begin{lemma}\label{lemma:spheres-are-pb}
  The sphere context $S^{n}$ is obtained as the following pullback
  \[
    \begin{tikzcd}
      S^n \ar[r]\ar[d]\ar[dr,phantom,"\lrcorner" very near start] & D^n\ar[d,"\pi"] \\
      D^n\ar[r,"\pi"] & S^{n-1}
    \end{tikzcd}
  \]
\end{lemma}

\paragraph{The syntactic category of $\Glob$.}
We now characterize the syntactic category of $\Glob$. This is an
important step in order to study the models of the theory, since
understanding precisely the syntactic category gives good insights on
the functors mapping out of it. Interestingly, in all the cases we
study here, it always turn out that the syntactic category is dual to
the category of finitely presented objects that we are studying, in
accordance with the Gabriel-Ülmer duality~\cite{gabriel2006lokal}. In
order to prove this result, we introduce a functor that we denote $V$
(the $V$ stands for ``variable''), that we describe as follows.

\begin{defi}
  We define a functor $V : \Syn{\Glob} \to \op{\FinGSet}$, which to any context $\Gamma = (x_i:A_i)$, then associates
  \[
  \pa{V\Gamma}_n = \setof{x_i}{\dim(A_i) = n} = \set{\textnormal{derivable terms
      of dimension $n$ in $\Gamma$}}
  \]
  and to any substitution $\GD \vdash \sub{x_i:t_i} :\GG$ associates the map
  \[
  \begin{array}{ccccc}
    V\gamma & : & V\Gamma & \to & V\Delta \\
    & & x_i & \mapsto & t_i
  \end{array}
  \]
  or equivalently, we require the equation $(V\Gg) x = V(x[\Gg])$.
\end{defi}

\begin{lemma}
  The functor $V$ is well-defined.
\end{lemma}
\begin{proof}
  For $x$ of type $A$ in $\Gamma$, with $\dim(A) = n+1$, by definition of the
  dimension, $A$ is of the form $A = \Hom{}yz$, for two derivable terms $y$ and
  $z$, with $\dim_\Gamma(y) = \dim_\Gamma(z) = n$. We therefore have
  $y,z \in \pa{V\Gamma}_n$, and we define $s(x) = y$ and $t(x) = z$. The
  derivation rule for $A$ implies that $y$ and $z$ have the same type, thus
  $s(y) = s(z)$ and $t(y) = t(z)$, which proves that the globular relations are
  satisfied, and that $V\Gamma$ is indeed a globular set.

  Let $\Delta\vdash\gamma : \Gamma$ be a substitution, and write
  $\Gamma = (x_i : A_i)$, then the substitution $\gamma$ is of the form
  $\gamma = \sub{x_i:t_i}$, where $t_i$ is a derivable term in the context
  $\Delta$, \ie $t_i\in V\Delta$. Suppose that $x$ is of type $\Hom{}yz$ in
  $\Gamma$, then $x[\Gg] = \Hom{}{y[\Gg]}{z[\Gg]}$ in $\Delta$. This means that
  as an element of $F\Delta$, $x[\Gs]$ satisfies $s(x[\Gg]) = y[\Gg]$ and
  $t(x[\Gg]) = z[\Gg]$, or in other words, $s((F\Gg) x) = (F\Gg)(sx)$ and
  $t((F\Gg)x) = (F\Gg)(tx)$.  Hence $F\Gg$ defines a morphism of globular sets.
\end{proof}

\begin{thm}
  \label{thm:syn-glob}
  The functor $V$ is part of an equivalence of categories
  $\Syn\Glob\equivto\op{\FinGSet}$.
\end{thm}
\begin{proof}
  We first show that $V$ is full and faithful. Consider two
  substitutions $\Gg$ and $\Gd$ such that $V\Gg = V\Gd$. This implies
  in particular that for all variables $x$ in $\GG$,
  $(V\Gg) x = (V\Gd) x$, thus $x[\Gg] = x[\Gd]$. By
  Lemma~\ref{lemma:funext-c-system}, this proves that $\Gg = \Gd$,
  hence $V$ is faithful. Dually, consider two contexts $\GG$ and
  $\GD$, where $\GD = (x_i:A_i)_{0\leq i~\leq l}$, together with a
  morphism of globular sets $f : V\GG \to V\GD$. Then one can define
  the substitution $\Gg_f = \sub{x_i~: f(x_i)}_{1\leq i~\leq l}$. We
  check by induction on the length $l$ of $\GD$ that this produces a
  well-defined substitution $\Gg_f$ such that $V(\Gg_f) = f$. If
  $l = 0$ then $\GD = \emptycontext$ and $\Gg_f = \sub{}$, then the
  rule \regle{es} gives a derivation of
  $\GG\vdash\sub{}:\emptycontext$. If $\GD = (\GD',x_{l+1}:A_{l+1})$,
  then the natural inclusion $V\GD'\hookrightarrow V\GD$ induces by
  composition a map $f':V\GD'\to V\GG$. By induction hypothesis, we
  have $\GG\vdash \Gg_{f'}:\GD'$, and since $\GD$ is a context, we
  also have $\GD\vdash A_{l+1}$. Moreover, if $A_{l+1} = \Obj$, then
  $\GG\vdash f(x_{l+1}):\Obj$ since $f$ preserves the dimension, and
  otherwise $A_{l+1} = \Hom{}yz$, and
  $\GG\vdash f(x_{l+1}) : \Hom{}{f(y)}{f(z)}$ since $f$ is a morphism
  of globular sets. In both cases, this proves that
  $\GG\vdash f(x_{l+1}) : A_{l+1}[\Gg_{f'}]$. By application of the
  rule \regle{se}, this proves that
  $\GG\vdash\sub{\Gg_{f'},x_{l+1}\mapsto f(n_{l+1})}:\GD$. Since
  $\Gg_f = \sub{\Gg_{f'},x_{l+1}\mapsto f(x_{l+1})}$, this proves that
  $\Gg_f$ is well defined, and by definition it satisfies $V\Gg = f$.
  Hence the functor $V$ is full.

  Moreover, $V$ is essentially surjective. Indeed, considering a finite globular
  set $X$, we show by induction on the number of elements of $X$ that we can
  construct a context $\GG$ such that $V\GG = X$. If $X$ is the empty globular
  set, then $\GG = \emptycontext$ is well defined by the rule \regle{ec},
  otherwise, if $X$ is not empty, consider an element $x$ of maximal dimension
  in $X$ and consider the globular set $Y$ obtained by removing this element
  from $X$. By induction, the context $\GD$ constructed from $Y$ is well-defined.
  Moreover, if $x$ is of dimension $0$, then define $A~= \Obj$ and we have
  $\GD\vdash A$, and otherwise, we have $\GD\vdash s x : B$ and $\GD\vdash t x :
  B$ since both $s x$ and $t x$ are parallel elements in $Y$, and define $A~=
  \Hom{}{s~x}{t x}$, this shows that $\GD\vdash A$. In both cases, we have
  $\GD\vdash A$, and the rule $\regle{ce}$ applies to prove that
  $\GD,x:A\vdash$. Moreover $V(\GD,x:A)$ is obtained from $V\GD$ by adding one
  element $x'$ of the same dimension as $x$, and such that $s x' = s x$ and $t
  x' = t x$ if this dimension is not $0$. Since by induction $V\GD = Y$, we deduce
  $V(\GD,x:A) = X$. This construction is not canonical, and there are in general
  many contexts $\GG$ such that $V\GG=X$, but the fact that we can construct one
  shows that  $V$ is essentially surjective. Since the functor $V$ is fully
  faithful and essentially surjective, it is an equivalence of categories.
\end{proof}

We can give an alternative description of $V$ in the light of
Lemma~\ref{lemma:up-disk-glob}. Indeed a term of dimension $n$ in $\GG$ is
simply a substitution $\GG\to D^n$, hence $V(\GG)_n =
\Syn{\Glob}(\GG,D^n)$. Consider the generalized nerve functor
$\Syn{\Glob}(\_,D^\bullet):\op{\Syn{\Glob}}\to\GSet$ associated to the inclusion
$D^\bullet:\op\G\to\Syn\Glob$. This functor can be seen as a functor
$\Syn{\Glob}\to \op{\GSet}$. By the previous remark, it coincides with $V$ on
objects, and hence it restricts to a functor
$\Syn{\Glob}(\GG,D^\bullet)\to\FinGSet$. Moreover, for any variable
$\GG\vdash x:A$ and any substitution $\GD\vdash\Gg:\GG$, we have the equalities
\begin{align*}
  \Gc_x\circ\Gg &= \Gc_{x[\Gg]} \\
  V(\Gg)(x) &= V(x[\Gg])
\end{align*}
which show that the functors $V$ and $\Syn{\Glob}(\_,D^\bullet)$ coincide on
morphisms. From now on, we thus identify $V$ with the generalized nerve functor
$\Syn{\Glob}(\_,D^\bullet)$, and use this point of view in more involved
situations of interest.

\begin{rem}
  \label{rem:proof-disks}
  Under the equivalence of categories of \cref{thm:syn-glob}, the globular set
  $D^n$ corresponds exactly to the context $D^n$, and the globular set $S^n$
  corresponds to the context $S^n$. This justifies the choice of the same
  notations for the contexts and the globular sets.
\end{rem}

\subsection{Models of the type theory $\Glob$.}
\label{sec:models-glob}
We can use the characterization of the syntactic category of $\Glob$ obtained in
previous section in order to study its models. This relies heavily on the fact
that $\Syn{\Glob}\equivto \op\FinGSet$ is the free finite limit completion of
the category $\op\G$. It is helpful for to start with a small discussion on Kan
extensions and their properties.

\paragraph{Properties of Kan extensions.} We first recall a few important
properties of the right Kan extensions. These are known results, on which our
constructions rely. Given a functor \(G : \C\to\D\) and an object \(d \in \D\),
there is a coma-category \(d \downarrow G\) equipped with the forgetful functor
\(\Pi_{d} : d\downarrow G \to \C\). Given a functor \(F : \C\to\E\) such that
for all \(d\), the limit of the diagram \(F\Pi_{d}\) exists, then it is a
classical result~\cite[th.6.2.1 and 6.3.7]{riehl2017category} that the right Kan
extension \(\Ran_{G}F\) exists and it pointwise, \ie it is given by the formula
\[
  (\Ran_{G}F)d = \lim\pa{d\downarrow G
    \overset{\Pi_d}{\to} \C \overset{G}{\to}\E}
\]
Define the nerve functors \(N_{G} : \op\D \to \widehat{\op\C}\) by
\(N_{G}(d) = \D(d,G\_{})\) and \(N_{F} : \op\E \to \widehat{\op\C}\) by
\(N_{F}(e) = \E(e,F\_{})\).
\begin{lemma}\label{lemma:ran-V}
  The pointwise right Kan extension is uniquely characterized by the existence
  of a natural isomorphism
  $\E(e,(\Ran_GF)d) \isoto \widehat{\C}(N_{G}d,N_{F}e)$.
\end{lemma}
\begin{proof}
  The set of cones of apex \(e\) over the diagram \(G\Pi_{d}\) is naturally
  isomorphic to \(\widehat{\C}(N_{G}d, N_{F}e)\) (see~\cite[lemma
  6.3.8]{riehl2017category}). Under this isomorphism, this is the universal
  property of the limit.
\end{proof}
\begin{lemma}\label{lemma:ran-continuous}
  If \(N_{G}\) sends finite limits to the corresponding finite colimits, and if
  the pointwise right Kan extension $\Ran_{G}F$ exists, then it preserves
  finite limits.
\end{lemma}
\begin{proof}
  Consider a finite diagram $A : I \to \C$, together with its limit $\lim A$.
  Then, for any object~$e$ in $\E$, we have the following isomorophisms, by
  Lemma~\ref{lemma:ran-V}, continuity of the Hom-functor and assumption on
  \(N_{G}\)
  \begin{align*}
    \E(e,(\Ran_G F)(\lim A)) & \isoto \widehat\C(N_{G}(\lim A),N_{F}e)\\
                             & \isoto \widehat\C(\colim(N_{G}\circ\op A), N_Fe)\\
                             & \isoto \lim(\widehat\C(N_{G}\circ\op A, N_Fe))\\
                             & \isoto \lim\E(e,\Ran_G(F\circ A))\\
                             & \isoto \E(e, \lim \Ran_G(F\circ A))
  \end{align*}
  This shows that $(\Ran_G F)(\lim A)$ satisfies the characterization of
  $\lim \Ran_{G}(F\circ A)$ given in Lemma~\ref{lemma:ran-V}.
\end{proof}
\begin{lemma}\label{lemma:ran-extension}
  If \(G\) is fully faithful and the right Kan extension \(\Ran_{G}F\) exists,
  then the universal natural transformation \(\epsilon : (\Ran_G F)G \isoto G\)
  is a natural isomorphism.
\end{lemma}
\begin{proof}
  If \(G\) is fully faithful, then \(N_{G}(Gc) = \D(Gc,G\_{}) = \C(c,\_{})\) is
  a representable presheaf over \(\op{\C}\). By the characterization given by
  Lemma~\ref{lemma:ran-V} and the Yoneda lemma,
  \begin{align*}
    \E(e,(\Ran_{G}F)Gc) \isoto \widehat{\op{\C}}(N_{G}(Gc),N_{F}e) \isoto (N_Fe)_c \isoto \E(e,F c)
  \end{align*}
  Since this isomorphism is natural in $e$, it shows that $(\Ran_{G}F)Gc$ is the
  limit of the diagram with a single point $Fc$, hence
  $(\Ran_{G}F)Gc\isoto F c$.
\end{proof}

\begin{lemma}\label{lemma:extend-natural-isos}
  If \(G\) is fully faithful, and the right Kan extension \(\Ran_{G}F\) exists,
  then for every natural transformation
  \(\alpha : \Ran_{G}F \Rightarrow \Ran_{G}F\) such that the restriction
  \(\alpha_{G}\) is a natural isomorphism, \(\alpha\) is a natural isomorphism.
\end{lemma}
\begin{proof}
  Consider the universal natural transformation
  \(\epsilon : (\Ran_{G}F)G \Rightarrow G\). By Lemma~\ref{lemma:ran-extension},
  it is an isomorphism. Then consider the composition
  \(\epsilon(\alpha_{G})^{-1} : (\Ran_{G}F)G \Rightarrow G\). By universality of
  the Kan extension, there exists \(\beta : \Ran_{G}F \Rightarrow \Ran_{G}F\)
  such that \(\epsilon\beta_{G} = \epsilon(\alpha_{G})^{-1}\). We now show that
  \(\beta\) is the inverse of \(\alpha\). We have the following equalities of
  diagrams
  \[
    \begin{tikzcd}[column sep=large]
      \D \ar[r] \ar[r, "\Ran_{G}F", bend left = 70] \ar[r, bend left = 30]
      \ar[r, bend left = 15, phantom, "\Downarrow_{\alpha}"] \ar[r, bend left =
      45, phantom, "\Downarrow_{\beta}"] \ar[rd, phantom,
      "\overset{\epsilon}{\Rightarrow}"{sloped, near start}]& \E \\
      \C\ar[u,"G"]\ar[ur,"F"'] & \phantom{A}
    \end{tikzcd}
    =
    \begin{tikzcd}[column sep=large]
      \D \ar[r] \ar[r, "\Ran_{G}F", bend left = 70] \ar[r, bend left = 30,
      phantom,"\Downarrow_{\id{}}"] \ar[rd, phantom,
      "\overset{\epsilon}{\Rightarrow}"{sloped, near start}]& \E \\
      \C\ar[u,"G"]\ar[ur,"F"'] & \phantom{A}
    \end{tikzcd}
    \qquad\qquad
    \begin{tikzcd}[column sep=large]
      \D \ar[r] \ar[r, "\Ran_{G}F", bend left = 70] \ar[r, bend left = 30]
      \ar[r, bend left = 45, phantom, "\Downarrow_{\alpha}"] \ar[r, bend left =
      15, phantom, "\Downarrow_{\beta}"] \ar[rd, phantom,
      "\overset{\epsilon}{\Rightarrow}"{sloped, near start}]& \E \\
      \C\ar[u,"G"]\ar[ur,"F"'] & \phantom{A}
    \end{tikzcd}
    =
    \begin{tikzcd}[column sep=large]
      \D \ar[r] \ar[r, "\Ran_{G}F", bend left = 70] \ar[r, bend left = 30,
      phantom,"\Downarrow_{\id{}}"] \ar[rd, phantom,
      "\overset{\epsilon}{\Rightarrow}"{sloped, near start}]& \E \\
      \C\ar[u,"G"]\ar[ur,"F"'] & \phantom{A}
    \end{tikzcd}
  \]
  By universal property of the Kan extension, this shows that
  \(\beta\alpha = \id{}\) and \(\alpha\beta = \id{}\).
\end{proof}

\paragraph{Application to the type theory.}
We now come back to \(\Syn{\Glob}\), and consider extensions along the disk
functor \(D^{\bullet} : \op\G \to \Syn{\Glob}\). Up to equivalence of
Theorem~\ref{thm:syn-glob}, this functor is the Yoneda embedding, so it is fully
faithful, and its nerve is the functor \(V\) defined in \secr{syn-glob}, which
associates to each context the globular set of its variables. By
Theorem~\ref{thm:syn-glob}, this functor is an equivalence of categories
\(\op{\Syn{\Glob}} \isoto \FinGSet\), hence it sends limits onto colimits. Given
a category \(\C\) and a functor \(F : \op\G\to \C\) we denote its nerve
\(T_{F}\), and in the case where \(\C\) is a category with families, this
functors gives a class of terms.
\begin{thm}
  \label{thm:universal-completion}
  Consider a finitely complete category $\C$ together with a functor
  $F : \op\G\to\C$, then the following pair of functors defines an
  equivalence of categories
  \[
  \begin{array}{ccrclcc@{\qquad\qquad}ccrcl}
    \_\circ D^\bullet & : & \Fun[flim]{\Syn{\Glob}}{\C} &\simeq
    &\Fun{\op\G}{\C} & : & \Ran_{D^\bullet}
  \end{array}
  \]
\end{thm}
\begin{proof}
  Since \(\C\) is finitely complete, the right Kan extension exists.
  Lemma~\ref{lemma:ran-extension} shows that for any functor $F:\op\G\to\C$, we
  have a natural transformation
  $\Ran_{D^\bullet}(F\circ D^\bullet)\circ D^{\bullet} \isoto F\circ
  D^{\bullet}$. Conversely, we show that there is a natural isomorphism
  $\Ran_{D^\bullet}(F\circ D^\bullet) \isoto F$, \ie that every functor
  preserving finite limits is isomorphic to the Kan extension of its
  restriction. Lemma~\ref{lemma:ran-extension} shows that $F$
  and $\Ran_{D^\bullet}(F\circ D^\bullet)$ coincide on all the disk objects.
  Moreover, any object $\GG$ is a finite limit of disk objects, by
  Theorem~\ref{thm:syn-glob} alongside with the fact that every presheaf is a
  colimit of representables. Since both $F$ and
  $\Ran_{D^\bullet}(F\circ D^\bullet)$ coincide on disk objects and preserve
  finite limits, they coincide on $\GG$, hence they are naturally isomorphic.
\end{proof}
\noindent Considering the Kan extension of the disk
\(D^{\bullet} : \op\G\to \Syn{\Glob}\) along itself, this theorem implies that
\(\Ran_{D^{\bullet}}{D^{\bullet}}\) is the identity functor.
Lemma~\ref{lemma:ran-V} then states that
$\Syn{\Glob}(\GG,\GD) \isoto \widehat{\G}(V\GD,V\GG)$, that is, substitutions
from $\GG$ to $\GD$ are given by the data of a variable of $\GG$ for every
variable of $\GD$ in a way that is compatible with the source and targets.
In~\secr{catt} we present a type theory whith term constructors, for which a
substitution associates a term to every variable. We discuss a of a way to
generalize this result to that theory in
\secr{algebraic-transformations-and-substitutions}. Note that
Theorem~\ref{thm:syn-glob} characterizes the syntactic category \(\Syn{\Glob}\)
as the opposite of finite globular sets. Since all the representable disks
\(D^{n}\) are themselves finite, the finite globular sets are exactly the
universal cocompletion of the category \(\G\) by finite colimits and
Theorem~\ref{thm:universal-completion} is the universal property of this
characterization. This is a standard construction (see for
instance~\cite[Theorem 5.37]{kelly1982basic}), but the tools that we have
introduced serve as preparatory work for the study of the models of \CaTT.

\paragraph{Models of the theory $\Glob$.}
We can now characterize the models of the type theory \(\Glob\). This
characterization of the models relies the following lemma

\begin{lemma}\label{lemma:ran-extension-pullbacks}
  Given a functor \(F : \Syn{\Glob} \to \C\) which preserves the terminal object
  and sends pullbacks along the display maps \(\set{\pi : D^{n} \to S^{n-1}}\),
  then \(F\) is naturally isomorphic to the pointwise right Kan extension of its
  restriction \(F \isoto \Ran_{D^{\bullet}}FD^{\bullet}\).
\end{lemma}
\begin{proof}
  We prove this property by induction on the context. Specifically, we show that
  for every object \(\Gamma\) of \(\C\) and every context \(\Delta\), we have
  the isomorphism \(\C(\Gamma,F\Delta) \isoto\widehat\G(V\Delta,T_{F}\Gamma)\),
  which characterizes the right Kan extension. We first prove this property for
  the disk objects. This is given by the Yoneda lemma: since \(VD^{n}\) is the
  representable presheaf in \(\widehat\G\), we have
  \(\widehat{\G}(VD^{n},T_{F}\Gamma) \isoto (T_{F}\Gamma)_{n} \isoto
  \C(\Gamma,FD^{n})\). We now prove this property on all spheres by induction on
  the dimension:
  \begin{itemize}
  \item The sphere \(S^{-1}\) is the empty context \(\emptycontext\), which is
    terminal in \(\Syn{\Glob}\), hence \(FS^{-1}\) is terminal in \(\C\).
    Moreover, \(VS^{-1}\) is the empty presheaf since there are no variable in
    the empty context, and thus it is initial. This proves that
    \(\C(\Gamma,FS^{-1}) = \widehat{\G}(VS^{-1},T_{F}\Gamma) = \set{\bullet}\).
  \item Assume the property for the sphere \(S^{n-1}\). The sphere \(S^{n+1}\)
    is obtained as a pushout as follows
    \[
      \begin{tikzcd}
        S^{n}\ar[r]\ar[d]\ar[dr,phantom,"\lrcorner" very near start] & D^{n}\ar[d] \\
        D^{n}\ar[r] & S^{n-1}
      \end{tikzcd}
    \]
    The equivalence is then shown by the following computation, using the
    inductive hypothesis as well as the preservation of this pushout square by
    \(F\), and the continuity properties of the involved hom-functors, and the
    fact that \(V\) is an equivalence of categories.
    \begin{align*}
      \C(\Gamma,FS^{n})
      & \isoto \C(\Gamma,F(\lim(D^{n} \to S^{n-1} \leftarrow
        D^{n})))\\
      & \isoto \lim(\C(\Gamma, FD^{n}) \to \C(\Gamma,FS^{n-1}) \leftarrow
        \C(\Gamma,FD^{n})) \\
      & \isoto \lim (\widehat{\G}(VD^{n},T_{F}\Gamma) \to
        \widehat{\G}(VS^{n-1},T_{F}\Gamma) \leftarrow
        \widehat{\G}(VD^{n},T_{F}\Gamma)) \\
      & \isoto \widehat{\G} (\colim (VD^{n} \leftarrow VS^{n-1} \to VD^{n}),
        T_{F}\Gamma) \\
      & \isoto \widehat{\G} (V(\lim(D^{n} \to S^{n-1}\leftarrow
        D^{n})),T_{F}\Gamma)\\
      & \isoto \widehat{\G}(VS^{n},T_{F}\Gamma)
    \end{align*}
  \end{itemize}
  We now prove the desired property by induction on the contexts.
  \begin{itemize}
  \item A context of length \(0\) is the empty context, which is also the sphere
    \(S^{-1}\), for which we have already proven the property.
  \item The context \((\Delta,A)\) is obtained as the following pullback
    \[
      \begin{tikzcd}
        (\Delta,A) \ar[r] \ar[d]\ar[dr,phantom,"\lrcorner"very near start ]& D^{n}\ar[d] \\
        \Delta\ar[r,"\chi_{A}"] & S^{n-1}
      \end{tikzcd}
    \]
    Then, by a computation similar to the case of the sphere, using induction,
    the preservation of this pullback by \(F\), and continuity property of the
    hom-functors, we have
    \begin{align*}
      \C(\Gamma,F(\Delta,A))
      & \isoto \lim (\widehat{\G}(V\Delta,T_{F}\Gamma) \to
        \widehat{\G}(VS^{n-1},T_{F}\Gamma) \leftarrow
        \widehat{\G}(VD^{n},T_{F}\Gamma)) \\
      & \isoto \widehat{\G}(V(\Delta,A),T_{F}\Gamma) \tag*{\qedhere}
    \end{align*}
  \end{itemize}
\end{proof}

\begin{restatable}{thm}{modelsglob}
\label{thm:models-glob}
  There is an equivalence of categories $\Mod{\Syn\Glob}\equivto \GSet$.
\end{restatable}
\begin{proof}
  Theorem~\ref{thm:universal-completion} shows that
  \(\GSet \equivto \Fun[flim]{\Syn{\Glob}}{\Set}\), so it suffices to prove that
  the models are equivalent to that category. Define the pair of functors
  \[
    \begin{tikzcd}
      \Mod{\Syn\Glob} \ar[r, shift left] &
      \Fun[flim]{\Syn{\Glob}}{\Set} \ar[l, shift left]
    \end{tikzcd}
  \]
  associating to every morphism of categories with families
  \(F : \Syn\Glob \to \Set\) the right Kan extension, and to every finite limit
  preserving functor \(G : \Syn{\Glob} \to \Set\), the morphism of category with
  families that it defines by Lemma~\ref{lemma:models-cwf}. Then
  Lemmas~\ref{lemma:ran-extension} and~\ref{lemma:ran-extension-pullbacks} show
  that this defines an equivalence of categories.
\end{proof}

\noindent
Note that this proof consists in two parts:
\begin{enumerate}
\item restate the models as being equivalent to the functors
  preserving finite limits,
\item use standard categorical machinery to show that these are equivalent to
  globular sets.
\end{enumerate}
The first step abstracts away the category with families structure, to restate
the problem as a plain category theory one. This approach is contingent to
recognizing the syntactic category to be $\op\FinGSet$ (\cref{thm:syn-glob}) and
does not generalize. For this reason, we give a reformulation of the proof,
using an initiality theorem for the category $\Syn{\Glob}$, which provides a
better account of the category with families structure.

\subsection{Globular categories with families.}
\label{sec:glob-cwf}
We now introduce the notion of \emph{globular categories with
  families}, which are particular categories with families that share
a lot of structural properties with the category $\Syn{\Glob}$.

\begin{defi}
  A globular category with families is a category with families $\C$
  equipped with two families of objects $(S^{n-1})_{n\in\N}$ and
  $(D^n)_{n\in\N}$ and a family of types $U_n\in\Ty_{S^{n}}$ such that
  the following equations are satisfied
  \begin{align*}
    S^{-1} &= \emptycontext & \\
    S^{n} &= (D^{n}, U_{n-1}[\partial_n]) & D^{n} &= (S^{n-1},U_{n-1})
  \end{align*}
  where $\partial_n$ denotes the projection map
  $\partial_n = \pi_{S^{n-1}, U_{n-1}}: D^{n}\to S^{n-1}$.
\end{defi}

We suppose given a globular category with families $\C$, we denote
$\uhemisphere_n: S^n \to D^n$ the display map given by the category with
families. Moreover, we also define another map $\lhemisphere_n: S^n\to D^n$ by
\(\lhemisphere_n = \sub{\partial_n\uhemisphere_n, p_{n}}\), where $p_{n}$
denotes the universal term of $S^{n}$ of type
$U_{n-1}[\partial_n\uhemisphere_n]$ obtained by context comprehension. With the
help of these maps, we define a pair of maps
\begin{align*}
  \begin{array}{ccrcl}
    s_n & : & D^{n+1} & \to & D^n \\
    s_n & = & \multicolumn{3}{l}{\uhemisphere_n\partial_{n+1}}
  \end{array}
  &&
  \begin{array}{ccrcl}
    t_n & : & D^{n+1} & \to & D^n \\
    t_n & = & \multicolumn{3}{l}{\lhemisphere_n\partial_{n+1}}
  \end{array}
\end{align*}

\begin{lemma}
  The maps $s_n$ and $t_n$ satisfy the globular relations
  \eqref{eq:glob-rel}.
\end{lemma}
\begin{proof}\belowdisplayskip=-12pt
  Using the properties of substitution extension in categories with
  families, we have the following computations.
  \begin{align*}
    s_it_{i+1}
    &= \uhemisphere_i\partial_{i+1}\lhemisphere_{i+1}\partial_{i+2}
    & t_it_{i+1}
    &= \lhemisphere_i\partial_{i+1}\lhemisphere_{i+1}\partial_{i+2} \\
    &= \uhemisphere_i\partial_{i+1}\sub{\partial_{i+1}\uhemisphere_{i+1},p_{i+1}}\partial_{i+2}
    &
    &= \lhemisphere_i\partial_{i+1}\sub{\partial_{i+1}\uhemisphere_{i+1},p_{i+1}}\partial_{i+2} \\
    &= \uhemisphere_i\partial_{i+1}\uhemisphere_{i+1}\partial_{i+2}
    &
    &= \lhemisphere_i\partial_{i+1}\uhemisphere_{i+1}\partial_{i+2} \\
    &= s_is_{i+1}
    &
    &= t_is_{i+1}
  \end{align*} \qedhere
\end{proof}

\noindent This shows that given a category with families $\C$, we can
construct a functor $G_\C : \op\G\to\C$, by setting $G_\C(n) = D^n$,
$G_\C(\sigma_n) = s_n$ and $G_\C(\tau_n) = t_n$. We call it the
\emph{induced globular structure} of the globular category with families
$\C$.

\begin{defi}
  A morphism of globular category with families between $\C$ and $\D$
  is a morphism $f$ of categories with families between them, along with
  a natural transformation between their induced globular structures
  as follows
  \[
    \begin{tikzcd}
      \C \ar[r,"f"]\ar[dr,phantom,"\Rightarrow"{sloped, near start}] & \D \\
      \op\G \ar[u,"G_\C"]\ar[ur,"G_\D"'] & \phantom{A}
    \end{tikzcd}
  \]
  We denote $\gCwF$ the category of globular categories with families
  defined this way. We also define a $2$-morphism between two morphisms
  of globular categories with families $(f, \alpha)$ and $(g,\beta)$
  to be a natural transformation $\gamma: f \Rightarrow g$ which
  commutes with the natural transformations of the induced globular
  structures, that is such that we have the following equality
  \[
    \begin{tikzcd}
      \C
      \ar[rr, "f"]
      \ar[rrd, phantom, "\overset{\alpha}{\Rightarrow}"{sloped,
      near start}]
      && \D \\
      \op\G \ar[u,"G_{\C}"]\ar[urr, "G_{\D}"'] && \phantom{A}
    \end{tikzcd}
    \quad=\quad
    \begin{tikzcd}
      \C
      \ar[rr,"g" description]
      \ar[rr, bend left = 50, "f"]
      \ar[rr, bend left = 25, "\Downarrow_{\gamma}", phantom]
      \ar[rrd, phantom, "\overset{\beta}{\Rightarrow}"{sloped,
      near start}]
      && \D \\
      \op\G \ar[u,"G_{\C}"]\ar[urr, "G_{\D}"'] && \phantom{A}
    \end{tikzcd}
  \]
  We denote $\gCwFh$ the resulting $2$-category.
\end{defi}

\paragraph{Initiality of $\Syn{\Glob}$.}
We now reformulate \Cref{lemma:ran-extension} about Kan extensions in terms of
the notion of globular category with families. This provides a local form of
$2$-categorical initiality of the category $\Syn{\Glob}$ in the $2$-category of
globular categories with families.

\begin{lemma}\label{lemma:ran-exists}
  For every globular category with families \(\C\),
  \(\Ran_{D^{\bullet}}G_{\C} : \Syn{\Glob} \to \C\) exists and is a pointwise.
\end{lemma}
\begin{proof}
  We construct by induction a functor \(r : \Syn{\Glob} \to \C\) such that
  \(r(D^{n}) = G_{\C}(n)\), and \(r\) preserves the pullbacks along the
  projection maps \(\pi : D^{n} \to S^{n-1}\) for all \(n\in\N\), and preserves
  the terminal object. We define by induction on \(\Gamma\), an object
  \(r(\Gamma)\) of \(\C\) such that \(r(D^{n}) = G_{\C}(n)\) and
  \(r(S^{n-1}) = S^{n-1}\), along with, for every morphism
  \(m : \Gamma \to S^{n}\) (\resp \(m : \Gamma \to D^{n}\)), a corresponding
  morphism \(m' : r(\Gamma) \to S^{n}\) (\resp \(m' : r(\Gamma) \to D^{n}\)),
  such that \(\partial_{n}' = \partial_{n}\), \(\smile_{n}' = \smile_{n}\),
  \(\frown_{n}' = \frown_{n}\) and \(\id{S^{n}}' = \id{S^{n}}\).
  \begin{itemize}
  \item For the empty context, \(\emptycontext\), we define \(r(\emptycontext)\)
    to be the terminal object \(\emptycontext\) in the category \(\C\). Since
    \(\emptycontext = S^{-1}\) both in \(\Syn{\Glob}\) and in \(\C\), this gives
    the commutation property that we want. Moreover, the only map from the empty
    context to a disk or a sphere context is the identity map
    \(\id{\emptycontext} = \emptycontext \to \emptycontext\), and we define
    \(\id{\emptycontext}' = \id{\emptycontext}\).
  \item For a context of the form \((\Gamma,A)\), where \(\dim A = k\), we
    define \(r(\Gamma,A) = (r(\Gamma), U_{k}[\chi_{A}'])\). Moreover, given a
    morphism \(\chi_{x} = (\Gamma,A) \to D^{n}\), either the variable \(x\) is a
    variable in \(\Gamma\), in which case \(\chi_{x}\) factors as
    \(\chi_{y}\pi_{\Gamma,A}\) (where \(y\) denotes the same variable \(x\) but
    seen as variable in \(\Gamma\)), and we chose
    \(\chi_{x}' = \chi_{y}'\pi_{r(\Gamma),U_{k}[\chi_{A}']}\), or \(x\) is the
    last variable in \(\Gamma\), in which case we have the pullback square on
    the left, and we chose \(\chi_{x}'\) as displayed on the pullback square on
    the right.
    \[
      \begin{tikzcd}
        (\Gamma,A) \ar[d]\ar[r,"\chi_{x}"]\ar[dr,phantom,"\lrcorner" very near start] & D^{k+1}\ar[d,"\partial_{k+1}"] \\
        \Gamma \ar[r,"\chi_{A}"]& S^{k}
      \end{tikzcd}
      \qquad\qquad
      \begin{tikzcd}
        r(\Gamma,A) \ar[d]\ar[r,"\chi_{x}'"]\ar[dr,phantom,"\lrcorner" very near start] & D^{k+1}\ar[d,"\partial_{k+1}"] \\
        r(\Gamma) \ar[r,"\chi_{A}'"]& S^{k}
      \end{tikzcd}
    \]
    Given a map \(m : (\Gamma,A) \to S^{n}\), if \(n = -1\), then, m is the
    unique map to the terminal object, and we chose \(m'\) to be the unique map
    to the terminal object in \(\C\), otherwise,
    \(S^{n} = (D^{n},U_{n-1}[\partial_{n}])\), and we have
    \(m = \sub{\frown_{k}m,t}\). We then chose
    \[
      m' = \sub{(\frown_{k}m)',p_{S^{n-1},U_{n-1}}[\chi_{t}']}.
    \]

    In the case where \((\Gamma,A) = D^{k+1}\), then \(\Gamma = S^{k}\) and
    \(A = U_{k}\). Then \(\chi_{U_{k}} = \id{S^{k}}\), so by induction
    \(\chi_{U_{k}}' = \id{S^{k}}\), and thus,
    \(r(D^{k+1}) = (S^{k},U_{k}) = D^{k+1}.\) Moreover, given the map
    \(\partial_{k+1} : D^{k+1} \to S^{k}\), we have
    \((\frown_{k}\partial_{k+1})' = \frown_{k}'\partial_{k+1}\), and
    \(\chi_{p_k[\partial_{k+1}]}' = \chi_{p_{k}}'\partial_{k+1}\). Moreover,
  \(\chi_{p_{k}} = \smile_{k}\). By induction, this shows that we have
  \begin{align*}
    \partial_{k+1}'
    & =
      \sub{(\frown_{k}\partial_{k+1})',p_{S^{k},U_{k}}[\chi_{p_k[\partial_{k+1}]}']}\\
    &= \sub{\frown_{k}\partial_{k+1},p_{S^{k},U_{k}}[\smile_{k}\partial_{k+1}]}\\
    &= \sub{\frown_{k}\partial_{k+1},p_k[\partial_{k+1}]}\\
    & = \partial_{k+1}
  \end{align*}

  In the case where \((\Gamma,A) = S^{k+1}\), then \(\Gamma = D^{k+1}\) and
  \(A = U_{k}{\partial_{k+1}}\), so \(\chi_{A} = \partial_{k+1}\), and by
  induction \(\chi'_{A}=\partial_{k+1}\), and \(r(\Gamma) = D^{k+1}\). This
  shows that \(r(S^{k+1}) = (D^{k+1},U_{k}[\partial_{k+1}]) = S^{k+1}\).
  Moreover, the map \(\frown_{k+1}\) is the display map, characterising the
  previous to last variable in \(S^{n+1}\), hence \(\frown_{k+1}'\) is defined
  to be \(\id{D^{k+1}}\frown_{k+1} = \frown_{k+1}\). We have the following
  square defines a pullback both in \(\Syn{\Glob}\) and in \(\C\), showing the
  equality for \(\smile_{k+1}' = \smile_{k}\)
  \[
    \begin{tikzcd}
      S^{k+1} \ar[d]\ar[r,"\frown_{k+1}"]\ar[dr,phantom,"\lrcorner" very near start] & D^{k+1}\ar[d,"\partial_{k+1}"] \\
      D^{k+1} \ar[r,"\partial_{k+1}"]& S^{k}
    \end{tikzcd}
  \]
  Finally, note that we have \(\chi_{p_{k+1}} = \smile_{k+1}\), and so by
  definition and by induction hypothesis, we have
  \begin{align*}
    \id{S^{k+1}}' &= \sub{\frown_{k+1},p_{k+1}}' \\
                  &=\sub{\frown_{k+1}',p_{S^{k},U_{k}}[\smile_{k+1}']} \\
                  &=\sub{\frown_{k+1},p_{S^{k},U_{k}}[\smile_{k+1}]} \\
                  &=\sub{\frown_{k+1},p_{k+1}} \\
                  &=\id{S^{k+1}}\\
  \end{align*}
\end{itemize}
We now define for a substitution \(\gamma : \Delta\to\Gamma\), the map
\(r(\gamma) : r(\Delta)\to r(\Gamma)\).
\begin{itemize}
\item For \(\gamma = \sub{}\) the empty substitution,
  \(\Gamma = \emptycontext\), so \(r(\Gamma) = \emptycontext\) is terminal, and
  we define \(r(\gamma)\) to be the unique map \(r(\Delta)\to r(\Gamma)\).
\item For the map \(\sub{\gamma,t} : \Delta \to (\Gamma,A)\), denote
  \(n = \dim A\). Define
  \(r(\sub{\gamma,t}) = \sub{r(\gamma),p_{S^{n},U_{n}}[\chi_{t}']}\)
\end{itemize}
This assignment is functorial and for every map \(m : \Gamma \to S^{n}\) or
\(m : \Gamma\to D^{n}\), we have \(r(m) = m'\). Moreover, by definition, \(r\)
preserves the terminal object and the pullbacks along the maps
\(\set{\partial_{n}}\), hence by Lemma~\ref{lemma:ran-extension-pullbacks}, we
then have \(r \isoto \Ran_{D^{\bullet}}G_{\C}\).
\end{proof}

\begin{lemma}\label{lemma:ran-definition-glob}
  For a globular category with families $\C$, there is a morphism of category
  with families
  \[
    (\Ran_{D^{\bullet}}G_{\C}, \epsilon(\C)) : \Syn{\Glob} \to \C
  \]
  where the transformation
  $\epsilon(\C) : \Ran_{D^{\bullet}}G_{\C} \circ D^{\bullet} \to G_{\C}$ is a
  natural isomorphism.
\end{lemma}
\begin{proof}
  the Kan extension $\Ran_{D^{\bullet}} G_{\C}$ exists by
  \cref{lemma:ran-exists}. \cref{lemma:ran-continuous} shows that
  $\Ran_{D^{\bullet}} G_{\C}$ preserves the limits, and hence by
  \cref{lemma:models-cwf}, it defines a unique morphism of categories with
  families. Let
  $\epsilon(\C): \Ran_{D^{\bullet}}G_{\C} \circ D^{\bullet} \Rightarrow G_{\C}$
  be the universal natural transformation obtained as the Kan extension.
  Lemma~\ref{lemma:ran-extension} then shows that $\epsilon(\C)$ is a natural
  isomorphism.
\end{proof}

\begin{thm}[local initiality of the syntactic category]
  \label{thm:weak-initiality-glob}
  The morphism of globular categories with families
  $(\Ran_{D^{\bullet}}G_{\C}, \epsilon(\C)) : \Syn{\Glob} \to \C$ is a terminal
  object in the category $\gCwFh(\Syn{\Glob},\C)$.
\end{thm}
\begin{proof}
  This is exactly the universal property of the right Kan extension. Indeed,
  consider a morphism of globular category with families
  $(F, \alpha) : \Syn{\Glob} \to \C$. The universal property of the right Kan
  extension lets us construct a natural transformation
  $\beta : F \Rightarrow \Ran_{D^{\bullet}}G_{\C}$ such that we have the
  following equality:
  \[
    \begin{tikzcd}
      \Syn{\Glob} \ar[rr, "F"] \ar[rrd, phantom,
      "\overset{\alpha}{\Rightarrow}"{sloped, very near start}]
      && \C \\
      \G \ar[u,"D^{\bullet}"]\ar[urr, "G_{\C}"'] && \phantom{A}
    \end{tikzcd}
    \quad=\quad
    \begin{tikzcd}
      \Syn{\Glob} \ar[rr,"\Ran_{D^{\bullet}}G_{\C}"] \ar[rr, bend left = 60,
      "F"] \ar[rr, bend left = 40, "\Downarrow_{\beta}", phantom] \ar[rrd,
      phantom, "\overset{\epsilon(\C)}{\Rightarrow}"{sloped, very near start}]
      && \C \\
      \G \ar[u,"D^{\bullet}"]\ar[urr, "G_{\C}"'] && \phantom{A}
    \end{tikzcd}
  \]
  Thus, $\beta$ is a natural transformation satisfying
  $\beta : (F,\alpha) \to (\Ran_{D^{\bullet}}G_{\C},\epsilon(\C))$.
\end{proof}

\paragraph{Models of the theory $\Glob$.}
Using the machinery of globular categories with families, we can give
a new proof that the models of $\Glob$ are the globular sets, in a way
that is more generalisable to more complicated cases. For this, we
consider the forgetful functor $\forget : \gCwF \to \CwF$, and we
consider, for a given category with families $\C$, the category
$\fiber \C$ whose objects are the globular categories with families
$\D$ such that $\mathcal{U}(\D) = \C$, morphisms are the morphisms of
globular categories with families which project onto $\id\C$ by
$\mathcal{U}$. Note that since $\mathcal{U}$ is an isofibration, this
construction is invariant by equivalence of categories. Intuitively,
$\fiber \C$ is the category of choices of disk and sphere objects in
$\C$ in a way that is compatible with the structure of category with
families.

\begin{prop}\label{prop:models-gcwf}
  Consider a category with families $\C$, there is an equivalence of
  categories $\CwFh(\Syn\Glob,\C) \simeq \fiber \C$.
\end{prop}
\begin{proof}
  We build a pair of functors
  \[
    \begin{tikzcd}
      \CwFh(\Syn{\Glob},\C) \ar[r, shift left, "\ind"]
      & \ar[l,shift left, "\Ran_{D^{\bullet}}"]\fiber \C
    \end{tikzcd}
  \]
  and show that they define an equivalence of categories.

  \emph{Definition of the functor $\ind$.} Consider a functor of
  category with families $F : \Syn{\Glob}\to \C$, we define a globular category
  with families structure on $\C$ that we call $\ind(F)$ by choosing the disk
  and sphere objects to be $F(D^{n})$ and $F(S^{n-1})$, so $(F,\id{})$ defines a
  morphism of globular categories with families $\Syn{\Glob}\to\ind(F)$. Given
  two morphisms of categories with families $F,G : \Syn{\Glob}\to \C$ and a
  $2$-cell $\alpha : F \Rightarrow G$, we define an induced morphism of globular
  category with families
  $\ind(\alpha) = (\id\C,\alpha_{D^{\bullet}}) : \ind(F) \to \ind(G)$.

  \emph{Definition of the functor $\Ran_{D^{\bullet}}$.} As the notation
  suggests, this functor is constructed as a right Kan extension. Given an
  object $\D$ in $\fiber \C$, $\Ran_{D^{\bullet}} G_{\D} : \Syn{\Glob} \to \C$
  exists and is a morphism of categories with families by
  \cref{lemma:ran-definition-glob}. Consider a morphism
  $(\id{\C},\alpha) : \D \to \D'$ in $\fiber \C$, we have
  $\alpha\epsilon(\D) \in\Cath((\Ran_{D^{\bullet}}G_{\D})D^{\bullet}, G_{\D'})$.
  We then define \(\Ran_{D^{\bullet}}(\alpha)\) by universal property of the Kan
  extension \(\Ran_{D^{\bullet}}G_{\D'}\), to be the unique map such that
  \(\epsilon(\D')\Ran_{D^{\bullet}}(\alpha) = \alpha\epsilon(\D)\).

  \emph{Equivalence $\ind\circ\Ran_{D^{\bullet}} \simeq \id{}$.}
    Consider an object $\D$ in $\fiber\C$, then
    \cref{lemma:ran-definition-glob} shows that we have a morphism of
    globular category with families
    $(\Ran_{\D}G_{\D},\epsilon(\D)) : \Syn{\Glob} \to \D$, with
    $\epsilon(\D) : (\Ran_{\D}G_{\D})\circ D^{\bullet} \Rightarrow
    G_{\D}$ a natural isomorphism. We then have the following
    commutative diagram of globular categories with families
    \[
      \begin{tikzcd}
        \Syn{\Glob}
        \ar[r,"\pa{\Ran_{D^{\bullet}}G_{\D},\epsilon(\D)}"]
        \ar[rd, "\pa{\Ran_{D^{\bullet}}G_{\D},\id{}}"']
        & \D \\
        & \ind(\Ran_{D^{\bullet}}G_{\D})\ar[u,"\pa{\id{},\epsilon(\D)}"']
      \end{tikzcd}
    \]
    Since $\epsilon(\D)$ is a natural isomorphism, $(\id{},\epsilon(\D))$ is an
    isomorphism in the category $\fiber \C$, whose inverse is
    $(\id{},\epsilon(\D)^{-1})$. This therefore shows that we have a family of
    isomorphisms
    $(\id{},\epsilon(\D)) : \ind(\Ran_{D^{\bullet}}G_{\D}) \simeq \D$. This
    family is natural in $\D$, as witnessed by the equality
    $\epsilon(\D')\Ran_{D^{\bullet}}(\alpha)_{D^{\bullet}} =
    \alpha_{D^{\bullet}}\epsilon(\D)$ characterizing
    $\Ran_{D^{\bullet}}(\alpha)$.

    \emph{Equivalence $\Ran_{D^{\bullet}}\circ\ind \simeq \id{}$.} A
    morphism of categories with families $F : \Syn{\Glob} \to \C$ defines a
    morphism of globular categories with families
    $(F, \id{}) : \Syn{\Glob} \to \ind(F)$. Then by
    \cref{thm:weak-initiality-glob}, we have a natural transformation
    $\alpha(F) : F \Rightarrow \Ran_{D^{\bullet}}G_{\ind(F)}$ obtained by
    universality of the Kan extension. Since \(\epsilon(\ind(F))\) is an
    isomorphism, so is \(\alpha(F)_{D^{\bullet}}\). Consider the isomorphism
    \(\gamma: F \isoto \Ran_{D^{\bullet}}(G_{\ind(F)})\) obtained by
    Lemma~\ref{lemma:ran-extension-pullbacks}. Then
    \((\alpha(F)\gamma^{-1})_{D^{\bullet}}\) is a natural isomorphism,
    Lemma~\ref{lemma:extend-natural-isos} then shows that
    \(\alpha(F)\gamma^{-1}\) is a natural isomorphism, hence so is
    \(\alpha(F)\). We now show that the family \(\alpha(F)\) is natural in
    \(F\): Given two morphisms of categories with families
    \(F,G : \Syn{\Glob} \to \C\), we consider the two following diagram, whose
    compositions are both equal to \(\ind(\beta)\), using the equations that
    characterize \(\epsilon\) and \(\Ran_{D^{\bullet}}(\ind(\beta))\).
    \[
     \begin{tikzcd}[column sep=large]
        \Syn{\Glob} \ar[rr] \ar[rr, "F", bend left =
        70] \ar[rr, bend left = 30] \ar[rr, bend left = 15, phantom,
        "{\scriptscriptstyle\alpha(G)}"] \ar[rr, bend left = 50,
        phantom, "{\scriptscriptstyle\beta}"] \ar[rrd, phantom,
        "{\scriptscriptstyle\epsilon(\ind(G))}"{very near start}]
        && \C \\
        \op\G\ar[u,"D^{\bullet}"]\ar[urr,"\ind(F)"'] && \phantom{(A)}
      \end{tikzcd}=
     \begin{tikzcd}[column sep=large]
       \Syn{\Glob} \ar[rr] \ar[rr, "F", bend left = 70] \ar[rr, bend left = 30]
       \ar[rr, bend left = 50, phantom, "{\scriptscriptstyle\alpha(F)}"] \ar[rr,
       bend left = 15, phantom,
       "{\scriptscriptstyle\Ran_{D^{\bullet}}(\ind(\beta))}"] \ar[rrd, phantom,
       "{\scriptscriptstyle\epsilon(\ind(G))}"{very near start}]
       && \C \\
       \op\G\ar[u,"D^{\bullet}"]\ar[urr,"\ind(F)"'] && \phantom{(A)}
      \end{tikzcd}
    \]
    By universality of the Kan extension, this shows the equation
    \(\alpha(G)\beta = \Ran_{D^{\bullet}}(\ind(\beta))\alpha(F)\), which is
    exactly the naturality of \(\alpha\). \qedhere

\end{proof}

\paragraph{Set-theoretic models of $\Glob$.}
Applying the previous result lets us give a second proof of the characterization
of the models of the theory $\Glob$.

\begin{prop}\label{prop:set-fiber-glob}
  There is an equivalence of categories $\fiber \Set \simeq \GSet$.
\end{prop}
\begin{proof}
  We define a functor $\mathcal{M} : \GSet \to \fiber \Set$, as
  follows: for a globular set $G : \op\G\to \Set$, we consider
  $\mathcal{M}(G)$ to be the globular category with families on $\Set$
  obtained by defining
  \begin{align*}
    D^n &= \G(n) = \GSet(\Yoneda(n), G) & S^{n} &=
                                                  \GSet(\partial\Yoneda(n),G)
    & U_n &= (i_n^{1}\langle x\rangle)_{x\in \Yoneda(n)}
  \end{align*}
  where $\partial\Yoneda(n)$ is the globular set obtained by removing
  its top dimensional cell to $\Yoneda(n)$ and
  $i_n : \partial\Yoneda(n) \to \Yoneda(n)$ is the inclusion morphism
  and $i_{n}^{-1}\langle x\rangle$ is the preimage of $i_{n}$ at $x$.
  By definition, the globular structure induced by $\mathcal{M}(G)$ on
  $\Set$ is exactly $G$. Thus, $\mathcal{M}$ sends a morphism of
  globular sets $\alpha : G \to G'$ to $\alpha$ seen as a morphism in
  $\fiber \Set$. We show that $\mathcal{M}$ defines an equivalence of
  categories, by showing that it is fully faithful and essentially
  surjective.

  Consider an object $M$ in $\fiber \Set$, it is a globular category
  with families, so it induces a globular structure $G_M$ on $\Set$,
  such that $\mathcal{M}(G_M) = M$. Hence $\mathcal{M}$ is essentially
  surjective. Moreover, considering two globular sets $G$, $G'$, the
  moprhisms $\mathcal{M}(G) \to \mathcal{M}(G')$ in $\fiber\Set$ are
  by definition exaclty the same as the morphisms of globular sets
  $G \to G'$. Hence $\mathcal{M}$ is fully faithful. $\mathcal{M}$ is
  thus an equivalence of categories.
\end{proof}
\modelsglob*
\begin{proof} \belowdisplayskip=-12pt
  By \cref{prop:models-gcwf}
  and \cref{prop:set-fiber-glob} we have the following equivalences of
  categories
  \begin{align*}
    \Mod{\Syn\Glob}=\CwFh(\Syn\Glob, \Set) \simeq \fiber\Set \simeq
    \GSet
  \end{align*} \qedhere
\end{proof}



\section{The Grothendieck-Maltsiniotis definition of $\omega$-categories}
\label{sec:gm-def}
This entire section is a quick presentation of the definition of weak
$\omega$-categories given by Maltsionitis~\cite{maltsiniotis}, based
on the definition of a weak $\omega$-groupoid introduced by
Grothendieck~\cite{pursuing-stacks}. We introduce here the notions on
which the type theory $\CaTT$ relies, as well as the notations. For a
more in-depth study of this definition, we refer the reader to the
original article by Maltsionitis~\cite{maltsiniotis} or the PhD thesis
of Ara~\cite{ara2010groupoides}.

\subsection{Pasting schemes.}
\label{sec:pasting-schemes}
We define a subcategory of globular sets, called the \emph{pasting
  schemes}. These are meant to represent composable situations in a
globular set, and thus serve as the arities of the operations in
$\omega$-categories.

\paragraph{Globular sums.}
Consider a category $\C$. A \emph{globular structure} on a
category~$\C$ consists in a functor~$F:\G\to\C$. When given such a
structure, we often denote respectively by $D^n$, $\sigma_i$ and
$\tau_i$ the images under $F$ of $n$, $\sigma_i$ and $\tau_i$. When
there is no ambiguity, we may write $\sigma$ and $\tau$, leaving the
index implicit, moreover, we also write $\sigma$ (\resp $\tau$) to
denote a composite of maps of the form~$\sigma$ (\resp $\tau$). In the
category $\C$, a \emph{globular sum} is a colimit of a diagram of the
form
\[
  \begin{tikzcd}[column sep = small]
    D^{i_1} & & D^{i_2} & & \ar[d,phantom,"\cdots"] & & D^{i_k} \\
    & D^{j_1} \ar[ur,"\sigma"']\ar[ul,"\tau"] & & D^{j_2} \ar[ul,"\tau"] & \phantom{x} & D^{j_{k-1}} \ar[ur,"\sigma"'] &
  \end{tikzcd}
\]
In this diagram, we always assume that the iterated sources $\Gs$ and
the iterated targets $\Gt$ are not identity, so that we always have
the inequality $i_k>j_k<i_{k+1}$. Given a non-canonical diagram (\ie a
diagram which can contain identities), one can contract away all the
identity morphisms without changing the colimit of the diagram. It is
be useful to encode such a colimit by its \emph{table of dimensions}
\[
  \left(
    \begin{array}{ccccccc}
      i_1 & & i_2 & & \cdots & & i_k\\
          & j_1 & & j_2 & \cdots & j_{k-1} &
    \end{array}
  \right)
\]
Dually, if a category $\C$ is endowed with a contravariant functor
$F:\op{\G}\to\C$, called a \emph{contravariant globular structure}, we
denote respectively by $D^n$, $s$ and $t$ the images by $F$ of $n$,
$\sigma$ and $\tau$. We call a \emph{globular product} a limit of the
diagram of the following form, which we can also encode with a table
of dimensions.
\[
  \begin{tikzcd}[column sep = small]
    D^{i_1} \ar[dr,"t"']& & D^{i_2} \ar[dl,"s"]\ar[dr,"t"'] & & \ar[d,phantom,"\cdots"] & & D^{i_k} \ar[dl, "s"] \\
    & D^{j_1} & & D^{j_2} & \phantom{x} & D^{j_{k-1}} &
  \end{tikzcd}
\]
If $\C$ has a globular structure and $\catD$ has a contravariant
globular structure, we say that a globular sum in $\C$ and a globular
product in $\catD$ are \emph{dual} to each other if they share the
same table of dimensions.

\paragraph{The category of pasting schemes.}
The category $\GSet$ is equipped with a globular structure given by
the Yoneda embedding $\Yoneda : \G\to\GSet$, defined by
$\Yoneda(n) = \GSet(\_,n)$. In this situation we call all the globular
sets that are obtained as globular sums the \emph{pasting schemes},
and we denote by $\COH_0$ the full subcategory of $\GSet$ whose
objects are the pasting schemes. Note that since the globular sum
diagrams are finite, the pasting schemes are finite globular sets. A
few examples and counter-examples of pasting schemes are depicted in
Figure~\ref{fig:ps-order}, using the diagrammatic notation for finite
globular sets.

\paragraph{A relation characterizing the pasting schemes.}
Apart from tables of dimensions, there are several ways of parametrizing pasting
schemes using combinatorial structures, such as Batanin
trees~\cite{batanin1998monoidal}. In fact these trees also assemble into a
category, which can be proved to be equivalent to the category
$\COH_0$~\cite{ara2010groupoides, berger2002cellular, joyal1997disks}. Other
combinatorial descriptions of pasting schemes are also possible, such as Dyck
words, or non-decreasing parking functions, as well as inductive definitions. We
refer the reader to~\cite{benjamin2020type} for a brief presentation of these
views. We focus here on a characterization due to Finster and
Mimram~\cite{catt}, using a binary relation.

\begin{defi}
  Consider a globular set $G$, we introduce the relation $\triangleleft$ on its
  set of cells to be the transitive closure of the relation generated, for every
  cell $x$ of $G$, by
  \[
  s(x) \triangleleft x \triangleleft t(x)
  \]
\end{defi}

\noindent
This relation can be used to characterize the pasting schemes among all the
finite globular sets:

\begin{thm}[Finster, Mimram~\cite{catt}]
  \label{thm:ps-|>}
  The pasting schemes are exactly the non-empty finite globular sets
  such that $\triangleleft$ is total and antisymmetric, that is, when
  we have
  \[
  x\neq y \iff (x\triangleleft y \text{ or } y\triangleleft x)
  \]
  We also say in this case that the globular set is $\triangleleft$-linear.
\end{thm}

\noindent
We refer the reader to the original article~\cite{catt} for a proof of this
theorem, and illustrate the relation $\triangleleft$ on a few examples, pasting
schemes and non pasting schemes, in Figure~\ref{fig:ps-order}.

\begin{figure}[!h]
  \centering
  \begin{tabular}{|c|c|c|}
    \hline
    globular set & relation $\triangleleft$ & pasting scheme? \\
    \hline\hline
    $
    \begin{tikzcd}[ampersand replacement =\&]
      x \ar[r,"f"] \& y \ar[r,"g"] \& z
    \end{tikzcd}
    $
    &
    $x\triangleleft f\triangleleft y \triangleleft g \triangleleft z$
    &
    yes\\
    \hline
    $
    \begin{tikzcd}[ampersand replacement =\&]
      x \ar[r,"f"] \& y \& z\ar[l,"g" above]
    \end{tikzcd}
    $
    &
    $
    \begin{tikzcd}[ampersand replacement =\&, row sep = tiny, column sep = tiny]
      x\ar[r, phantom, "\triangleleft"] \& f\ar[rd, phantom, "\triangleleft"{sloped}] \& \\
      \& \& y \\
      z\ar[r, phantom, "\triangleleft"] \& g\ar[ru, phantom, "\triangleleft"{sloped}]\&
    \end{tikzcd}
    $
    &
    no \\
    \hline
    $
    \begin{tikzcd}[ampersand replacement = \&]
      x \ar[r, bend left = 60, "f"]\ar[r, bend right = 60, "f''"below]\ar[r, "f'"description]
      \ar[r, bend left = 25, phantom, "\Downarrow{\scriptstyle\alpha}"]
      \ar[r, bend right = 25, phantom, "\Downarrow{\scriptstyle\beta}"]
      \& y\ar[r, "g"] \& z
    \end{tikzcd}
    $
    &
    $x \triangleleft f\triangleleft \alpha \triangleleft f' \triangleleft \beta \triangleleft f'' \triangleleft y \triangleleft g \triangleleft z$
    & yes \\
    \hline
    $
    \begin{tikzcd}[ampersand replacement = \&]
      x \ar[r, bend left = 80, "f"]\ar[r, bend right = 80, "g'"below]\ar[r, bend left = 15, "f'"{description}] \ar[r, bend right = 15, "g"{description}]
      \ar[r, bend left = 40, phantom, "\Downarrow{\scriptstyle\alpha}"]
      \ar[r, bend right = 40, phantom, "\Downarrow{\scriptstyle\beta}"]
      \& y
    \end{tikzcd}
    $
    &
    $
    \begin{tikzcd}[ampersand replacement =\&, row sep = tiny, column sep = tiny]
      \& f\ar[r, phantom, "\triangleleft"] \& \alpha\ar[r, phantom, "\triangleleft"]\& f' \ar[rd, phantom, "\triangleleft"{sloped}] \& \\
      x\ar[ru, phantom, "\triangleleft"{sloped}]\ar[rd, phantom, "\triangleleft"{sloped}] \& \& \& \& y \\
      \& g\ar[r, phantom, "\triangleleft"] \& \beta \ar[r,phantom,"\triangleleft"] \&g' \ar[ru, phantom, "\triangleleft"{sloped}]\&
    \end{tikzcd}
    $
    & no\\ \hline
    $
    \begin{tikzcd}
      x\ar[loop, "f"]
    \end{tikzcd}
    $
    &
    $
    \begin{tikzcd}[ampersand replacement =\&, row sep = tiny, column sep = tiny]
      x \ar[r, phantom, "\triangleleft", shift left]\& f \ar[l, phantom, "\triangleright", shift left]
    \end{tikzcd}
    $
    &
    no
    \\\hline
  \end{tabular}
  \caption{Globular sets and the relation $\triangleleft$}
  \label{fig:ps-order}
\end{figure}

The proof of \cref{thm:ps-|>} given in~\cite{catt} relies on constructing
a globular sum diagram associated to a $\triangleleft$-linear globular set.
Since this association reaches all the globular sums and is injective, this
also proves the following result.
\begin{lemma}\label{lemma:unique-gsum}
  Every pasting scheme can be written as a globular sum in a unique way.
\end{lemma}

\paragraph{Source and target of pasting schemes.}
A pasting scheme $X$ canonically comes equipped with a source and a target, that
are two distinguished sub globular sets of $X$ which are also pasting schemes.
Since the source and target are isomorphic globular sets, we will define a
unique object $\partial X$ along with the two inclusions which identify
$\partial X$ as a subobject of $X$ in two different ways.
\[
  \sigma_X,\tau_X : \partial X \to X
\]
When $X$ is given by a table of dimensions as above, write
$i = \max(i_{1},\ldots,i_{n})$ the dimension of $X$. We then define
its boundary $\partial X$ to be given by the table
\[
  \pa{
    \begin{array}{ccccccc}
      \overline{i_1} & & \overline {i_2} & &\cdots & & \overline{i_k} \\
                     & j_1 & & j_2 & \cdots & j_{k-1} &
    \end{array}}
  \qquad\qquad
  \text{where }
  \overline{i_k} =
  \begin{cases}
    i_k&\text{ if $i_m < i$}\\
    i-1&\text{ if $i_m = i$}
  \end{cases}
\]
\cref{fig:boundary} shows an example of a pasting scheme along with
its boundary, represented as pasting schemes as well as as diagrams.
\begin{figure}
  \centering
  \begin{tabular}{|c|c|c|}
    \hline
    & dimension table & diagram reprepsentation \\
    \hline
    $X$
    &
      $\begin{pmatrix}
        2 & & 1 \\
        & 0 &
      \end{pmatrix}$ & $\begin{tikzcd}[ampersand replacement=\&] \bullet \ar[r,
        bend left] \ar[r, bend right] \ar[r, phantom, "\Downarrow"] \&
        \bullet\ar[r] \&\bullet
      \end{tikzcd}$ \\
    \hline $\partial X$ & $\begin{pmatrix}
      1 & & 1 \\
      & 0 &
    \end{pmatrix}$ & $\begin{tikzcd}[ampersand replacement=\&] \bullet \ar[r] \&
      \bullet\ar[r] \&\bullet
    \end{tikzcd}$ \\
    \hline
  \end{tabular}
  \caption{A pasting scheme and its boundary}
  \label{fig:boundary}
\end{figure}
Note that the definition of the boundary of a pasting scheme may
produce tables that do not strictly comply with the definition of
globular sums, as presented before, since it is possible to have the
equality
\[
  \overline{i_m} = j_m = \overline{i_{m+1}} = i-1
\]
However, when it is the case we will chose the corresponding iterated sources
and target to be the identity maps (\ie the map iterated $0$ times). We can then
normalize with the following rewriting rule, that does not change the colimit
and thus exhibits $\partial X$ as a pasting scheme
\[
  \pa{
    \begin{array}{ccccc}
      \cdots & i-1 & & i-1 & \cdots \\
      \cdots & & i-1 & & \cdots
    \end{array}
  } \quad \rightsquigarrow \quad \pa{
    \begin{array}{ccc}
      \cdots & i-1 & \cdots \\
      \cdots &  & \cdots
    \end{array}
  }
\]
Now, we can define the two inclusion maps $\sigma_X$ and $\tau_X$ to induced by
the families
\begin{align*}
  \overline{\sigma_{i_m}} &: D_{\overline{i_m}} \longrightarrow D_{i_m}
  &
    \overline{\tau_{i_m}} &: D_{\overline{i_m}} \longrightarrow D_{i_m}
  \\
  \overline{\sigma_{i_m}} &=
                            \begin{cases}
                              \id{} &\text{if $i_m < i$} \\
                              \sigma_{i-1} &\text{if $i_m = i$}
                            \end{cases}
  &
    \overline{\tau_{i_m}} &=
                            \begin{cases}
                              \id{} & \text{if $i_m < i$} \\
                              \tau_{i-1} & \text{if $i_m = i$}
                            \end{cases}
\end{align*}
Note that there is a subtlety whenever there are two or more successive cells of
maximal dimension $n$ composed in dimension $n-1$. In this case we have to
renormalize the dimension of table of $\partial X$ in order to remove multiple
successive instances of $n-1$. Defining $\sigma_X$ and $\tau_X$ in this case
requires to handle carefully this renormalization, as illustrated in the
following example:
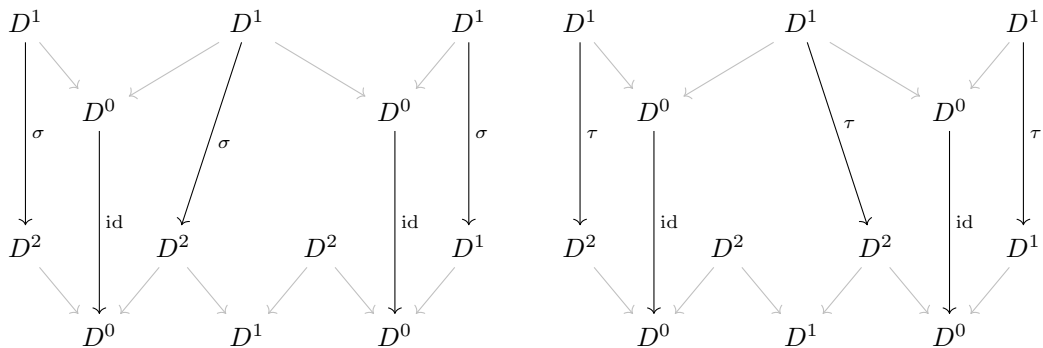
\[
  \begin{tikzcd}[column sep = tiny]
    D^1\ar[ddd,"\sigma"]\ar[dr, gray!50] & & & D^1 \ar[dddl,"\sigma"]\ar[dll, gray!50] \ar[drr, gray!50]& & & D^1 \ar[ddd,"\sigma"] \ar[dl, gray!50]\\
    & D^0\ar[ddd, "\id{}"] & & & & D^0\ar[ddd,"\id{}"] & \\ \\
    D^2\ar[dr, gray!50] & & D^2\ar[dl, gray!50]\ar[dr, gray!50] & & D^2 \ar[dr, gray!50]\ar[dl, gray!50] & & D^1\ar[dl, gray!50] \\
    & D^0 & & D^1 & & D^0 &
  \end{tikzcd}
  \qquad
  \begin{tikzcd}[column sep = tiny]
    D^1\ar[ddd,"\tau"]\ar[dr, gray!50] & & & D^1 \ar[dddr,"\tau"]\ar[dll, gray!50] \ar[drr, gray!50]& & & D^1 \ar[ddd,"\tau"] \ar[dl, gray!50]\\
    & D^0\ar[ddd,"\id{}"] & & & & D^0\ar[ddd,"\id{}"] & \\ \\
    D^2\ar[dr, gray!50] & & D^2\ar[dl, gray!50]\ar[dr, gray!50] & & D^2 \ar[dr, gray!50]\ar[dl, gray!50] & & D^1\ar[dl, gray!50] \\
    & D^0 & & D^1 & & D^0 &
  \end{tikzcd}
\]
exhibiting the diagram on the left respectively as source and target
of the one on the right.
\[
  \begin{tikzcd}
    \cdot\ar[r]&\cdot\ar[r]&\cdot\ar[r]&\cdot
  \end{tikzcd}
  \qquad\qquad\qquad
  \begin{tikzcd}
    \cdot\ar[r, bend left]\ar[r, bend right]\ar[r,
    phantom,"\Downarrow"]&\cdot\ar[r,bend left=60]\ar[phantom,r,bend
    left=30,"\Downarrow"]\ar[r]\ar[phantom,r,bend
    right=30,"\Downarrow"]\ar[r,bend right=60]&\cdot\ar[r]&\cdot
  \end{tikzcd}
\]
By convention, in the case of the pasting scheme $D_0$, we chose $\partial D_0$
to be the empty globular set, which is not a pasting scheme.

\paragraph{Characterization of the source and target via the relation $\triangleleft$.}
The notions of source and target are defined for all the pasting schemes, and
are closely related to the relation $\triangleleft$ defined above. Given a
pasting scheme $X$ and two parallel cells $x$ and $y$ in $X$, we denote by
$X(x,y)$ the set of cells with source $x$ and target $y$. Then the relation
$\triangleleft$ on the entire pasting scheme $X$ is a preorder, and therefore
also induces a preorder on the set $X(x,y)$. We thus define two sub-globular
sets of the pasting scheme $X$ of dimension $n$, denoted $\partial^-X$ and
$\partial^+X$ as follows
\begin{align*}
  \text{For $k < n-1 $} & &(\partial^-X)_k &= X_k &(\partial^+X)_k &= X_k \\
  \text{For all $x,y\in X_{n-2}$}& &\partial^-X (x,y) &= \min X(x,y) &\partial^+X (x,y) &= \max X(x,y)\\
  \text{For $k \geq n $} & &(\partial^-X)_k &= \emptyset &(\partial^+X)_k &= \emptyset
\end{align*}
Where the $\max$ and the $\min$ are respectively the maximal and minimal
elements for the preorder $\triangleleft$.

\begin{prop}\label{prop:src-tgt-|>}
  The globular set $\partial^-X$ (\resp $\partial^+X$) is the image of the
  source morphism $\sigma_X:\partial X\to X$ (\resp target morphism
  $\tau_X:\partial X\to X$).
\end{prop}
\begin{proof}
  One can check these images, by definition morphisms $\sigma_X$ and $\tau_X$,
  since it removes the variables of maximal dimension $n$, and keeps the
  variables of dimension $n-2$. Proving the equality in dimension $n-1$ requires
  a careful handling of the subtlety that appears in the case of several
  successive cells dimension $n$ composed in dimension $n-1$.
\end{proof}

\paragraph{Maps in the category $\COH_0$.}
The characterization of pasting schemes using the relation
$\triangleleft$ allows us show that the maps in the category $\COH_0$
are very restricted.

\begin{lemma}
  \label{lemma:maps-ps}
  Any map $f : X\to Y$ in the category $\COH_0$ is injective.
\end{lemma}
\begin{proof}
  A map of globular sets has to preserve the relation $\triangleleft$, since it
  preserves the source and target. Consider two distinct elements $x$ and $y$ in
  the pasting scheme $X$, then Theorem~\ref{thm:ps-|>} proves that either
  $x\triangleleft y$ or $y\triangleleft x$, hence we have either
  $f(x)\triangleleft f(y)$ or $f(y) \triangleleft f(x)$, which by applying
  Theorem~\ref{thm:ps-|>} again shows that $f(x) \neq f(y)$.
\end{proof}

\begin{lemma}
  \label{lemma:aut-ps}
  A pasting scheme has no non-trivial automorphism.
\end{lemma}
\begin{proof}
  Consider a pasting scheme $X$ together with an automorphism $f : X \to
  X$. Suppose that there exists an element $x\in X$ such that $x\neq f(x)$.
  Then by Theorem~\ref{thm:ps-|>}, we have either $x \triangleleft f(x)$ or
  $f(x) \triangleleft x$: we suppose that we are in the first case, the second
  one being similar. Since $f$ preserves the relation $\triangleleft$, this
  provides us with an infinite chain
  \[
  x \triangleleft f(x) \triangleleft f(f(x)) \triangleleft f(f(f(x))) \triangleleft \cdots
  \]
  which is impossible since $X$ has only finitely many elements. The
  automorphism $f$ is thus necessarily the identity.
\end{proof}

\subsection{Globular extensions and globular theories.}
In order to define weak $\omega$-categories, we rely on the notion of a
\emph{coherator} which is a category whose objects are the arities of the
operations expected in $\omega$-categories and the morphisms encode the
algebraic operations that they should have. It can be thought of as an analogue of Lawvere
theories in the dependently sorted case. Recall that in Lawvere theories, one
requires the set of objects to be freely generated by the finite products of
a single object, the coherator satisfies an analogous condition for the dependently
sorted case, which is captured by the notion of \emph{globular theory}

\paragraph{Globular extensions.}
A category $\C$ with a globular structure $F$ is called a
\emph{globular extension} when all the globular sums exist in $\C$.
Given two globular extensions $F : \G\to\C$ and $G : \G\to\D$, a
morphism of globular extensions is a functor $H : \C\to\D$ such that
$H\circ F = G$, which preserves globular sums. Dually, a category
equipped with a contravariant globular structure and which has all
globular products is called a \emph{contravariant globular extension},
the notion of morphism contravariant globular extensions is dual to
that of globular extensions.

\paragraph{The universal globular completion.}
There is a canonical functor $\G\to\COH_0$, sending an object~$n$ to the
disk~$D^n$, which is an object of $\COH_0$ as being obtained as the globular sum
corresponding to the table of dimensions $(n)$. This exhibits $\COH_0$ as the
completion of $\G$ under globular sums: we sometimes say that it is the
universal \emph{globular completion} of~$\G$.

\begin{prop}
  The category $\COH_0$ is the universal globular extension: for any globular
  extension $F:\G\to\C$, there is an essentially unique morphism of globular
  extensions $\COH_0\to\C$.
\end{prop}
\begin{proof}
  Consider a globular extension $F:\G\to\C$ together with a morphism of globular
  extensions $f : \COH_0\to\C$. Then an object $X$ in $\COH_0$ decomposes as a
  globular sum induced by a table of dimensions
  \[
    \pa{
      \begin{array}{ccccccc}
        i_1 & & i_2 & & \cdots & & i_n \\
            & j_1 & & j_2 & \cdots & j_{n-1} &
      \end{array}}
  \]
  By definition, $f(X)$ is the globular sum of the same diagram in $\C$, hence
  $f$ is determined up to natural isomorphism. Conversely, we can define $f(D^n)
  = F n$ and extends this definition to all the pasting schemes while preserving
  the globular sums since by \cref{lemma:unique-gsum} every pasting scheme is
  written as a globular sum in a unique way.
\end{proof}

\noindent
Dually $\op{\COH_0}$ is a \emph{globular cocompletion}: for every contravariant
globular extension $\op{\G}\to\C$, there is an essentially unique morphism
$\op{\COH_0}\to\C$.

\paragraph{The category of globular extensions.}
Globular extensions are characterized by the fact that they have globular sums,
and the globular sums factorize through the category $\COH_0$. We can thus use
the universality of the category~$\COH_0$ in order to characterize the category
of globular extensions as follows.

\begin{lemma}
  \label{lemma:cat-glob-ext}
  The universal property of the category $\COH_0$ induces an equivalence of
  categories between the category of globular extensions and the full
  subcategory of the coslice category $\COH_0\backslash\Cat$ whose objects are
  the functors preserving the globular sums.

  Dually, there is an equivalence of categories between the category of
  contravariant globular extensions and the full subcategory of the coslice
  category $\op{\COH_0}\backslash\Cat$, whose objects are the functors
  preserving the globular products.
\end{lemma}
\begin{proof}
  By the universal property of $\COH_0$, a globular extension $F : \G\to\C$
  induces a morphism of globular extensions $\COH_0\to\C$, which is an object of
  the coslice $\COH_0\backslash\Cat$ preserving globular sums, and this
  assignment is functorial.
  Conversely, consider a functor $F : \COH_0 \to \C$ preserving globular sums.
  Then, by precomposition by the canonical functor $\G\to\COH_0$, it induces a
  globular structure on~$\C$, and $F$ is a morphism from of globular structures
  from $\COH_0$ to~$\C$. Any globular sum diagram for this structure in $\C$
  factorizes through $F$. Since $\COH_0$ has all the globular sums, this diagram
  has a globular sum in $\COH_0$, and since $F$ preserves those, this diagram
  has a globular sum in $\C$, hence $\C$ has all the globular sums and is a
  globular extension. Moreover, consider a commutative triangle of the form
  \[
    \begin{tikzcd}
      &\ar[dl,"F"']\op\G\ar[dr,"G"]&\\
      \C\ar[rr,"f"']&&\D
    \end{tikzcd}
  \]
  with $F$ and $G$ preserving the globular sums. Then, by the previous
  statement, $\C$ has all the globular sums, which all factor through $F$. Since
  $f\circ F = G$ preserves the globular sums, it follows that necessarily $f$
  preserves the globular sums and thus defines a morphism of globular
  extensions. This proves the equivalence of categories.
\end{proof}

\paragraph{Globular theories.}
Given a globular extension $\G\to\C$, by universality of the globular
completion, there exists a unique morphism of globular extensions
$F : \COH_0\to\C$. The functor $\G\to\C$ is called a \emph{globular
  theory} if the induced functor $F$ is faithful and bijective on the
isomorphism classes of objects. Whenever it is the case, we can up to
equivalence identify $\COH_0$ as a subcategory of $\C$. A
\emph{morphism of globular theories} is just a morphism of the
underlying globular extensions. A morphism $f$ of a globular theory
$\C$ is said to be \emph{globular} if it is in $\COH_0$. Dually, a
contravariant globular extension $\op{\G}\to\C$ is called a
\emph{contravariant globular theory} if $\op{\C}$ is a globular
theory.

\subsection{Weak $\omega$-categories.}
We have introduced the notion of globular theory, which plays the role
of Lawvere theories, in the case where we have dependent sorts with
the dependency described by the category of globes and the arities are
given by pasting schemes. There are various such theories, and we now
introduce the one we will be interested in for weak
$\omega$-categories. As it is often the case for higher structures,
there is not a single theory of weak $\omega$-categories, but several
of them, called \emph{coherators}. We introduce here one such
coherator.

\paragraph{Admissible pair of arrows.}
Let $\G\to\C$ be a globular extension, two arrows $f,g : D^i \to X$ in~$\C$ are
said to be \emph{parallel} when
\begin{align*}
  f\circ\sigma &= g\circ\sigma& f\circ\tau &= g\circ\tau
\end{align*}
If $\C$ is a globular theory, then an arrow $f$ of $\C$ is said to be
\emph{algebraic}, when for every decomposition $f = g\circ f'$, with $g$
globular, then $g$ is an identity. A pair of parallel arrows $f,g : D^i\to X$ is
called an \emph{admissible pair} if either both $f$ and $g$ are algebraic, or
there exists a decomposition $f = \sigma_X\circ f'$ and $g = \tau_X\circ g'$,
with $f'$ and $g'$ algebraic.
\begin{defi}
  Given an admissible pair of maps $f,g: D^i\to X$, we call a \emph{lift} a map
  $h : D^{i+1} \to X$ such that $h\circ\sigma = f$ and $h\circ\tau = g$
  \[
    \begin{tikzcd}
      D^{i+1} \ar[dr, "h"]& \\
      D^i\ar[u,"\Gs", shift left] \ar[u,"\Gt"',shift right] \ar[r, "f", shift
      left]\ar[r,"g"', shift right]& X
    \end{tikzcd}
  \]
\end{defi}
We also say that an arrow is algebraic or that a pair is admissible in a
contravariant globular theory $\C$, to mean that it is the same in $\op\C$, and
a lift for an admissible is a lift in the opposite category in $\op\C$.

\paragraph{Cat-coherator.}
We introduce here the Batanin-Leinster cat-coherator, which is the one we will
be using for our type theory. For a more general definition of cat-coherators,
as well as other examples, see~\cite{maltsiniotis}. For the rest of this paper,
we will simply say cat-coherator to refer to the Batanin-Leinster
cat-coherator. The cat-coherator $\COH_\infty$ is defined to be the colimit
\[
  \COH_\infty = \colim (\COH_0 \to \COH_1 \to \COH_2 \to \cdots \to \COH_n \to \cdots)
\]
where the categories $\COH_n$ are defined by induction on $n$. Given
$n\in\N$, define $E_n$ to be the set of all pairs of admissible arrows
of $\COH_n$ that are not in $E_{n'}$ for any $n' < n$. Then we can
define $\COH_{n+1}$ to be the universal globular extension of $\COH_n$
obtained by formally adding a lift for each pair in~$E_n$. In other
words $\COH_{n+1}$ is the category such that, for each globular
extension $f : \COH_n \to \C$ equipped with the choice of a lift in
$\C$ for all the images by $f$ of the pairs of arrows in $E_{n}$,
$\tilde{f}$ which makes the following triangle commute
\[
  \begin{tikzcd}
    \COH_n \ar[r]\ar[rd,"f"'] & \COH_{n+1}\ar[d, dotted, "\tilde{f}"] \\
    & \C
  \end{tikzcd}
\]
and for every pair of arrows in $E_{n}$, sending the chosen lift of
the pair in $\COH_{n+1}$ to the chosen lift of its image in $\C$.

\paragraph{Weak $\omega$-categories.}
We define a \emph{weak $\omega$-category} to be functor
$F : \op{\COH_\infty} \to \Set$ which sends globular sums in
$\op{\COH_\infty}$ to the globular product on the opposite diagram, for the
globular structure on $\Set$ induced by $F$. Given $n\in\N$, the elements of the
set~$F D^n$ are called the \emph{$n$-cells} of the weak $\omega$-category. The
category $\omega$-$\Cat$ of weak $\omega$-categories is the full subcategory of
$\widehat{\COH_\infty}$ whose objects are the presheaves that are weak
$\omega$-categories.

\subsection{Identity and composition.}
In order to illustrate the above definition, we show that a weak
$\omega$-category $F : \op{\COH_\infty} \to \Set$ has identities on $0$-cells
and composites of composable $1$-cells. We refer the reader
to~\cite{maltsiniotis, ara2010groupoides} for more examples of the same nature.
More advanced examples of operations are presented in~\secr{catt-examples},
where they are described in a type theoretic style.

\paragraph{Identities on $0$-cells.}
The pair of maps $(\id{D^0},\id{D^0})$ is admissible. Hence, there exists a lift
\[
  \begin{tikzcd}
    D^1 \ar[dr,dotted, "\iota"]& \\
    D^0 \ar[r, shift left, "\id{D^0}"]\ar[r,shift right, "\id{D^0}"'] \ar[u,
    shift left]\ar[u, shift right] & D^0
  \end{tikzcd}
\]
Given a $0$-cell $x\in FD^0$, its \emph{identity $1$-cell} $i(x) \in FD^1$ is
$i(x) = F\iota(x)$. Moreover, by definition, we have $s(i(x)) = t(i(x)) = x$ as
expected for the identity $1$-cell on $x$.

\paragraph{Composition of $1$-cells.}
Consider the globular sum given as $D^1\coprod_{D^0} D^1$. There are two
canonical maps $\iota_1,\iota_2 : D^1 \to D^1\coprod_{D^0} D^1$, and we consider
the admissible pair
\[
  (\iota_1\sigma, \iota_2\tau) : D^0\to D^1\coprod_{D^0}D^1
\]
which provides the lift
\[
  \begin{tikzcd}
    D^1 \ar[dr,dotted, "c"]& \\
    D^0 \ar[r, shift left, "\iota_1\sigma"]\ar[r,shift right, "\iota_2\tau"']
    \ar[u, shift left]\ar[u, shift right] & D^1\coprod_{D^0}D^1
  \end{tikzcd}
\]
A pair of composable $1$-cell is the same as an element
$(f,g)\in F(D^1\coprod_{D^0}D^1)$, and the element $F(c)(f,g)$ in $F D^1$
defines the composition $f\cdot g$. By definition, $s(f\cdot g) = s(f)$ and
$t(f\cdot g) = t(g)$, as expected for the composition.


\section{Type theory for weak $\omega$-categories}
\label{sec:catt}
Our aim is now to extend the type theory $\Glob$ presented in \secr{glob}, by
adding term constructors corresponding to the algebraic structure that one need
to add to globular sets in order to obtain weak $\omega$-categories. We call the
resulting theory $\CaTT$ and motivate its introduction by following the ideas of
the Grothendieck-Maltsiniotis definition of weak $\omega$-categories recalled in
\secr{gm-def}.

\subsection{Ps-contexts.}
We have proved in \cref{thm:syn-glob} that the syntactic category of the
theory~$\Glob$ is equivalent to the opposite of the category of finite globular
sets. The Grothendieck-Maltsiniotis definition of weak $\omega$-categories
strongly relies on a particular class of such finite globular sets, namely the
pasting schemes, obtained as globular sums. In order to translate this
definition in a type theory, it is useful to transfer this notion of pasting
scheme in a type theoretic framework.

\paragraph{Recognition algorithm.}
We introduce a new kind of judgment to the theory, that we denote
\[
\GG\vdashps
\]
A context $\GG$ such that the judgment $\GG\vdashps$ is derivable is called a
\emph{ps-context}. It intuitively corresponds to a situation where the
context~$\GG$ is a pasting scheme, as formally shown in
\cref{thm:ps-are-ps}.
In order to define this judgment by induction, we also introduce an auxiliary
judgment
\[
\GG\vdashps x:A
\]
where the variable $x$ is called the \emph{dangling variable}. We require these
judgments to be subject to the following inference rules:
\[
  \begin{array}{|cl@{\qquad}c|}
    \hline
    \inferrule{\null}{(x:\Obj)\vdashps x:\Obj}{\regle{pss}} & & \inferrule{\GG\vdashps f: \Hom Axy}{\GG\vdashps y:A}{\regle{psd}} \\[1.5em]
    \inferrule{\GG\vdashps x:A}{\GG,y:A,f:\Hom Axy\vdashps f:\Hom Axy}{\regle{pse}} & \text{when $y,f\notin \Var\GG$} & \inferrule{\GG\vdashps x:\Obj}{\GG\vdashps}{\regle{ps}}\\[4ex]
    \hline
  \end{array}
\]
Note that every derivation of the judgment $\GG\vdashps$ starts with the rule
$\regle{pss}$ and ends with the rule $\regle{ps}$, with an equal number of
applications of the rules $\regle{pse}$ and $\regle{psd}$ in between.

\paragraph{An example of a derivation.}
In order to understand how a derivation of this judgment works, we have
illustrated in Figure~\ref{fig:Gamma-vdashps} the derivation of $\GG\vdashps$
where $\GG$ is the context
\[
  \GG =
  (x:\Obj,y:\Obj,f_1:\Hom{}xy,f_2:\Hom{}xy,\alpha:\Hom{}{f_1}{f_2},z:\Obj,
  g:\Hom{}yz)
\]
which corresponds to the globular set
\[
  \begin{tikzcd}
    x\ar[r, "f_1", bend left]\ar[r, "f_2"', bend right]\ar[r,phantom,
    "\Downarrow\scriptstyle\alpha"] & y \ar[r,"g"]& z
  \end{tikzcd}
\]
We follow the step-by-step derivation of the judgment $\GG\vdashps$, and give
a graphical representation of the globular corresponding globular set being
constructed, where we encircle the dangling variable on the judgment.
\begin{figure}
  \centering
  \begin{tabular}{l@{\qquad\qquad}cl}
    \begin{tikzcd}
      \phantom{x}&\emptyset
    \end{tikzcd}
    &
    &
    \\
    &\hrulefill&\regle{pss}
    \\
    \begin{tikzcd}
      \mtextcircled{$x$}
    \end{tikzcd}
    &
    $x:\Obj \vdashps x:\Obj$
    &
    \\
    &\hrulefill&\regle{pse}
    \\
    \begin{tikzcd}[ampersand replacement =\&]
      x \ar[r,"\mtextcircled{$f_1$}"]\& y
    \end{tikzcd}
    &
    $(x:\Obj,y:\Obj,f_1:\Hom{}xy)\vdashps f_1:\Hom{}xy$
    &
    \\
    &\hrulefill&\regle{pse}
    \\
    \begin{tikzcd}[ampersand replacement =\&, row sep = large]
      x \ar[r, bend left = 40, "f_1"] \ar[r, bend right = 40, "f_2"below] \ar[r,phantom,"\Downarrow{\mtextcircled{$\Ga$}}"] \& y
    \end{tikzcd}
    &
    \makecell[l]{$(x:\Obj, y:\Obj,f_1:\Hom{}xy,$ \\
      $f_2:\Hom{}xy , \alpha:\Hom{}{f_1}{f_2}) \vdashps
      \alpha:\Hom{}{f_1}{f_2}$}
    &
    \\
    &\hrulefill&\regle{psd}
    \\
    \begin{tikzcd}[ampersand replacement = \&]
      x \ar[r, bend left = 40, "f_1"] \ar[r, bend right = 40, "\mtextcircled{$f_2$}"below] \ar[r,phantom,"\Downarrow\Ga"] \& y
    \end{tikzcd}
    &
    \makecell[l]{$(x:\Obj, y:\Obj,f_1:\Hom{}xy,$ \\
      $f_2:\Hom{}xy , \alpha:\Hom{}{f_1}{f_2}) \vdashps
      f_2:\Hom{}xy$}
    &
    \\
    &\hrulefill&\regle{psd}
    \\
    \begin{tikzcd}[ampersand replacement = \&]
      x \ar[r, bend left = 40, "f_1"] \ar[r, bend right = 40, "f_2"below] \ar[r,phantom,"\Downarrow\Ga"] \& \mtextcircled{$y$}
    \end{tikzcd}
    &
    \makecell[l]{$(x:\Obj, y:\Obj,f_1:\Hom{}xy,$ \\
      $f_2:\Hom{}xy , \alpha:\Hom{}{f_1}{f_2}) \vdashps
      y:\Obj$}
    &
    \\
    &\hrulefill&\regle{pse}
    \\
    \begin{tikzcd}[ampersand replacement = \&]
      x \ar[r, bend left = 40, "f_1"] \ar[r, bend right = 40, "f_2"below] \ar[r,phantom,"\Downarrow\Ga"] \& y \ar[r,"\mtextcircled{$g$}"]\& z
    \end{tikzcd}
    &
    \makecell[l]{$(x:\Obj, y:\Obj,f_1:\Hom{}xy,$ \\
      $f_2:\Hom{}xy , \alpha:\Hom{}{f_1}{f_2}$ \\
      $z:\Obj,g:\Hom{}yz) \vdashps
      g:\Hom{}yz$}
    &
    \\
    &\hrulefill&\regle{psd}
    \\
    \begin{tikzcd}[ampersand replacement = \&]
      x \ar[r, bend left = 40, "f_1"] \ar[r, bend right = 40, "f_2"below] \ar[r,phantom,"\Downarrow\Ga"] \& y \ar[r,"g"]\& \mtextcircled{$z$}
    \end{tikzcd}
    &
    \makecell[l]{$(x:\Obj, y:\Obj,f_1:\Hom{}xy,$ \\
      $f_2:\Hom{}xy , \alpha:\Hom{}{f_1}{f_2}$ \\
      $z:\Obj,g:\Hom{}yz) \vdashps
      z:\Obj$}
    &
    \\
    &\hrulefill&\regle{ps}
    \\
    \begin{tikzcd}[ampersand replacement = \&]
      x \ar[r, bend left = 40, "f_1"] \ar[r, bend right = 40, "f_2"below] \ar[r,phantom,"\Downarrow\Ga"] \& y \ar[r,"g"]\& z
    \end{tikzcd}
    &
    $\GG\vdashps$
    &
  \end{tabular}
  \caption{Derivation of the judgment $\GG\vdashps$}
  \label{fig:Gamma-vdashps}
\end{figure}

The rules that we have given for recognizing ps-contexts do in particular
recognize usual contexts in the theory $\Glob$, in other words, the following
holds~\cite{catt,benjamin2020type}:

\begin{prop}
  The following rules are admissible
  \[
    \inferrule{\GG\vdashps x:A}{\GG\vdash}
    \qquad\qquad
    \inferrule{\GG\vdashps x:A}{\GG\vdash A}
    \qquad\qquad
    \inferrule{\GG\vdashps x:A}{\GG\vdash x:A}
    \qquad\qquad
    \inferrule{\GG\vdashps}{\GG\vdash}
  \]
\end{prop}
\begin{proof}
  The admissibility of the first three of these rules can be shown by mutual
  induction, and the admissibility of the last one is then a consequence of the
  former, see~\cite{benjamin2020type} for a detailed proof.
\end{proof}

\paragraph{The category of ps-contexts.}
Note that the notion of ps-context is not invariant under equivalence:
a context can be isomorphic to a ps-context without being a ps-context
itself. For the sake of simplicity, we consider the subcategory
$\Syn{\ps}$ of $\Syn{\Glob}$, whose objects are exactly the
ps-contexts, not including the contexts that are invariant under
equivalence.

\paragraph{The correspondence between ps-contexts and pasting schemes.}
We now show that ps-contexts correspond to pasting schemes. In order to do so,
we use the following useful lemma, which involves the functor $V : \Syn\Glob \to
\op\FinGSet$, introduced in \secr{glob} and defined by $V\GG =
\Syn{\Glob}(\GG,D^\bullet)$, and the relation $\triangleleft$, recalled in
\secr{pasting-schemes}.

\begin{prop}
  \label{prop:ps-linear}
  For every ps-context $\GG\vdashps$, the globular set $V\GG$ is
  $\triangleleft$-linear.
\end{prop}
\begin{proof}
  The proof requires the introduction of subtle invariants beforehand.
  One can check by induction on the derivation tree, that whenever a
  judgment of the form $\GG\vdashps x:A$ is derivable, there is no
  variable $f$ whose type in $\GG$ is $\Hom{}yz$ where $y$ is an
  iterated target of $x$ (\ie{} there exists a sequence of terms
  $x=x_0,x_1,x_2,\ldots,x_n = y$ with
  $\GG\vdash x_i : \Hom{}{a}{x_{i+1}}$ for a variable $a$). As a
  consequence, in this situation, every relation of the form
  $x\triangleleft y$ with $x,y\in V\Gamma$ is such that $y$ is an
  iterated target of~$x$. Using this fact, it can be shown by
  induction that the set $V\GG$ is $\triangleleft$-linear.

  Now, suppose fixed a ps-context $\GG\vdashps$. We first show that if we have
  $a,b$ in $\GG$ such that $a\neq b$ then necessarily $a\triangleleft b$ or $b
  \triangleleft a$, by induction on the form of the pasting scheme.
  \begin{itemize}
  \item For the pasting scheme $(x : \Obj)$, the statement is vacuously true
    since there are no two disjoint variables.
  \item For a ps-context of the form $\GG = (\GG',y:A,f:\Hom{}xy)$, we
    distinguish different cases.
    \begin{itemize}
    \item If both $a$ and $b$ are in $\GG'$ then by induction, either
      $a\triangleleft b$ or $b\triangleleft a$ in $\GG'$, and thus the same
      holds in $\GG$.
    \item If $a$ is in $\GG'$, but not $b$, either $b = y$ or $b = f$, then by
      induction, either $a = x$, or $x\triangleleft x$ or $x\triangleleft a$. In
      the first two cases, since we have $x\triangleleft b$, the transitivity
      shows that $a\triangleleft b$. We can thus assume that $x\triangleleft a$.
      In this case, by the fact that we have proved, $a$ is an iterated target
      of $x$. Since $y$ is parallel to $x$ and $y$ is a target of $f$, in either
      case, $a$ is also an iterated target of $b$, which shows that
      $b\triangleleft a$.
    \item If $b$ is in $\GG'$ but not $a$, the situation is symmetric to the
      previous case.
    \item If neither $a$ nor $b$ is in $\GG'$, then necessarily they are $f$ and
      $y$, and we have $f\triangleleft y$.
    \end{itemize}
  \end{itemize}
  Conversely, we show that for every ps-context $\GG$, we never have
  $x\triangleleft x$. In order to prove this, we first note that whenever we
  have a relation of the form $a\triangleleft b$ in the ps-context
  $(\GG,y:A,f:\Hom{}xy)$ with $a$ and $b$ variables of $\GG$, we also have the
  same relation in $\GG$. Indeed, considering the chain of generating relations
  $a\triangleleft a_1 \triangleleft\cdots\triangleleft b$, it suffices to prove
  that there is a chain completely included in $\GG$. If it is not the case,
  that means that there are occurrences of the form $s\triangleleft y
  \triangleleft t$ or $x\triangleleft f \triangleleft y\triangleleft t$ with $s$
  the source of $y$ and $t$ its target (these are the only possibilities because
  of the fact that $y$ can never be a source). In the first case, one can
  replace the occurrence with $s\triangleleft x \triangleleft t$ and in the
  second case, one can replace it with $s\triangleleft t$, in order to obtain a
  chain proving $a\triangleleft b$ in $\GG$. Proving that there is no variable
  $x$ such that $x\triangleleft x$ in ps-context $\GG$ is then a straightforward
  induction over the derivation tree of the judgment $\GG\vdashps$.
\end{proof}

We now prove the converse, that any $\triangleleft$-linear globular set
corresponds to a ps-context. In order to do this, we introduce the notion of
\emph{locally maximal element} of a pasting scheme as an element $x$ such that
there is no variable $y$ such that $s(x)\triangleleft y \triangleleft x$ or
$x\triangleleft y \triangleleft t(x)$. Alternatively, the locally maximal
elements are the elements corresponding to the peaks in the decomposition as a
globular sum.

\begin{ex}
  In the globular set
  \[
    \begin{tikzcd}
      x\ar[r,bend left,"f"]\ar[r,bend
      right,"g"']\ar[r,phantom,"\scriptstyle\phantom\alpha\Downarrow\alpha"]&y
    \end{tikzcd}
  \]
  we have that $\alpha$ is maximal but $f$ is not maximal because
  $f\triangleleft\alpha\triangleleft y=t(f)$.
\end{ex}

\noindent
In order to prove the result, we use the following lemma:

\begin{lemma}
  \label{lemma:>-target}
  Consider a globular set $G$ with two elements $x,y$ such that $x\triangleleft
  y$ and $\dim x > \dim y$. Then $t(x) \triangleleft y$ or $t(x) = y$.
\end{lemma}
\begin{proof}
  Suppose that $x\triangleleft y$. By definition of the relation $\triangleleft$
  there exists a sequence of elements $x=x_0\triangleleft
  x_1\triangleleft\ldots\triangleleft x_n=y$ such that, for every index~$i$,
  $x_i=s(x_{i+1})$ or $x_{i+1}=t(x_i)$. We reason by induction on the length $n$
  of this sequence.
  \begin{itemize}
  \item If $n=1$, then necessarily, either $y = t(x)$ or $x = s(y)$, and the
    condition on the dimensions implies that we have $y = t(x)$.
  \item Suppose that the result holds for all chains of length at most $n-1$.
    Note that either $x_1 = t(x)$, or $x = s(x_1)$. The first case gives the
    result immediately. In the second case, the induction shows that we have a
    relation $t(x_1) \triangleleft y$, given by a chain of length less than
    $n-1$, so applying again the induction hypothesis proves that $t(t(x_1))
    \triangleleft y$. And we conclude by using the fact that $t(t(x_1)) =
    t(s(x_1)) = t(x)$.\qedhere
  \end{itemize}
\end{proof}

\begin{prop}
  \label{prop:linear-ps}
  For any $\triangleleft$-linear non-empty finite globular set $G$,
  there exists a unique (up-to $\alpha$-equivalence) ps-context $\GG$
  such that $V\GG = G$.
\end{prop}
\begin{proof}
  We construct the context $\GG$ inductively and then prove that it satisfies
  $\GG\vdashps x:A$, where~$V(x)$ is the greatest (for the relation
  $\triangleleft$) locally maximal element of $G$.
  \begin{itemize}
  \item If the globular set $G$ has a unique element, then this element is
    necessarily of dimension $0$ and we then associate the context $\GG =
    (x:\Obj)$, where the derivation of $\GG\vdashps x:\Obj$ is given by the
    rule~\regle{pss}.
  \item If $G$ has more than one element, write $a$ for the greatest
    locally maximal element of $G$. We can consider the globular
    set~$G'$ obtained by removing $a$ and $t(a)$ from $G$: indeed, by
    definition of locally maximal element, there is no element whose
    source or target is $a$, hence $a$ can safely be removed.
    Moreover, any element $x$ distinct from $a$ and whose target is
    $t(a)$ satisfies $x\triangleleft t(a)$, thus either it is $a$ or
    it compares to $a$ by linearity. Since $a$ is locally maximal, we
    then cannot have $a\triangleleft x$, so we necessarily have
    $x\triangleleft a$ and thus $x\triangleleft s(a)$.
    Lemma~\ref{lemma:>-target} then applies to show that
    $t(x) \triangleleft s(a)$. Since we have $t(x) = t(a)$ and also
    $s(a) \triangleleft t(a)$, this implies in particular that
    $t(a) \triangleleft t(a)$, which contradicts the linearity of $G$.
    So any element whose target is $t(a)$ is necessarily $a$, and
    since $a$ is the greatest locally maximal element, there cannot be
    any element whose source is $t(a)$. Hence, after removing $a$, one
    can also remove $t(a)$ safely. In fact, this analysis shows that
    the resulting globular set $G'$ is still a non-empty finite
    $\triangleleft$-linear set, so by induction, one can construct a
    context $\GG'$ such that $V(\GG') = G'$ and $\GG'\vdashps x:A$,
    where $x$ is the greatest locally maximal element in $G'$.
    \begin{itemize}
    \item Either the greatest locally maximal element of $G'$ is
      $s(a)$. In this case, we define the context
      $\GG = (\GG',t(a):A,a:\Hom{}{s(a)}{t(a)})$, we then have
      $V(\GG) = G$ by definition, and the rule~$\regle{pse}$ gives a
      derivation of $\GG\vdashps a:\Hom{}{s(a)}{t(a)}$.
    \item Or the greatest locally maximal element $x$ of $G'$ is such that
      $x\triangleleft s(a)$. In this case, since $x$ is locally maximal, $s(a)$
      is necessarily an iterated target of $x$, and we write $n$ for the number of
      iterations. Then applying the rule~\regle{psd} $n$ times gives a
      derivation of $\GG\vdash s(a):B$. We define $\GG =
      (\GG',t(a):B,a:\Hom{}{s(a)}{t(a)})$, in such a way that we have $V(\GG) =
      G$ and $\GG\vdashps a:\Hom{}{s(a)}{t(a)}$ obtained from the previous
      derivation by applying the rule~\regle{pse}.
    \end{itemize}
    Since $a$ is the greatest locally maximal variable, these are the two only
    cases, and in both cases, we constructed a suitable preimage.\qedhere
  \end{itemize}
\end{proof}

The two previous propositions, together with Theorem~\ref{thm:syn-glob} finally
allows us to conclude:

\begin{thm}
  \label{thm:ps-are-ps}
  There is an isomorphism of categories
  \[
    \Syn\ps \isoto \op{\COH_0}
  \]
\end{thm}
\begin{proof}
  We have already proved, by Proposition~\ref{prop:ps-linear}, that
  the functor $V$ induces a functor $\Syn{\ps} \to \op{\COH_0}$.
  Moreover, we have shown by Theorem~\ref{thm:syn-glob} that $V$ is
  fully faithful, so the restriction is also fully faithful, and
  Proposition~\ref{prop:linear-ps} shows that this restriction defines
  a bijection on objects (note that the objects of $\Syn{\ps}$ are
  assumed to be quotiented by $\alpha$-equivalences), hence it is an
  isomorphism categories.
\end{proof}

We illustrate in Figure~\ref{fig:ps-|>} the correspondence between the
ps-contexts and the $\triangleleft$-linear contexts with our previous example of
derivation, showing how we construct $\triangleleft$ to be a preorder.

Note that the notion of ps-context is not invariant under isomorphism in the
category $\Syn{\Glob}$. As an example, one can consider the two following $\GG$
and $\GG'$ which are isomorphic, as they only differ from the order of the
variables, but the context $\GG$ is a ps-context whereas the context $\GG'$ is
not:
\begin{align*}
  \GG &= (x:\Obj,y:\Obj,f:\Hom {}xy,z : \Obj,g:\Hom{}yz) \\
  \GG' &= (x:\Obj, y:\Obj, z:\Obj, f:\Hom{}xy, g:\Hom{}yz)
\end{align*}
Thus one can understand the notion of ps-context as a recognition algorithm for
a particular representative of a context in each equivalence classes of contexts
corresponding to a pasting scheme.

\begin{figure}
  \centering
  \begin{tabular}{l@{\qquad\qquad}cl}
    \begin{tikzcd}
      \phantom{x}&\emptyset
    \end{tikzcd}
    &
    &
    \\
                 &\hrulefill&\regle{pss}
    \\
    \begin{tikzcd}
      \mtextcircled{$x$}
    \end{tikzcd}
                 &
                   $x:\Obj \vdashps x:\Obj$
    &
    \\
                 &\hrulefill&\regle{pse}
    \\
    \begin{tikzcd}[ampersand replacement =\&, column sep = tiny, row sep = tiny]
      \& \mtextcircled{$f_1$} \ar[rd, "\triangleleft"sloped, phantom]\& \\
      x \ar[ru,"\triangleleft"sloped, phantom]\& \& y
    \end{tikzcd}
                 &
                   $(x:\Obj,y:\Obj,f_1:\Hom{}xy)\vdashps f_1:\Hom{}xy$
    &
    \\
                 &\hrulefill&\regle{pse}
    \\
    \begin{tikzcd}[ampersand replacement =\&, column sep = tiny, row sep = tiny]
      \& \& \mtextcircled{$\alpha$}\ar[rd, "\triangleleft"sloped, phantom] \& \& \\
      \& f_1 \ar[ru, "\triangleleft"sloped, phantom]\& \& f_2 \ar[rd, "\triangleleft"sloped, phantom] \& \\
      x \ar[ru,"\triangleleft"sloped, phantom]\& \& \& \& y
    \end{tikzcd}
                 &
                   \makecell[l]{$(x:\Obj, y:\Obj,f_1:\Hom{}xy,$ \\
    $f_2:\Hom{}xy , \alpha:\Hom{}{f_1}{f_2}) \vdashps
    \alpha:\Hom{}{f_1}{f_2}$}
                 &
    \\
                 &\hrulefill&\regle{psd}
    \\
    \begin{tikzcd}[ampersand replacement =\&, column sep = tiny, row sep = tiny]
      \& \& \alpha\ar[rd, "\triangleleft"sloped, phantom] \& \& \\
      \& f_1 \ar[ru, "\triangleleft"sloped, phantom]\& \& \mtextcircled{$f_2$} \ar[rd, "\triangleleft"sloped, phantom] \& \\
      x \ar[ru,"\triangleleft"sloped, phantom]\& \& \& \& y
    \end{tikzcd}
                 &
                   \makecell[l]{$(x:\Obj, y:\Obj,f_1:\Hom{}xy,$ \\
    $f_2:\Hom{}xy , \alpha:\Hom{}{f_1}{f_2}) \vdashps
    f_2:\Hom{}xy$}
                 &
    \\
                 &\hrulefill&\regle{psd}
    \\
    \begin{tikzcd}[ampersand replacement =\&, column sep = tiny, row sep = tiny]
      \& \& \alpha\ar[rd, "\triangleleft"sloped, phantom] \& \& \\
      \& f_1 \ar[ru, "\triangleleft"sloped, phantom]\& \& f_2 \ar[rd, "\triangleleft"sloped, phantom] \& \\
      x \ar[ru,"\triangleleft"sloped, phantom]\& \& \& \& \mtextcircled{$y$}
    \end{tikzcd}
                 &
                   \makecell[l]{$(x:\Obj, y:\Obj,f_1:\Hom{}xy,$ \\
    $f_2:\Hom{}xy , \alpha:\Hom{}{f_1}{f_2}) \vdashps
    y:\Obj$}
                 &
    \\
                 &\hrulefill&\regle{pse}
    \\
    \begin{tikzcd}[ampersand replacement =\&, column sep = tiny, row sep = tiny]
      \& \& \alpha\ar[rd, "\triangleleft"sloped, phantom] \& \& \\
      \& f_1 \ar[ru, "\triangleleft"sloped, phantom]\& \& f_2 \ar[rd, "\triangleleft"sloped, phantom] \& \& \mtextcircled{$g$}\ar[rd,"\triangleleft"sloped, phantom] \& \\
      x \ar[ru,"\triangleleft"sloped, phantom]\& \& \& \&
      y\ar[ru,"\triangleleft"sloped, phantom] \& \& z
    \end{tikzcd} &
                   \makecell[l]{$(x:\Obj, y:\Obj,f_1:\Hom{}xy,$ \\
    $f_2:\Hom{}xy , \alpha:\Hom{}{f_1}{f_2}$ \\
    $z:\Obj,g:\Hom{}yz) \vdashps g:\Hom{}yz$} &
    \\
                 &\hrulefill&\regle{psd}
    \\
    \begin{tikzcd}[ampersand replacement =\&, column sep = tiny, row sep = tiny]
      \& \& \alpha\ar[rd, "\triangleleft"sloped, phantom] \& \& \\
      \& f_1 \ar[ru, "\triangleleft"sloped, phantom]\& \& f_2 \ar[rd, "\triangleleft"sloped, phantom] \& \& g\ar[rd,"\triangleleft"sloped, phantom] \& \\
      x \ar[ru,"\triangleleft"sloped, phantom]\& \& \& \&
      y\ar[ru,"\triangleleft"sloped, phantom] \& \& \mtextcircled{$z$}
    \end{tikzcd} &
                   \makecell[l]{$(x:\Obj, y:\Obj,f_1:\Hom{}xy,$ \\
    $f_2:\Hom{}xy , \alpha:\Hom{}{f_1}{f_2}$ \\
    $z:\Obj,g:\Hom{}yz) \vdashps z:\Obj$} &
    \\
                 &\hrulefill&\regle{ps}
    \\
    \begin{tikzcd}[ampersand replacement =\&, column sep = tiny, row sep = tiny]
      \& \& \alpha\ar[rd, "\triangleleft"sloped, phantom] \& \& \\
      \& f_1 \ar[ru, "\triangleleft"sloped, phantom]\& \& f_2 \ar[rd, "\triangleleft"sloped, phantom] \& \& g\ar[rd,"\triangleleft"sloped, phantom] \& \\
      x \ar[ru,"\triangleleft"sloped, phantom]\& \& \& \&
      y\ar[ru,"\triangleleft"sloped, phantom] \& \& z
    \end{tikzcd}
                 &
                   $\GG\vdashps$
    &
  \end{tabular}
  \caption{$\triangleleft$-linearity of a ps-context}
  \label{fig:ps-|>}
\end{figure}

\paragraph{Uniqueness of derivation.}
The following results rely on a more detailed analysis of the allowed derivation
trees and show that these rules enjoy good computational properties.

\begin{prop}
  Given a context $\Gamma$, the derivability of the judgment $\GG\vdashps$ is
  decidable, and when this judgment is derivable, it has a unique derivation.
\end{prop}
\begin{proof}
  The proof is more subtle than it may appear at first glance, as one cannot
  just use a straightforward induction to prove this result. Indeed, any
  derivation of the judgment $\GG\vdash$ is obtained from a derivation of the
  judgment $\GG\vdashps x:\Obj$, but there is no guarantee a priori that this
  the variable $x:\Obj$ is the same for all possible derivations. However, in
  the proof of Proposition~\ref{prop:ps-linear} we have characterized the
  variable $x$ in a judgment of the form $\GG\vdashps x:A$ as an iterated target
  of the greatest locally maximal variable. This proves that whenever we have
  two derivations of the form $\GG\vdashps x:A$ and $\GG\vdashps y:B$ with
  $\dim A = \dim B$, then necessarily $x=y$. Moreover,
  Proposition~\ref{prop:ps-linear} also characterizes the judgments
  $\GG\vdashps x:A$ obtained from the rule~\regle{pse} as those when $x$ is
  locally maximal in $\GG$, all the other ones are obtained from the
  rule~\regle{psd}. These two facts together combine allow for a straightforward
  proof by induction on the structure of the derivation trees, that each
  judgment of this form has a single derivation.
\end{proof}

\paragraph{Source and target of a ps-context.}
The ps-contexts come equipped with a notion of source and target, which mirror
the corresponding operations on pasting scheme, already presented in
\secr{gm-def}. Following the proofs that we have given, one could already figure
out how to define these: indeed, it suffices to use the correspondence of
ps-contexts and pasting schemes in order to compute the source, or the target of
a pasting scheme, and then use the correspondence in the other direction to get
back a ps-context. We give here a direct computation by induction on the syntax
of a ps-context of this process. We define for all $i\in\N$ the
\emph{$i$-source} of a ps-context $\GG$ induction on the length of $\GG$, by
setting $\partial^-_i(x:\Obj) = (x:\Obj)$ and
\[
  \partial^-_i(\GG,y:A,f:\Hom{}xy) =\left\{
    \begin{array}{l@{\quad}l}
      \partial^-_i\GG & \text{if $\dim A~\geq i-1$} \\
      \partial^-_i\GG,y:A,f:\Hom{}xy & \text{otherwise}
    \end{array} \right.
\]
and similarly the \emph{$i$-target} of $\GG$ is defined by $\partial^+_i(x:\Obj)
= (x:\Obj)$, and
\[
  \partial^+_i(\GG,y:A,f:\Hom{}xy) =\left\{
    \begin{array}{l@{\quad}l}
      \partial^+_i\GG & \text{if $\dim A~\geq i$} \\
      \mathrm{drop}(\partial^+_i\GG),y:A~& \text{if $\dim A~= i-1$}\\
      \partial^+_i\GG,y:A,f:\Hom{}xy & \text{otherwise}
    \end{array} \right.
\]
where $\mathrm{drop}(\GG)$ is the context $\GG$ with its last variable removed.
One can check by induction on the derivation of the judgment $\GG\vdashps$ that
whenever $\GG$ is a ps-context of non-zero dimension, both $\partial^-_i\GG$ and
$\partial^+_i\GG$ are also ps-contexts. It is straightforward in the case of the
$i$-source, and for the $i$-target, it relies on the fact that whenever the
$\mathrm{drop}$ operator is used, immediately afterwards a variable of the same
type that the one that was removed is added. We denote $\src\GG =
\partial^-_{\dim \GG-1} \GG$ and $\tgt\GG = \partial^+_{\dim \GG -1}{\GG}$ and
call these the \emph{source} and \emph{target} of $\GG$.

\begin{lemma}
  For every ps-context $\GG$, the globular set $V(\src\GG)$ is exactly the
  sub-globular set $\src{V\GG}$ of $V\GG$, and similarly $V(\tgt\GG)$ is the
  sub-globular set $\tgt{V\GG}$.
\end{lemma}
\begin{proof}
  By definition, $V(\src\GG)$ contains the same elements as $V(\GG)$ in
  dimension up to $\dim\GG-2$, and is empty in dimensions $\dim\GG$ and higher.
  So by Proposition~\ref{prop:src-tgt-|>}, it suffices to check that in
  dimension $\dim\GG-1$, the globular set $V(\src\GG)$ contains exactly the
  minimal elements for the preorder $\triangleleft$ in $V(\GG)$, with source and
  target fixed. This is true by a straightforward induction. In the case of the
  target, it, it is similar, except one has to check that we only keep the
  maximal element. For the induction to work, we thus have to also show that in
  a derivation of the form $(\GG,y:A,f:\Hom{}xy)\vdashps$ with $\dim A = \dim
  \GG -2$ the last variable in the context $\partial^+\GG$ is the maximal
  element of type $A$ in $\GG$.
\end{proof}

\subsection{Operations and coherences.}
In order to translate the Grothendieck-Maltsiniotis definition of weak
$\omega$-categories in type theory, we extend the type theory $\Glob$ with term
constructors which correspond the operations present in those categories.

\paragraph{Signature of the theory.}
We extend the signature of the theory $\Glob$ with two term
constructors $\cohop$ and $\coh$, which correspond to the liftings
that are formally added in the Grothendieck-Maltsiniotis definition of
weak $\omega$-categories. Both of these constructors take as arguments
a context, a type and a substitution, in such a way that terms in the
theory are now either variables, or of the form $\cohop_{\GG,A}[\Gg]$
or $\coh_{\GG,A}[\Gg]$, with $\GG$ a context, $A$ a type and $\Gg$ a
substitution. We define the set of variables of a term constructed
this way as
\begin{align*}
  \Var{\cohop_{\GG,A}[\Gg]}&= \Var\Gg
  &
  \Var{\coh_{\GG,A}[\Gg]}&= \Var\Gg
\end{align*}
Importantly, the variables that appear in the ps-context $\Gamma$ are
not accounted for in the variables of the term constructed this way.
One can understand this by thinking of the term constructors $\cohop$
and $\coh$ as binders for these variables. We also need to extend the
action of substitutions on terms to these new terms. This has to be
defined together with the composition of substitution, as they are
mutually inductive notions:
\begin{align*}
  t[\sub{}] &= t & y[\sub{\Gg,x\mapsto u}] &=
  \begin{cases}
    u & \text{if $y = x$} \\
    y[\Gg] & \text{otherwise}
  \end{cases}
  \\
  \cohop_{\GG,A}[\Gg][\Gd] &= \cohop[\Gg\circ\Gd] & \coh_{\GG,A}[\Gg][\Gd] &= \coh_{\GG,A}[\Gg\circ\Gd] \\
  \Obj[\Gg] &= \Obj & (\Hom Atu)[\Gg] &= \Hom{(A[\Gg])}{(t[\Gg])}{(u[\Gg])}\\
  \sub{}\circ\Gg &=\sub{} & \sub{\Gd,x\mapsto t}\circ\Gg &= \sub{\Gd\circ\Gg, x\mapsto t[\Gg]}
\end{align*}

\paragraph{Rules for coherences.}
The introduction rules for these two term constructors are subject to two side
conditions, expressing the fact that some terms use all of the variables of a
context. In order to express these conditions in a more compact way, we write
$\Var{t:A}=\Var{t}\cup\Var{A}$ for the union of the set of variables of the term
$t$ and the set of variables of the type $A$. In this notation, it is always
implicit that the term $t$ is of type $A$ in the context we are considering. The
introduction rules for the term constructors $\cohop$ and $\coh$ are then given
as follows.
\begin{itemize}
\item For the constructor $\cohop$, the rule is
  \[
    \inferrule{\GG\vdashps\\
      \src\GG\vdash t:A\\
      \tgt\GG\vdash u:A\\
      \Delta\vdash\Gg:\Gamma} {\Delta\vdash\cohop_{\Gamma,\Hom Atu}[\Gg] :
      \Hom {A[\Gg]}{t[\Gg]}{u[\Gg]}} {\regle{op}}
  \]
  subject to the side conditions
  \begin{equation}
    \tag{$\mathrm{C_{\cohop}}$}
    \Var{t:A} = \Var{\src\Gamma}
    \qquad\text{and}\qquad
    \Var{u:A} = \Var{\tgt\Gamma}
  \end{equation}
\item For the constructor $\coh$, the rule is
  \[
    \inferrule{\Gamma\vdashps\\
      \Gamma\vdash t : A\\
      \Gamma\vdash u : A\\
      \Delta\vdash \gamma:\Gamma} {\Delta\vdash \coh_{\Gamma,\Hom Atu}[\gamma]:
      \Hom {A[\gamma]}{t[\gamma]}{u[\gamma]}} {\regle{coh}}
  \]
  subject to the side conditions
  \begin{equation}
    \tag{$\mathrm{C_{\coh}}$}
    \Var{t:A} = \Var{\Gamma}
    \qquad\text{and}\qquad
    \Var{u:A} = \Var{\Gamma}
  \end{equation}
\end{itemize}
Note that the rule \regle{coh} presented
here is slightly different from the one introduced in~\cite{catt}: it is
equivalent but makes the presentation closer to the conditions of Maltsiniotis'
definition of weak $\omega$-categories~\cite{maltsiniotis}. A detailed account
of the equivalence between the two presentations is given in~\cite[Section
3.5.1]{benjamin2020type}. We give in Figure~\ref{fig:rules-catt} a full summary
of all the rules of the theory $\CaTT$.
\begin{figure}
  \centering
  \begin{tabular}{|cc|}
    \hline
    \multicolumn{2}{|l|}{\emph{For contexts:}}\\
    $\inferrule{\null}{\emptycontext\vdash}{\regle{ec}}$ & $\inferrule{\GG\vdash A}{\GG,x:A\vdash}{\regle{ce}}$\quad when $x\notin\Var\GG$\\
    \multicolumn{2}{|l|}{\emph{For types:}}\\
    $\inferrule{\GG\vdash}{\GG\vdash\Obj}{\regle{$\Obj$-intro}}$ &
    $\inferrule{\GG\vdash A \\ \GG\vdash t:A \\ \GG\vdash u:A}{\GG\vdash \Hom Atu}{\regle{$\Hom{}{}{}$-intro}}$\\
    \multicolumn{2}{|l|}{\emph{For terms:}}\\
    $\inferrule{\GG\vdash\\(x:A)\in\GG}{\GG\vdash x:A}{\regle{var}}$ & \\[1.2em]
    \multicolumn{2}{|c|}{
    $\inferrule{\GG\vdashps\\ \src\GG\vdash t:A\\ \tgt\GG\vdash u:A\\ \GD\vdash\Gg:\GG}
      {\GD\vdash\cohop_{\Gamma,\Hom Atu}[\Gg] : \Hom {A[\Gg]}{t[\Gg]}{u[\Gg]}} {\regle{op}}$ \quad when $\Vop$} \\[1.2em]
    \multicolumn{2}{|c|}{
   $\inferrule{\GG\vdashps\\ \GG\vdash t : A\\ \GG\vdash u : A\\ \GD\vdash \Gg:\GG}
                {\Delta\vdash \coh_{\GG,\Hom Atu}[\Gg]: \Hom {A[\Gg]}{t[\Gg]}{u[\Gg]}} {\regle{coh}}$ \quad when $\Vcoh$}\\
    \multicolumn{2}{|l|}{\emph{For substitutions:}}\\
    $\inferrule{\GD\vdash}{\GD\vdash\sub{}:\emptycontext}{\regle{es}}$ & $\inferrule{\GD\vdash\Gg:\GG \\ \GG,x:A\vdash \\ \GD\vdash t:A[\Gg]}{\GD\vdash\sub{\Gg,x\mapsto t}:(\GG,x:A)}{\regle{se}}$\\
    \hline
  \end{tabular}
  \caption{Derivation rules of the theory $\CaTT$.}
  \label{fig:rules-catt}
\end{figure}

\paragraph{Interpretation.}
These rules are to be understood as follows. A derivable judgment
$\Gamma\vdash t:A$ can be thought of as a given composite of various cells that
are supposed to be known in the context $\GG$, and in the case of a ps-context,
adding the side condition $\Var{t:A} = \Var{\GG}$ enforces that the composite
uses all the cells of $\Gamma$. This intuition is made more formal in
\secr{syn-catt}, where we show that the contexts of this theory are finite
polygraphs for weak $\omega$-categories. In the light of this identification, a
term $\GG\vdash t:A$ is a cell in the free category generated by the
polygraph~$\GG$. The use of the two rules can be detailed as follows.
\begin{itemize}
\item Rule \regle{op}. Given a pasting scheme $\Gamma$ and a way to
  compose entirely its source and its target encoded as the terms
  $\src\GG\vdash t:A$ and $\tgt\GG\vdash u:A$ satisfying the condition~$\Vop$,
  this rule provides a way to compose entirely $\Gamma$. The result of this
  composition goes from the specified composition of the source to the specified
  composition of the target, and is encoded as the term
  $\GG\vdash\cohop_{\GG,\Hom{}tu}[\id\GG]:\Hom{}tu$.
\item Rule \regle{coh}. Given two ways of composing entirely the
  pasting scheme $\Gamma$, encoded as a pair of terms $\GG\vdash t:A$
  and $\GG\vdash u:A$ satisfying the condition~$\Vcoh$, the rule
  provides a cell between these two compositions, encoded as the term
  $\GG\vdash\coh_{\GG,\Hom{}tu}[\id{\GG}] : \Hom{}tu$. It turns out
  that this rule only produces invertible cells, and thus it can be
  reformulated as: ``any two ways of composing entirely a pasting
  scheme are weakly equivalent'', or by adopting a more topological
  view it expresses that the space of ways to compose a pasting scheme
  is contractible.
\end{itemize}

\subsection{Some examples of derivations.}
\label{sec:catt-examples}
We provide some examples of derivations that one may compute in $\CaTT$, using
the actual syntax implemented in the tool~\cite{gitcatt}. This software relies
on the fact that the derivability of the judgments in the theory $\CaTT$ is
decidable and provides an algorithm to decide it. The inputs from the user are
interpreted by the system as typing judgments, and the software accepts an input
whenever it is able to find a derivation for the corresponding judgement. This
implementation follows the convention introduced by Finster and
Mimram~\cite{catt} and does not distinguish between the term constructors
$\cohop$ and $\coh$, assuming a single term constructor with two rules that are
mutually exclusive. As a result, all the new constructions are introduced with
the keyword \texttt{coh}, followed by a name to identify it. Then comes a list
of arguments which is the description of a ps-context followed by a column and a
type. For instance the following line
\begin{verbatim}
coh id (x : *) : x -> x
\end{verbatim}
defines a coherence called \texttt{id}, which corresponds to the construction
$\coh_{(x:\Obj):\Hom{}xx}$. Note that this expression is not a complete term, as
it lacks a substitution. Implicitly, we may assume that we have in fact defined
the term
\[
  (x:\Obj) \vdash \coh_{(x:\Obj),\Hom{\Obj}xx}[\id{(x:\Obj)}] : \Hom\Obj xx
\]
The derivation of this judgments is then guaranteed by the software (in this
example, it follows from an application of the rule \regle{coh}). We can then
use the admissibility of the action of substitutions (given by
Lemma~\ref{lemma:cat-derivation}) to define the term
$\coh_{(x:\Obj),\Hom{}xx}[\Gg]$ for any substitution. Thus further references to
this coherence just have to specify the substitution $\Gg$ towards the context
$(x:\Obj)$. We encode such a substitution as a list of arguments, for instance
one may write \verb?id y? to refer to the term identity, in a context containing
a variable \verb?y? of type \verb?*?. In general, we only specify some of the
argument for instance, considering the following declaration defining
composition
\begin{verbatim}
coh comp (x : *) (y : *) (f : x -> y) (z : *) (g : y -> z) : x -> z
\end{verbatim}
one needs to write only \verb?comp f g? instead of \verb?comp x y f z g? when
referring to it, as the terms \verb?x?, \verb?y? and \verb?z? can be inferred
from the data of \verb?f? and \verb?g?. \cref{lemma:glob-prods-catt} proves that
it suffices to provide the terms corresponding to the locally maximal variables
of the target ps-context, and the software implements an elaboration mechanism
that builds a full substitution out of only these arguments. Thus it can detect
automatically which argument should be left implicit and allows the user to
write shorter terms. Other examples of declarations one may define in \CaTT{}
include
\begin{itemize}
\item left unitality and its inverse
\begin{verbatim}
coh unitl (x : *) (y : *) (f : x -> y) : comp (id x) f -> f
\end{verbatim}
\begin{verbatim}
coh unitl- (x : *) (y : *) (f : x -> y) : f -> comp (id x) f
\end{verbatim}
\item right unitality and its inverse
\begin{verbatim}
coh unitr (x : *) (y : *) (f : x -> y) : comp f (id y) -> f
\end{verbatim}
\begin{verbatim}
coh unitr- (x : *) (y : *) (f : x -> y) : f -> comp f (id y)
\end{verbatim}
\item associativity and its inverse
\begin{verbatim}
coh assoc (x : *) (y : *) (f : x -> y) (z : *)
          (g : y -> z) (w : *) (h : z -> w)
          : comp f (comp g h) -> comp (comp f g) h
\end{verbatim}
\begin{verbatim}
coh assoc- (x : *) (y : *) (f : x -> y) (z : *)
           (g : y -> z) (w : *) (h : z -> w)
           : comp (comp f g) h -> comp f (comp g h)
\end{verbatim}
\item vertical composition of $2$-cells
\begin{verbatim}
coh vcomp (x : *) (y : *) (f : x -> y) (g : x -> y)
          (a : f -> g) (h : x -> y) (b : g -> h) : f -> h
\end{verbatim}
\item horizontal composition of $2$-cells
\begin{verbatim}
coh hcomp (x : *) (y : *) (f : x -> y) (f' : x -> y) (a : f -> f')
          (z : *) (g : y -> z) (g' : y -> z) (b : g -> g')
          : comp f g -> comp f' g'
\end{verbatim}
\item left whiskering
\begin{verbatim}
coh whiskl (x : *) (y : *) (f : x -> y) (z : *) (g : y -> z)
           (g' : y -> z) (b : g -> g') : comp f g -> comp f g'
\end{verbatim}
\item right whiskering
\begin{verbatim}
coh whiskr (x : *) (y : *) (f : x -> y) (f' : x -> y)
           (a : f -> f') (z : *) (g : y -> z) : comp f g -> comp f' g
\end{verbatim}
\end{itemize}
We also provide a syntax to define arbitrary compositions of the above
declarations in an arbitrary context. The corresponding keyword is \verb?let?
followed with an identifier and a context, the symbol \verb?=?, and a full
definition of the term using previously defined terms and declarations. For
instance, the following term defines the squaring of an endomorphism
\begin{verbatim}
let sq (x : *) (f : x -> x) = comp f f
\end{verbatim}
Note that the context associated to the keyword \verb?coh? is necessarily a
ps-context, whereas any context can be associated to the keyword \verb?let?.

\subsection{Properties of the theory $\CaTT$.}
In order to reason and prove results about $\CaTT$, we mostly reason
by induction on its terms. We thus first need to study some of the
properties of the syntax which, even though quite simple, will prove
quite useful in the following.

\paragraph{Preservation of the basic properties.}
The first thing that one can check about this theory is that the term
constructor are nice enough, so that the basic properties established for
$\Glob$ still hold for this new type theory.

\begin{lemma}
  \label{lemma:cat-derivation}
  All the properties of Lemma~\ref{lemma:glob-derivation} still hold in $\CaTT$,
  and every derivable judgment in $\CaTT$ has exactly one derivation.
\end{lemma}
\begin{proof}
  These results are proved as in the case of $\Glob$, by mutual induction on the
  derivation trees of the various judgments. The added term constructors make
  things a slightly more involved than in the case with only variables, and some
  of the properties that could be proved on a syntactical level in the theory
  $\Glob$ only hold for derivable judgments in the theory $\CaTT$. Apart for
  these technical subtleties, the generalization is straightforward.
\end{proof}

\paragraph{The syntactic category.}
As for the theory $\Glob$, the identity substitution $\id\GG$ associated to a
context $\GG$ is always derivable, as well as the composition of derivable
substitutions, using the action of substitution on raw terms. Moreover, all the
results that we have stated for the theory $\Glob$ still hold for the
theory~$\CaTT$: it is in particular the case for Proposition~\ref{prop:cwf},
Proposition~\ref{prop:category} and Proposition~\ref{prop:with-families}. This
shows that the derivable contexts of the theory $\CaTT$ assemble into a
category, whose morphisms are the derivable substitution. We write $\Syn{\CaTT}$
for this category and call it the \emph{syntactic category} of the
theory~$\CaTT$. The aforementioned results imply that~$\Syn{\CaTT}$ is equipped
with a canonical structure of category with families, where $\Ty^\GG$ is the set
of derivable types in the context $\GG$, and $\Tm^\GG_A$ is the set of terms of
type $A$ in the context $\GG$.

\paragraph{Inclusion of $\Syn{\Glob}$.}
The theory $\CaTT$ contains the theory $\Glob$ as a subtheory: anything that can
be derived using variables (in $\Glob$) only can still be derived using
variables and term constructors (in $\CaTT$). In particular, any valid context
$\GG$ in the theory $\Glob$ is also a valid context in the category $\CaTT$:
this is in particular the case for the disk contexts $D^n$ and the sphere
contexts $S^{n-1}$. Moreover, this induces a functor between the syntactic
categories $\Syn{\Glob}\to\Syn{\CaTT}$. It is immediate that this functor is a
morphism of categories with families, by taking any type (\resp any term) in the
theory $\Glob$ to the same type (\resp to the same term) in the theory $\CaTT$.

\paragraph{Familial representability of types.}
Central in the study of $\Glob$ was the \cref{lemma:up-disk-glob}, which
establishes that the family $S^\bullet$ familially represents the functor
$\Ty$. We have noted that its proof does not really depend on the extra terms
which are present in the theory, and thus immediately extends to the case
of~$\CaTT$:

\begin{lemma}
  \label{lemma:up-disk-catt}
    For any natural number $n$, the map
  \[
    \begin{array}{rcl}
      \Syn{\CaTT}(\GG,S^{n-1}) & \to & \setof{A \in \Ty^\GG}{\dim(A) = n-1} \\
      \Gg & \mapsto & U_{n}[\Gg]
    \end{array}
  \]
  is an isomorphism natural in $\GG$. Given a type $A$ of dimension $n-1$, we
  denote the associated substitution
  \[
    \Gc_A : \GG \to S^{n-1}
  \]
  Moreover, the maps
  \[
    \begin{array}{rcl}
      (\Syn{\CaTT}/S^{n-1})(\GG\xrightarrow{\Gc_A}S^{n-1},D^n\xrightarrow{\pi} S^{n-1}) & \to & \Tm^\GG_A \\
      \Gg & \mapsto & d_{2n}[\Gg]
    \end{array}
  \]
  are isomorphisms, natural in~$\Gamma$
  (the source is a hom-set in the slice category of $\Syn{\CaTT}$ over
  $S^{n-1}$). Given a term $t\in\Tm^\GG_A$ of type $A$, we denote the associated
  substitution over $\Gc_A$ by $\Gc_t:\GG\to D^{n}$, in such a way that the
  following diagram commutes
  \[
    \begin{tikzcd}
      \GG \ar[dr,"\Gc_A"']\ar[r,"\Gc_t"] & D^n\ar[d,"\pi"] \\
      & S^{n-1}
    \end{tikzcd}
  \]
\end{lemma}

\paragraph{Depth of a term.}
In order to study the theory $\CaTT$, we often reason by structural induction on
terms. In order to justify that these inductions are well-founded, we introduce
the notion of \emph{depth} of a term and of a substitution. It is the natural
number $\depth(t)$ (\resp{} $\depth(\gamma)$) defined by induction on the term $t$
(\resp{} substitution $\gamma$) by
\begin{align*}
  \depth(x) &= 0 & \depth(\coh_{\Gamma,A}[\Gg]) &= 1+ \depth(\Gg) \\
  \depth(\sub{}) &= 0 & \depth(\sub{\Gg,t\mapsto u})&= \max(\depth(\Gg),\depth(u))
\end{align*}
Informally, the depth of a term expresses how many nested term constructors are
needed to write it, and similarly for substitution. It should not be confused
with the notion of ``coherence depth'', introduced in \secr{syn-catt}.

\paragraph{Terms in the empty context.}
An important property that we can prove on the theory $\CaTT$, using induction
on the depth of terms, is that the there is no way to build a term in the empty
context.

\begin{lemma}
  \label{lemma:empty-ctx-catt}
  In the theory $\CaTT$ there is no term derivable in the empty context.
\end{lemma}
\begin{proof}
  We prove this result by induction on the depth of the term. First note that no
  variable is derivable in the empty context. A term of depth $d+1$ in the empty
  context has to be constructed using a substitution $\emptycontext \to\GD$ of
  depth $d$, where $\GD$ is a ps-context. Since $\GD$ is non-empty, such a
  substitution has to be built out of terms that are derivable in the empty
  context. Since the substitution is of depth at most $d$, these terms are of
  depth at most $d$ also, and by induction there is no such term, hence there is
  no such substitution. This proves that there is no term of depth $d+1$ in the
  context $\emptycontext$.
\end{proof}

\paragraph{Variables of the characteristic substitution.}
Using Lemma~\ref{lemma:up-disk-catt}, we can slightly reformulate the side
conditions of the rules~$\regle{op}$ and~$\regle{coh}$, which involve
expressions of the form $\Var{t:A}$.

\begin{lemma}
  \label{lemma:vars-char}
  Consider a context $\GG\vdash$, together with a term $\GG\vdash t:A$, then the
  following sets of variables are equal:
  \begin{align*}
    \Var A &= \Var{\Gc_A}
    &
    \Var{t:A} &= \Var{\Gc_t}
  \end{align*}
\end{lemma}
\begin{proof}
  We prove these two results by mutual induction, on the dimension of~$A$.
  \begin{itemize}
  \item If $\dim(A)=0$, then necessarily $A=\Obj$. We have $\Gc_{\Obj} = \sub{}$
    and $\Var\Obj = \Var{\sub{}} = \emptyset$.
  \item If $\dim(A)>0$, we can write $A=\Hom Btu$, and we have
    $\Gc_{A} = \sub{\Gc_t,d_{2n+1}\mapsto u}$ with
    $\dim(B)+1=\dim(A)$. Moreover, we have by definition
    \begin{align*}
      \Var A &= \Var B\cup \Var t\cup\Var u \\
      &= \Var {t:B} \cup \Var u \\
      \intertext{and on the other hand, we have}
      \Var{\Gc_A} &= \Var{\Gc_t}\cup\Var u
    \end{align*}
    The induction case for terms then shows $\Var A = \Var{\Gc_A}$.
  \item For $A$ of arbitrary dimension, we have
    $\Gc_t = \sub{\Gc_A,d_{2n}\mapsto t}$, and thus
    \begin{align*}
      \Var{t:A} &= \Var A\cup\Var t \\
      \Var{\Gc_t} &= \Var{\Gc_A} \cup\Var t
    \end{align*}
    The induction case for types then shows $\Var{t:A} = \Var{\Gc_t}$.\qedhere
  \end{itemize}
\end{proof}

\paragraph{Globular set of variables of a term.}
The contexts in the theory $\CaTT$ coming from the theory $\Glob$ play a
particular role in the theory, and we call them \emph{globular contexts}. They
are recognizable by the fact that they are built out only from variables, and as
we have shown in the definition of the functor $V$, their variables form into a
globular set. For instance, of the two following contexts, the first one is a
globular context, whereas the second one is not.
\begin{align*}
  &(x : \Obj,y:\Obj,z:\Obj,f:\Hom{}xy,g:\Hom{}zy)
  &
  &(x : \Obj, \alpha: \Hom{}{\texttt{id }x}{\texttt{id }x})
\end{align*}

\begin{lemma}
  \label{lemma:vars-tm-glob}
  Consider a globular context $\GG$ in the theory $\CaTT$ together with a
  derivable term $\GG\vdash t:A$. For every variable $x$ in the set $\Var{t:A}$,
  its source and target also belong to this set. This equips the set $\Var{t:A}$
  with a structure of a globular set which is a globular subset of $V\GG$.
\end{lemma}
\begin{proof}
  We prove this result by induction on the depth of the term $t$.
  \begin{itemize}
  \item Since $\GG$ is a context in $\Syn{\Glob}$, if the term $t$ is of depth
    $0$, then it is a variable $t=x$ and the map $\Gc_x : \GG\to D^n$ defines a
    map in $\Syn{\Glob}$. Then $\Var{\Gc_x}$ is the set of elements of the image
    of the map $V(\Gc_x) : V(D^n)\to V(\GG)$, so it is stable under source and
    target and is naturally a globular subset of $V(\GG)$. The result is then
    given by Lemma~\ref{lemma:vars-char}.
  \item If the term $t$ is of depth $d+1$, it is of the form
    $t=\cohop_{\GD,B}[\Gg]$ or $t = \coh_{\GD,B}[\Gg]$, with $\Gg$ a
    substitution of depth at most $d$. Consider a variable $x\in\Var{t:A}$ and
    denote respectively by~$y$ and~$z$ its source and target in
    $\GG$. Necessarily we have $x\in\Var\Gg$, and thus there exists a variable
    $x'$ in $\GD$ such that $x\in\Var{x'[\Gg]}$. Then consider the variables
    $y'$ and $z'$ that are respectively the source and target of $x'$ in $\GD$,
    in such a way that we have $\GD\vdash x':\Hom{}{y'}{z'}$. Then we have
    $\GG\vdash x'[\Gg] : \Hom{}{y'[\Gg]}{z'[\Gg]}$ with
    $x\in\Var{x'[\Gg]:\Hom{}{y'[\Gg]}{z'[\Gg]}}$ and $x'[\Gg]$ of depth at most
    $d$. By induction this proves that
    $y,z \in\Var{x'[\Gg] : \Hom{}{y'[\Gg]}{z'[\Gg]}}\subseteq\Var{t}$. Hence the
    source and target of $x$ belong to $\Var{t:A}$.\qedhere
  \end{itemize}
\end{proof}


\section{The syntactic categories associated to $\CaTT$}
\label{sec:syn-catt}
This section is dedicated to the study of the syntactic category
$\Syn{\CaTT}$. We have shown in Theorem~\ref{thm:ps-are-ps} that the subcategory
$\Syn{\ps}$ of the syntactic category $\Syn{\Glob}$ is equivalent to the
category $\COH_0$. We now show that adding the term constructors $\cohop$ and
$\coh$ allow us to recover exactly the missing pieces of information to obtain
weak $\omega$-categories: we exhibit a subcategory $\Syn{\ps,\infty}$ of the
category $\Syn{\CaTT}$ which is equivalent to the cat-coherator
$\op{\COH_\infty}$.

\subsection{A filtration in $\Syn{\CaTT}$.}
We consider the full subcategory $\Syn{\ps,\infty}$ of $\Syn{\CaTT}$, whose
objects are ps-contexts. Our aim is to exhibit this category as a colimit of the
form
\[
\Syn{\ps,\infty} = \colim\pa{\Syn{\ps,0}\to\Syn{\ps,1}\to\Syn{\ps,2}\to\ldots}
\]
that mimics the iterative construction of $\COH_\infty$ as a colimit of the
$\COH_n$ in the Grothendieck-Maltsiniotis definition of weak
$\omega$-categories.

\paragraph{Coherence depth.}
We introduce the notion of \emph{coherence depth} of a term, type or
substitution in order to construct the categories $\Syn{\ps,n}$. It is defined
inductively by
\begin{align*}
  \cd(v:A) &= \cd(A) & \cd(\cohop_{\Gamma,A}[\Gg]) &= \max(\cd(A) + 1,\cd(\Gg))\\
         & & \cd(\coh_{\Gamma,A}[\Gg]) &= \max(\cd(A) + 1,\cd(\Gg))\\
  \cd(\Obj) &= 0 & \cd(\Hom Atu) &= \max(\cd(A),\cd(t), \cd(u)) \\
  \cd(\sub{}) &= 0 & \cd(\sub{\Gg,x\mapsto t}) &= \max(\cd(\Gg),\cd(t))
\end{align*}
Note that this definition is distinct from the one of depth that we have
introduced in \secr{catt} for reasoning on syntax.

\paragraph{The filtration.}
We define the category $\Syn{\ps,n}$ to be the graph that has the same
objects as the category $\Syn{\ps,\infty}$ and whose morphisms are
substitutions of coherence depth at most $n$.
\Cref{lemma:bounded-cd} shows that this is actually a
subcategory of $\Syn{\ps,\infty}$. Note that in the case $n=0$, the
substitutions of coherence depth $0$ are the substitutions containing
only variables, and thus they are exactly the substitutions of
$\Syn{\Glob}$, \ie
\[
  \Syn{\ps,0}=\Syn{\ps}
\]
We can sum up the situation with the following diagram of inclusions
\[
  \begin{tikzcd}
    \Syn{\Glob}\ar[rrrr] &  &  &  & \Syn{\CaTT,\infty}\\
    \Syn\ps = \Syn{\ps,0} \ar[r]\ar[u] & \Syn{\ps,1}
    \ar[r] & \Syn{\ps,2} \ar[r] &
    \cdots\ar[r] & \Syn{\ps,\infty}\ar[u]\\
    \op{\G}\ar[u] & & & &
  \end{tikzcd}
\]
It is straightforward from the definition that $\Syn{\ps,\infty}$ is the colimit
of this sequence of morphisms of categories
\[
\Syn{\ps,\infty} = \colim\pa{\Syn{\ps} \to \Syn{ps,1}\to\Syn{ps,2}\to\cdots}
\]
Indeed, since all these functors are the identity on the objects, it amounts to
taking the colimit of the hom-sets, which define a filtration of sets:
\[
\set{\GD\vdash \Gg:\GG} = \bigcup_{n\in\N}\setof{\GD\vdash\Gg:\GG}{\cd(\Gg)\leq n}
\]

\paragraph{Properties of the coherence depth.}
The notion of coherence depth sometimes behaves awkwardly with respect
to the structure of the type theory. To illustrate this, consider the
context $(x:\Obj,\alpha : \Hom{}{\texttt{id }x}{\texttt{id }x})$:
although the term $\alpha$ is of coherence depth $0$, its type is of
coherence depth~$1$. This may be an issue when reasoning inductively
on the coherence depth, as one cannot consider all the terms and its
types (an example of this issue appears in \cref{lemma:reflectivity}).
However, we show below that such issues do not arise in globular
contexts, which are the only ones for which we are going to consider
coherence depths.

First note that the application of a substitution cannot increase the coherence
depth arbitrarily:

\begin{lemma}
  \label{lemma:bounded-cd}
  Given a substitution $\Gg$ we have
  \begin{itemize}
  \item for any type $A$, $\cd(A[\Gg]) \leq \max(\cd (A), \cd(\Gg))$,
  \item for any term $t$, $\cd(t[\Gg]) \leq \max(\cd (t), \cd(\Gg))$,
  \item for any substitution $\Gd$, $\cd(\Gd\circ\Gg) \leq
    \max(\cd(\Gd),\cd(\Gg))$.
  \end{itemize}
\end{lemma}
\begin{proof}
  We prove this result by mutual induction on the type, term and substitution.
  \begin{itemize}
  \item For the type $\Obj$, we have $\Obj[\Gg] = \Obj$, and hence
    $\cd(\Obj[\Gg]) = 0 \leq \max(0,\cd(\Gg))$.
  \item For the type $\Hom Atu$, we have
    \begin{align*}
      \cd((\Hom{A}{t}{u})[\Gg]) &= \cd(\Hom{A[\Gg]}{t[\Gg]}{u[\Gg]}) \\
                                &= \max(\cd(A[\Gg]),\cd(t[\Gg]),\cd(u[\Gg])) \\
                                &\leq \max(\cd(A), \cd(t),\cd(u),\cd(\Gg)) & \text{by induction} \\
                                &\leq \max(\cd(\Hom Atu),\cd(\Gg))
    \end{align*}
  \item For a variable $x$, we have $\cd(x[\Gg]) \leq \cd(\Gg)$ by definition of
    the coherence depth of a substitution.
  \item For the term $t = \cohop_{\GD,A}[\Gd]$, or for the term
    $t=\coh_{\GD,A}[\Gd]$, we have
    \begin{align*}
      \cd(t[\Gg]) &= \max(\cd(A)+1,\cd(\Gd\circ\Gg)) \\
                  &\leq \max(\cd(A)+1, \cd(\Gd),\cd\Gg) & \text{by induction} \\
                  &\leq \max(\cd(t),\cd(\Gg))
    \end{align*}
  \item For the substitution $\sub{}$, we have $\cd(\sub{}\circ\Gg) = 0 \leq
    \cd\Gg$.
  \item For the substitution $\sub{\Gd,x\mapsto t}$, we have
    \begin{align*}
      \cd(\sub{\Gd,x\mapsto t}\circ\Gg) &= \cd (\sub{\Gd\circ\Gg,x\mapsto t[\Gg]}) \\
                                        &=\max (\cd(\Gd\circ\Gg) , \cd(t[\Gg])) \\
                                        &\leq \max(\cd(\Gd),\cd(t),\cd(\Gg)) & \text{by induction} \\
                                        &\leq \max(\cd(\sub{\Gd,x\mapsto t}),\cd(\Gg))
    \end{align*}
  \end{itemize}
  From which we conclude.
\end{proof}

\begin{lemma}
  \label{lemma:cd-ty}
  In a globular context $\GG$, for every derivable term $\GG\vdash t:A$, we have
  $\cd(A)\le\cd(t)$.
\end{lemma}
\begin{proof}
  We distinguish between the case where $t$ is a variable and the case where $t$
  is obtained by application of a term constructors.
  \begin{itemize}
  \item If $t=x$ is a variable, and since it is derivable in a globular context,
    its type is derivable in the theory $\Glob$ and hence is of depth $0$.
  \item A term $t$ is not a variable, it is either of the form $t =
    \cohop_{\GD,B}[\Gd]$ or $t = \coh_{\GD,B}[\Gd]$, and in both cases we have
    $\cd(t) = \max(\cd(B)+1,\cd(\Gd))$, and the type $A$ is obtained as $A =
    B[\Gd]$. Lemma~\ref{lemma:bounded-cd} then shows that $\cd(B) \leq
    \max(\cd(B),\cd(\Gd)) \leq \cd(t)$.\qedhere
  \end{itemize}
\end{proof}

\begin{coro}
  \label{coro:cd-char}
  In a globular context $\GG$, for every type $\GG\vdash A$, we have $\cd(A) =
  \cd(\Gc_A)$ and for every term $\GG\vdash t:A$, we have $\cd(t) = \cd(\Gc_t)$.
\end{coro}
\begin{proof}
  We prove these two results by mutual induction on the dimension,
  \begin{itemize}
  \item For the type $\GG\vdash\Obj$, we have $\Gc_{\Obj} = \sub{}$, and by
    definition, $\cd(\Obj) = \cd(\sub{}) = 0$.
  \item For the type $\GG\vdash A$ of dimension $n \geq 0$, we can write $A=\Hom
    Btu$, and we have $\Gc_{A} = \sub{\Gc_t,d_{2n+1}\mapsto u}$. Applying
    Lemma~\ref{lemma:cd-ty} shows that $\cd(A) = \max(\cd(t),\cd(u))$. Moreover,
    by definition $\cd(\Gc_A) = \max(\cd(\Gc_t),\cd(u))$. The induction case for
    term then shows that $\cd(A) = \cd(\Gc_A)$.
  \item For a term $\GG\vdash t:A$ of dimension $n$, we have $\Gc_t =
    \sub{\Gc_A,d_{2n}\mapsto t}$, and we have by definition $\cd(\Gc_t) =
    \max(\cd(\Gc_A),\cd(t))$. The induction case for types together with
    Lemma~\ref{lemma:cd-ty} show that $\cd(\Gc_A) = \cd(A) \leq \cd(t)$ and
    hence $\cd(\Gc_t) = \cd(t)$.\qedhere
  \end{itemize}
\end{proof}

\subsection{Globular products in the category $\Syn{\CaTT}$.}
In order to show that $\Syn{\ps,\infty}$ dualizes the construction of the
category $\COH_\infty$, we characterize the globular products in this
category.

\paragraph{$\Syn{\CaTT}$ as a globular category with families.}
The inclusion functor $I : \Syn{\Glob} \to \Syn{\CaTT}$ induces a structure of
category with families on the category $\Syn{\CaTT}$, which coincides exactly
with the one given by Lemma~\ref{lemma:up-disk-catt}. Hence for this structure,
$I$ is a morphism of globular categories with families.
\begin{lemma}
  \label{lemma:glob-prods-catt}
  The inclusion functor $I : \Syn{\Glob}\to \Syn{\CaTT}$ preserves globular
  products.
\end{lemma}
\begin{proof}
  Since $I$ is a morphism of globular categories, it preserves the
  pullbacks along the display maps \(D^{n}\to S^{n-1}\). By
  \cref{lemma:ran-extension-pullbacks}, we have
  \(I \isoto \Ran_{D^{\bullet}} ID^{\bullet}\), and thus it preserves
  globular products because \(\Ran_{D^{\bullet}}ID^{\bullet}\) does,
  by \cref{lemma:ran-continuous}.
\end{proof}

\begin{lemma}
  \label{lemma:ps-glob-prods-catt}
  The inclusion functor $I : \Syn{\ps}\to\Syn{\ps,\infty}$ preserves globular
  products.
\end{lemma}
\begin{proof}
  Note that the inclusion of the full subcategory
  $\Syn{\ps,\infty}\hookrightarrow\Syn{\CaTT}$ reflects all limits. Moreover, by
  Lemma~\ref{lemma:glob-prods-catt}, the composite
  \[
  \Syn{\ps} \xrightarrow{I} \Syn{\ps,\infty}\hookrightarrow \Syn{\CaTT}
  \]
  preserves the globular products. Hence the functor $I$ also preserves the
  globular products.
\end{proof}

\paragraph{Reflexivity of the depth-bounded inclusion.}
There is a canonical functor $\Syn{\ps,n}\to\Syn{\ps,\infty}$, which consists in
forgetting that a substitution is of bounded coherence depth. In order to
understand the globular product in the categories $\Syn{\ps,n}$, it is useful to
study the behavior of this functor with respect to globular
products.

\begin{lemma}
  \label{lemma:reflectivity}
  The functor $\Syn{\ps,n}\to\Syn{\ps,\infty}$ reflects globular products.
\end{lemma}
\begin{proof}
  Consider an object $\GG$ which is a globular product in the category
  $\Syn{\ps,\infty}$, it suffices to show that is also a globular product in the
  category $\Syn{\ps,n}$. Any cone of apex $\GD$ over the diagram of $\GG$ in
  $\Syn{\ps,n}$ induces a cone over the diagram of $\GG$ in $\Syn{\ps,\infty}$,
  which by definition of a limit defines a unique substitution $\Gg:\GD\to\GG$,
  and it suffices to show that this substitution is in fact in $\Syn{\ps,n}$. By
  definition, all the maps $\Gc_{x[\Gg]}$ where $x$ is a maximal variable appear
  in the legs of the cone of apex $\GD$. Since these legs are chosen in the
  category $\Syn{\ps,n}$, this shows that for every locally maximal variable
  $\Gc_{x[\Gg]}$ is of depth at most $n$, and hence by
  Corollary~\ref{coro:cd-char}, $x[\gamma]$ is of depth at most $n$. Applying
  Lemma~\ref{lemma:cd-ty} ensures that all the iterated sources and targets of
  all the $x[\gamma]$ are of depth at most $n$, and since every variable of $\GG$
  is obtained as an iterated source or target of variables of dimension locally
  maximal in $\GG$, all the $x[\Gg]$ for every variable $x$ in $\GG$ is of depth
  at most $n$. By definition, this means that $\Gg$ is of depth at most $n$, and
  hence $\Gg$ is a substitution in $\Syn{\ps,n}$.
\end{proof}

\paragraph{Globular products in the category $\Syn{\ps,n}$.}
All the categories $\Syn{\ps,n}$, for $n\in\N\cup\set{\infty}$, have the same
objects, and there are more and more morphisms when $n$ increases. None of these
categories are equipped with a structure of category with families, since they
lack the possibility of extending the context by any type. However for $n = 0$
and $n=\infty$ we can exhibit them as full subcategories of categories with
families. Using these structure of category with families, we could prove that
the functor $\Syn{\ps}\to\Syn{\ps,\infty}$ preserves globular products
(\cref{lemma:ps-glob-prods-catt}), and we now use this result to study globular
products in all the categories $\Syn{\ps,n}$.

\begin{lemma}
  \label{lemma:glob-prods-cat}
  The functors $\Syn{\ps,0}\rightarrow \Syn{\ps,n}$ preserve globular products.
\end{lemma}
\begin{proof}
  We have the commutative triangle
  \[
    \begin{tikzcd}
      \Syn{\ps,n}\ar[r] & \Syn{\ps,\infty} \\
      \Syn{\ps,0} \ar[u]\ar[ru]
    \end{tikzcd}
  \]
  By Lemma~\ref{lemma:glob-prods-catt}, the functor
  $\Syn{\ps,0}\to\Syn{\ps,\infty}$ preserves the globular products and, by
  Lemma~\ref{lemma:reflectivity}, the functor $\Syn{\ps,n}\to\Syn{\ps,\infty}$
  reflects the globular products. This implies the the functor
  $\Syn{\ps,0}\to\Syn{\ps,n}$ preserves globular products.
\end{proof}

\begin{lemma}
  \label{lemma:n->n+1}
  The categories $\Syn{\ps,n}$ are contravariant globular extensions
  and the inclusion functors $\Syn{\ps,n}\rightarrow \Syn{\ps,n+1}$
  are morphisms of contravariant globular extensions.
\end{lemma}
\begin{proof}
  Lemma~\ref{lemma:glob-prods-cat} in conjunction with
  Lemma~\ref{lemma:cat-glob-ext} and Theorem~\ref{thm:ps-are-ps} shows that the
  functor $\Syn{\ps,0}\to\Syn{\ps,n}$ endows $\Syn{\ps,n}$ with a structure of
  contravariant globular extension. Moreover, Lemma~\ref{lemma:cat-glob-ext}
  lifts the commutative triangle
  \[
    \begin{tikzcd}
      \Syn{\ps,n}\ar[r] & \Syn{\ps,n+1}\\
      \Syn{\ps,0}\ar[u]\ar[ru]
    \end{tikzcd}
  \]
  into a morphism of contravariant globular extension
  $\Syn{\ps,n}\to\Syn{\ps,n+1}$.
\end{proof}

\paragraph{$\Syn{\ps,n}$ as a contravariant globular theory.}
Assembling altogether the results we have proved about the categories
$\Syn{\ps,n}$, with $n\in\N\cup\set{\infty}$, we have the following:

\begin{prop}
  For $n\in\N\cup\set{\infty}$, the category $\Syn{\ps,n}$ is equipped with a
  structure of a contravariant globular theory, and the functors
  $\Syn{\ps,n}\to\Syn{\ps,n+1}$ are morphisms of contravariant globular
  theories.
\end{prop}
\begin{proof}
  Lemma~\ref{lemma:n->n+1} already shows that $\Syn{\ps,n}$ is a contravariant
  globular extension, moreover note that $\Syn{\ps}$ and $\Syn{\ps,n}$ have the
  same objects, but $\Syn{\ps,n}$ has strictly more morphisms, and the functor
  $\Syn{\ps}\to\Syn{\ps,n}$ sends every object to itself and defines the
  inclusion of the morphisms. Hence it defines a contravariant globular theory.
  The same reasoning starting from Lemma~\ref{lemma:glob-prods-catt} shows that
  $\Syn{\ps,\infty}$ is also a contravariant globular theory. By
  Lemma~\ref{lemma:n->n+1} the functor $\Syn{\ps,n}\to\Syn{\ps,n+1}$ is a
  morphism of contravariant globular extensions, and hence it is also a morphism
  of contravariant globular theories.
\end{proof}

\subsection{Admissible pairs of substitutions.}
In the contravariant globular theory $\Syn{\ps,m}$, for
$m\in\N\cup\set{\infty}$, we consider a morphism $\Gx : \GD\to D^n$.
By Lemma~\ref{lemma:up-disk-catt}, such a morphism can be written as
$\Gx = \Gc_t$ for some term $t$ of type $A$ in $\Delta$. Using the
notations introduced in \secr{glob} for defining the disk and sphere
contexts, the term $t$ in $\GD$ can be recovered as $t = d_{2n}[\Gx]$
and the type of $t$ in $\GD$ is $U_n[\Gx]$. Note that $U_n$ contains
all the variables of $D^n$, except for the variable $d_{2n}$, and
hence $\Var{d_{2n}}\cup\Var{U_n} = \Var{D^n}$. This equality shows
\begin{align*}
  \Var\Gx &= \Var{d_{2n}[\Gx]}\cup\Var{U_n[\Gx]} \\
          &= \Var t\cup\Var A
\end{align*}
\begin{lemma}
  \label{lemma:algebraic-tm}
  Given a term $\GD\vdash t:A$, the morphism $\Gc_t : \GD\to D^n$ is algebraic
  in $\Syn{\ps,m}$ if and only if $\Var{t : A} = \Var{\GD}$
\end{lemma}
\begin{proof}
  First suppose that $\Var{t:A} = \Var{\GD}$, and consider a factorization of
  the form
  \[
    \begin{tikzcd}
      \GD\ar[rr, bend left, "\Gc_t"]\ar[r,"\Gg"'] & \GG \ar[r,"\Gc_u"'] & D^n
    \end{tikzcd}
  \]
  with $\Gg$ a globular substitution, \ie a substitution in
  $\Syn{\ps,0}$. Then we have a term $\GG\vdash u:B$ such that
  $B[\Gg] = A$ and $u[\Gg] = t$. The condition
  $\Var {t : A} = \Var\GD$ then implies in particular that
  $\Var\GD\subset\Var\Gg$. Note that under the correspondence of
  Theorem~\ref{thm:syn-glob}, $\Var{\Gg}$ is the set of elements in
  the image of the map $V(\Gg) : V\GG\to V\GD$, and the equation
  $\Var{\GD} \subset \Var{\Gg}$ then shows that the map $V(\Gg)$ is a
  surjective map of globular sets. By Lemma~\ref{lemma:maps-ps}, any
  map between two pasting schemes is injective so in particular that
  $V(\Gg)$ is an isomorphism, and Lemma~\ref{lemma:aut-ps} shows that
  then it is an identity. By Theorem~\ref{thm:ps-are-ps}, $V$ is an
  isomorphism of categories between $\Syn{\ps,0}$ and $\COH_{0}$.
  Hence $\Gg$ is an identity. This proves that $\Gc_t$ is an algebraic
  morphism. Conversely suppose that the morphism $\Gc_t$ is algebraic.
  Lemma~\ref{lemma:vars-tm-glob} shows that the set $\Var{t:A}$ can be
  viewed as a sub globular set of $V(\GD)$. By the equivalence of
  Theorem~\ref{thm:syn-glob} the inclusion $\Var{t:A}\to V(\GD)$
  provides a globular substitution $\GD\vdash\Gg:\GG$. Moreover, by
  definition, we have $\GG\vdash t:A$ and $A[\Gg] = A$, $t[\Gg] = t$.
  Hence by algebraicity of $\Gc_t$, this shows that $\Gg$ is an
  identity. This proves the inclusion $\Var(t:A)\to V(\GD)$ is the
  identity and thus $\Var{t:A} = V(\GD)$, which by forgetting the
  globular set structure implies $\Var{t :A} = \Var\GD$.
\end{proof}

\begin{lemma}\label{lemma:admissible-ty}
  The pairs of admissible morphisms in $\GG$ are classified by the
  types $\GG\vdash A$ satisfying either $\Vop$ or $\Vcoh$. For such a
  type $\GG\vdash A$, the terms $\GG\vdash t:A$ classify exactly the
  lifts of the corresponding admissible pair. More precisely, there is
  a natural isomorphisms between pairs of admissible morphisms in
  $\Gamma$ and types satisfying $\Vop$ or $\Vcoh$, as well as a
  natural isomorphism between terms of of such types and lifts of the
  corresponding isomorphisms.
\end{lemma}
\begin{proof}
  The types $\GG\vdash A$ of non-zero dimension, with $A$ of the form $\Hom
  Btu$, classify the pairs of terms $(t,u)$ of same type $B$, which are exactly
  the pairs of parallel maps $(\Gc_t,\Gc_u)$. Moreover, such a pair is
  admissible whenever we have one of the following.
  \begin{itemize}
  \item Both $\Gc_t$ and $\Gc_u$ are algebraic, which by
    Lemma~\ref{lemma:algebraic-tm} translates to the two conditions $\Var {t :
      B} = \Var \GG$ and $\Var{u : B} = \Var \GG$: this is exactly the condition
    $\Vcoh$.
  \item $\Gc_t$ factors through the source inclusion of $\GG$ as a algebraic
    morphism and $\Gc_u$ factors through the target as a algebraic morphism.
    Again, by Lemma~\ref{lemma:algebraic-tm}, these conditions translate to
    $\src\GG\vdash t:B$ with $\Var {t: B} = \Var{\src\GG}$ and $\tgt\GG\vdash
    u:B$ with $\Var {u : B} = \Var{\tgt\GG}$: this is the condition $\Vop$.
  \end{itemize}
  A~lift for such a admissible pair is a map $\Gx:\GG \to D^{\dim A + 1}$, such
  that we have both $s(\Gx) = \Gc_t$ and $t(\Gx) = \Gc_u$. In the category
  $\Syn{\CaTT}$ we can encode this data as a substitution $\chi_A:\Gamma\to
  S^{\dim A}$:
  \[
    \begin{tikzcd}
      \GG \ar[dr,"\Gc_A"]\ar[drr, bend left, "\Gc_t"]\ar[ddr,bend
      right,"\Gc_u"']
      \\
      & S^{\dim A}\ar[r,"s"]\ar[d,"t"']\ar[dr, phantom,"\lrcorner"very
      near start] & D^{\dim A}\ar[d] \\
      & D^{\dim A}\ar[r] & S^{\dim A -1}
    \end{tikzcd}
  \]
  A~lift thus amounts to a morphism $\Gx : \GG\to D^{\dim A +1}$ in
  $\Syn{\CaTT}$ which makes the following triangle commute:
  \[
    \begin{tikzcd}
      \GG \ar[r,"\Gx"]\ar[rd, "\Gc_A"'] & D^{\dim A+1}\ar[d,"\pi"]\\
      & S^{\dim A}
    \end{tikzcd}
  \]
  By Lemma~\ref{lemma:up-disk-catt}, these are classified by the terms
  $\GG\vdash t:A$ in the theory $\CaTT$.
\end{proof}

\subsection{Equivalence between $\Syn{\ps,\infty}$ and $\op{\COH_\infty}$.}
We now prove the main theorem, that the category $\Syn{\ps,\infty}$ is
equivalent to the opposite of the cat-coherator $\COH_\infty$. This result thus
identifies the cat-coherator $\COH_\infty$ as a full subcategory of the category
of the category with families $\Syn{\CaTT}$.
We define the set $F_n$ to be the set of all types $\GG\vdash \Hom Atu$ of
coherence depth exactly $n$ in a ps-context $\GG$, satisfying $\Vop$ or $\Vcoh$.
By Lemma~\ref{lemma:admissible-ty}, the family $F_n$ can be defined inductively
as the set of all pair of admissible maps in $\Syn{\ps,n}$ that do not belong to
any $F_{n'}$ for $n < n'$.
\begin{lemma}\label{lemma:tower-def-n}
  The inclusion $\Syn{\ps,n}\to\Syn{\ps,n+1}$ exhibits $\Syn{\ps,n+1}$
  as the universal coglobular extension of $\Syn{\ps,n}$ equipped with
  a lift for all pair of morphisms in $F_n$.
\end{lemma}
\begin{proof}
  By Lemma~\ref{lemma:n->n+1}, this functor is a morphism of coglobular
  theories. Moreover consider a admissible pair $(f,g) : \GG\to D^n$ in $F_n$
  corresponding to a type $\GG\vdash A$ in the ps-context $\GG$, which satisfies
  $\Vop$ or $\Vcoh$ and which is of depth $n$. We can derive a term $t$ by
  $\GG\vdash \cohop_{\GG,A}[\id\GG]:A$ if $A$ satisfies $\Vop$, or
  $\GG\vdash\coh_{\GG,A}[\id\GG]:A$ if $A$ satisfies $\Vcoh$, the term $t$ is
  then of coherence depth $n+1$. Hence $t$ defines a map $\chi_t$ in the
  category $\Syn{\ps,n+1}$, which by Lemma~\ref{lemma:admissible-ty} is a lift
  for the admissible pair $(f,g)$. We have thus proved that $\Syn{\ps,n+1}$ is a
  contravariant globular extension which contains a lift for all pairs in $F_n$.
  We now show that this extension is universal: consider another extension
  $F:\Syn{\ps,n}\to C$ that defines a lift for all the pairs in $F_n$, we show
  that there exists a unique $\tilde F$ that preserves the chosen lifts and
  makes the following diagram commute
  \[
    \begin{tikzcd}
      \Syn{\ps,n}\ar[d]\ar[r,"F"]&C\\
      \Syn{\ps,n+1}\ar[ur,dotted,"\tilde F"']
    \end{tikzcd}
  \]
  Indeed, the map $\tilde F$ is already defined on all objects of
  $\Syn{\ps,n+1}$, and all maps of coherence depth less than $n$, so that it
  coincides with $F$, so it suffices to show that there is a unique extension to
  the maps of coherence depth $n+1$. Since all the objects in $\Syn{\ps,n+1}$
  are globular products, it suffices to show this for the maps of the form
  $\GG\to D^n$. We can thus reformulate the condition by saying that it suffices
  to show that there is a unique map $\tilde F$ on terms, satisfying the
  condition $\tilde F(t[\Gg]) = \tilde Ft\circ \tilde F\Gg$. We proceed by
  induction on the depth, noticing that a term of coherence depth $n+1$ cannot
  be a variable, hence we have already defined a unique value for $\tilde F$ on
  terms of depth $0$, by our previous condition, and thus the induction is
  already initialized.
  \begin{itemize}
  \item For a term $\GD\vdash \cohop_{\GG,A}[\Gg]:A[\Gg]$ of depth $d + 1$, the
    value of $F$ is uniquely determined by $\tilde F(\cohop_{\GG,A}[\Gg]) =
    \tilde F(\cohop_{\GG,A}[\id\GG])\tilde F \Gg$, and since $\Gg$ is of depth
    $d$, by induction $\tilde F(\Gg)$ is defined, and $\tilde
    F(\cohop_{\GG,A}[\id\GG])$ is uniquely defined by the condition of
    preserving the lifts for the pairs in $F_n$.
  \item Similarly, for a term $\GD\vdash \coh_{\GG,A}[\Gg]:A[\Gg]$ of depth $d +
    1$, the value of $F$ is uniquely determined by $\tilde
    F(\cohop_{\GG,A}[\Gg]) = \tilde F(\cohop_{\GG,A}[\id\GG])\tilde F \Gg$, and
    since $\Gg$ is of depth $d$, by induction $\tilde F(\Gg)$ is defined, and
    $\tilde F(\coh_{\GG,A}[\id\GG])$ is uniquely defined by the condition of
    preserving the lifts for the pairs in $F_n$.
  \end{itemize}
  This proves that there exists a unique $\tilde F$ satisfying the condition,
  and hence $\Syn{\ps,n+1}$ is the universal coglobular extension obtained by
  adding a lift for all arrows in $F_n$ to $\Syn{\ps,n}$
\end{proof}

\noindent
This establishes a close correspondence between the categories $\Syn{\ps,n}$ and
$\COH_n$, and enables us to prove the following theorem.

\begin{thm}
  \label{thm:syn-ps}
  We have an equivalence of categories
  \[
    \Syn{\ps,\infty}\equivto\op{\COH_\infty}
  \]
\end{thm}
\begin{proof}
  By construction $\Syn{\ps,\infty}$ is obtained as the colimit of the
  inclusions of categories
  \[
    \op{\G}\to\Syn{\ps,0}\to\Syn{\ps,1}\to\cdots\to\Syn{\ps,n}\to\cdots\to\Syn{\ps,\infty}
    = \colim_n \Syn{\ps,n}
  \]
  It is therefore enough to prove that $\Syn{\ps,n}$ is equivalent to
  $\op{\COH_n}$, which we do by induction.
  \begin{itemize}
  \item We have already proved that $\Syn{\ps,0}$ is equivalent to $\op{\COH_0}$
    in Theorem~\ref{thm:ps-are-ps}.
  \item Suppose that $\Syn{\ps,k}$ is equivalent to $\op{\COH_k}$ for every
    $k\leq n$. Lemma~\ref{lemma:tower-def-n} shows that $\Syn{\ps,n+1}$ is the
    universal contravariant globular extension that adds a lift for each pair in
    the set $F_n$. Moreover, the set $F_n$ coincides with the set $E_n$ defined
    in \secr{gm-def} and, by definition, $\COH_{n+1}$ is the universal globular
    extension. Hence $\Syn{\ps,n+1}$ and $\op{\COH_{n+1}}$ satisfy the same
    universal property and are therefore equivalent.\qedhere
  \end{itemize}
\end{proof}


\section{Models of $\CaTT$}
\label{sec:models-catt}
This section is dedicated to the study of the models of the type theory
$\CaTT$ using tools that generalize the ones developed in
\secr{glob-cwf}. In particular, we prove an initiality result
analogous to \cref{thm:weak-initiality-glob} for the category
$\Syn{\CaTT}$. We then apply this result to characterize the
$\Set$-models of the theory and prove that they are equivalent to the
weak $\omega$-categories in the sense of Grothendieck-Maltsiniotis,
presented in \secr{gm-def}. We also give a detailed syntactic
interpretation of the construction that we develop here, showing that
although it uses abstract categorical machinery, it translates closely
the intuition coming from type theory.

\subsection{$\CaTT$-categories with families.}
In the case of the category $\Glob$, we have introduced the notion of
globular category with families, and proved that $\Syn{\Glob}$ is
initial among them (\cref{thm:weak-initiality-glob}), which implies
that we can compute the semantics of this theory in any category with
families. We further prove this result by defining the structure of a
\emph{$\CaTT$-category with families}, which plays an analogue role
for the theory $\Syn{\CaTT}$. We denote $D_P$ the functor
$\op\G\to\op{\COH_\infty}$, which defines the disk objects in the
category $\op{\COH_\infty}\simeq \Syn{\ps,\infty}$. It is the
corestriction of the functor $D^{\bullet}$ to ps-contexts.
\begin{defi}
  A $\CaTT$-category with families is a globular category with families
  $\C$ together with a functor $F : \op{\COH_\infty} \to \C$ sending
  globular sums to globular products, such that
  $G_{\C} = FD_{P}$.
\end{defi}

\noindent
Our main example of a $\CaTT$-category with families is the syntactic
category $\Syn{\CaTT}$. The associated functor, that we write
$P_\infty : \op{\COH_\infty}\to\Syn{\CaTT}$, is given by the inclusion
$\Syn{\ps,\infty}\to\Syn{\CaTT}$, together with the identification
given by Theorem~\ref{thm:syn-ps}. In fact a $\CaTT$-category with
families can be thought of as a category with families which supports
a type $\Obj$ along with all its iterated types $\Hom{}{}{}$, and for
which the term constructor $\cohop$ and $\coh$ exist, like in the
theory $\CaTT$. From now on, we use Theorem~\ref{thm:syn-ps}
implicitly to identify the categories $\Syn{\ps,\infty}$ and the
categories $\op{\COH_\infty}$ and we think of an object of
$\COH_{\infty}$ as a ps-context, and of a map
$\Gg\in \COH_{\infty}(\GG,\GD)$ as a substitution $\GD\vdash\Gg:\GG$
in $\CaTT$. Lemma~\ref{lemma:up-disk-catt}, lets us think of maps
$f:\COH_\infty(D^n,\GG)$ as terms in the ps-context $\GG$.

\paragraph{Morphisms of $\CaTT$-categories with families.}
A morphism between two $\CaTT$-categories with families
$F : \op{\COH_\infty} \to \C$ and $G : \op{\COH_\infty} \to \D$ is a
morphism of categories with families $f : \C\to\D$ together with a
natural transformation
\[
  \begin{tikzcd}
    \C \ar[r,"f"]\ar[dr, "{\Rightarrow}"{sloped, near
    start}, phantom] & \D \\
    \op{\COH_\infty} \ar[u,"F"]\ar[ru, "G"'] & \phantom{1}
  \end{tikzcd}
\]
We denote $\cCwF$ the category of $\CaTT$-categories with families defined
this way. We also define a $2$-cell of $\CaTT$-category with families
between two morphisms $(f,\alpha),(g,\beta) : \C\to \D$ to be a
natural transformation $\gamma : f \Rightarrow g$ such that
\[
  \begin{tikzcd}
    \C \ar[r,"f"]\ar[dr, "\overset{\alpha}{\Rightarrow}"{sloped, near
    start}, phantom] & \D \\
    \op{\COH_\infty} \ar[u,"F"]\ar[ru, "G"'] & \phantom{1}
  \end{tikzcd}
  =
   \begin{tikzcd}
     \C \ar[r,"g"{description}]
     \ar[r,"f",bend left = 50, "f"]
     \ar[r, bend left = 25, phantom, "\Downarrow_{\gamma}"]
     \ar[dr, "\overset{\beta}{\Rightarrow}"{sloped, near
     start}, phantom]
     & \D \\
    \op{\COH_\infty} \ar[u,"F"]\ar[ru, "G"'] & \phantom{1}
  \end{tikzcd}
\]
We denote $\cCwFh$ the $2$-category obtained this way.

\paragraph{The two nerve functors.}
Given a $\CaTT$-category with families $F : \op{\COH_\infty}\to\C$. We define
its associated nerve functor $N_{F}$.
\begin{align*}
  N_F : \C&\to\widehat{\COH_\infty}\\
  \GG&\mapsto\C(\GG,F\_)
\end{align*}
Recall that the nerve functor associated to $FD_{P}$ plays the
role of classifying terms, as per the theory of globular categories
with families. We denote it $T_{F}$.
\begin{align*}
  T_F : \C&\to\widehat\G\\
  \GG&\mapsto\C(\GG,FD_{P}\_)
\end{align*}
In the case of the $\CaTT$-category with families $\Syn{\CaTT}$, we simply
denote $N$ and $T$ these two functors. Our aim is to reproduce the arguments
given in \secr{glob-cwf} to show the equivalence between \(\CaTT\)-category
with families structures and morphism from the syntactic category
\(\Syn{\CaTT}\). In \secr{glob-cwf}, the construction relies on the fact
that the functor \(D^{\bullet}\) is fully faithful and its associated nerve
functor sends limits onto colimits. In our case, the functor \(P_{\infty}\) is
also fully faithful, but there is no reason for its nerve functor \(N\) to send
limits onto colimits. We introduce the concept of algebraic natural
transformation and show in \cref{lemma:extends-alg} that they satisfy a
condition of preservation of pullbacks along display maps as a workaround for
\(N\) not preserving limits.

\subsection{Algebraic natural transformations.}
Consider two $\CaTT$-categories with families given by $F:\op{\COH_\infty}\to\C$
and $G:\op{\COH_\infty}\to\D$, along with an object $\GG$ in $\C$ and an object
$\GD$ in $\D$. We define a notion of \emph{algebraic natural transformation}
between $T_G\GD$ and $T_F\GG$ in the category $\widehat\G$. This can be seen as
a compatibility condition, and might seem ad-hoc at first, but the reason why we
are interested in such transformations will be apparent in
Proposition~\ref{prop:alg<->nerve}, and we provide in
\secr{algebraic-transformations-and-substitutions} a discussion showing that
from the point of view of type theory, they are actually a very natural notion
to consider.

\paragraph{Induced nerve transformation.}
Consider two $\CaTT$-category with families
$F: \op{\COH_{\infty}}\to\C$ and $G : \op{\COH_{\infty}}\to\D$,
together with a natural transformation
$\eta\in\widehat{\G}(T_G\GD,T_F\GG)$. For every globular set $X$,
$\eta$ induces by composition, the following transformation natural in
$X$:
\begin{align*}
  \eta^\star : \widehat\G(X,T_G\GD)&\to \widehat\G(X,T_F\GG)\\
  \xi&\mapsto\eta\xi
\end{align*}
In the case where $X$ is of the form $V\GTH$ for a ps-context
$\GTH\in\COH_0$, \cref{prop:models-gcwf}, we have
$G\simeq \Ran_{D^{\bullet}}(GD^{\bullet})$, and the characterization
of Kan extensions given by Lemma~\ref{lemma:ran-V}, shows that we have
the following two natural isomorphisms:
\begin{align*}
  \widehat{\G}(V\GTH,T_G\GD)&\isoto\D(\GD,G\GTH) = (N_G\GD)_\GTH
  \\
  \widehat{\G}(V\GTH,T_F\GD)&\isoto\C(\GG,F\GTH) = (N_F\GG)_\GTH
\end{align*}
This construction is natural in $\Theta\in\COH_0$, hence $\eta^\star$
is a natural transformation
$\eta^{\star}\in \widehat{\COH_\infty}(N_{G}\GD,N_{F}{\GG})$. We thus
have constructed the following transformation, which is natural in
both $\GD$ and $\GG$
\[
\_^\star : \widehat{\G}(T_G\GD,T_F\GG) \to \widehat{\COH_0}(N_{G}\GD,N_{F}\GG)
\]
More explicitly, for $\Gg\in\D(\GD,GX)$, the transformation
$\eta^\star (\Gg)$ is characterized by
\[
 \text{For every $\Gc \in \COH_0(D^n,X)$, } (F\Gc)\eta^\star(\Gg) = \eta((G\Gc)\Gg)
\]
\begin{defi}
  A natural transformation $\eta\in\widehat{\G}(T_G\GD,T_F\GG)$ is
  \emph{algebraic} if
  \[
    \text{For every $\Gth \in\D(\GD,G\GTH)$ and
      $\Gc \in \COH_\infty(D^n,X)$, }\eta((G\Gc)\Gth) =
    (F\Gc)\eta^{\star}\Gth
  \]
  We write $\widehat{\G}(T_G\GD,T_F\GG)_{\alg}$ the set of algebraic
  natural transformations between $T_G\GD$ and $T_F\GG$
\end{defi}

\noindent
The algebraicity condition is a generalization of the defining
equality of $\eta^\star$, required to hold on $\COH_\infty$ instead of
only on $\COH_0$.

\paragraph{Algebraicity as a naturality condition.}
A natural transformation $\eta \in\widehat{\G}(T_G\GD,T_F\GG)$ induces
a natural transformation
$\eta^\star \in\widehat{\COH_0}(N_{G}\GD,N_{F}\GG)$. The following
result shows that when $\eta$ is algebraic, $\eta^{\star}$ satisfies a
stronger naturality condition. This is the main motivation for the
introduction of algebraicity.
\begin{lemma}
  \label{lemma:nt-alg-nerve}
  If $\eta\in\widehat{\G}(T_G\GD,T_F\GG)_{\alg}$ is an algebraic natural
  transformation, then $\eta^\star$ defines a natural transformation
  $\eta^\star\in\widehat{\COH_{\infty}}(N_G\GD,N_F\GG)$.
\end{lemma}
\begin{proof}
  We have already defined, for every element $\Gg$ of the presheaf $N_G\GD$ an
  element $\eta^\star(\Gg)$ of the presheaf~$N_F\GG$, and it is enough to verify
  that it induces a natural transformation between the presheaves~$N_G\GD$
  and~$N_F\GG$ over $\COH_\infty$. Consider an element $\Gg$ of $N_G(\GD)$, \ie
  $\Gg\in\D(\GD,GX)$, and recall that $\eta^\star (\Gg)$ is defined to be the
  transformation such that for every map $\Gc \in \COH_0(D^n,X)$, we have
  $(F\Gc)\eta^\star(\Gg) = \eta((G\Gc)\Gg)$. Given a map $f\in X\to Y$ in
  $\COH_\infty$, we have
  \begin{align*}
    (F\Gc)(Ff)\eta^\star(\Gg) &= F(\Gc f) \eta^\star(\Gg) \\
                                      &= \eta(G(\Gc f) \Gg) &&\text{by algebraicity}\\
                                      &= \eta((G\Gc)(Gf)\Gg) \\
                                      &= F\Gc\eta^\star((Gf)\Gg) &&\text{by definition of $\eta^\star$}
  \end{align*}
  Hence, for every variable $x$ derivable in the context $\GD$ in the theory
  $\Syn{\Glob}$, we have
  \[
    (F\Gc_x)(Ff)\eta^\star(\Gg) = (F\Gc)\eta^\star((Gf)\Gg)
  \]
  and thus we have the equality
  \[
    (Ff)\eta^\star(\Gg) = \eta^\star(Gf\Gg)
  \]
  This proves the commutativity of the square
  \[
    \begin{tikzcd}
      \D(\GD,GY) \ar[r,"\eta^\star_Y"]\ar[d,"Gf\circ\_"'] & \C(\GG,FY)\ar[d,"Ff\circ\_"] \\
      \D(\GD,GX)\ar[r,"\eta^\star_X"'] & \C(\GG,FX)
    \end{tikzcd}
    \qedhere
  \]
\end{proof}

\paragraph{Algebraicity of the nerve transformations.}
We now show the converse: Any given natural transformation
$\eta \in\widehat{\COH_{\infty}}(N_G\GD,N_F\GG)$ induces a natural
transformation $\eta\in\widehat{\G}(T_G\GD,T_F\GG)$ by restriction
along $D_{P}$. We show that this transformation is algebraic.

\begin{lemma}
  \label{lemma:restrict-nt}
  Consider a natural transformation
  $\eta \in\widehat{\COH_{\infty}}(N_G\GD,N_F\GG)$ along with its
  restriction $\overline\eta\in\widehat\G(T_G\GD,T_F\GG)$. The induced
  natural transformation
  ${\overline\eta}^\star \in\widehat{\COH_0}(N_G\GD,N_F\GG)$ coincides
  with~$\eta$.
\end{lemma}
\begin{proof}
  Consider an element $\Gd\in N_G\GD$, \ie $\Gd$ is a map $\Gd : \GD\to GX$ in
  the category $\D$ for some object $X$ of $\COH_0$. Then
  $\overline\eta^\star(\Gd)$ is defined to be the unique map such that for every
  map $\Gc : X\to D^n$ in $\op{\COH_0}$ we have
  $(F\Gc)\overline\eta^\star(\Gd) = \overline\eta((G\Gc)\Gd)$. The
  naturality of the transformation~$\eta$ ensures that
  $(F\Gc)\eta(\Gd) = \eta((G\Gc)\Gd)$ for every map $\Gc$ in the category
  $\COH_\infty$. In particular, this is satisfied for maps in $\COH_0$, and
  hence $\eta$ satisfies the defining property of $\overline\eta^\star$, and
  hence $\eta = \overline\eta^\star$.
\end{proof}

\begin{lemma}
  \label{lemma:nt-nerve-alg}
  For every natural transformation
  $\eta\in\widehat{\COH_\infty}(N_G\GD,N_F\GG)$, the induced natural
  transformation $\eta\in\widehat{\G}(T_G\GD,T_F\GG)$ is algebraic.
\end{lemma}
\begin{proof}
  By Lemma~\ref{lemma:restrict-nt}, the algebraicity condition rewrites as
  $\eta((Gf)\Gth) = (Ff)\eta(\Gth)$ for every map $\Gth\in\D(\GD,G\GTH)$
  and every map $f\in\COH_\infty(D^n,\GTH)$. This is given by the naturality of
  $\eta$ with respect to $\COH_\infty$.
\end{proof}

\paragraph{The equivalence.}
Combining Lemma~\ref{lemma:nt-alg-nerve} and
Lemma~\ref{lemma:nt-nerve-alg}, we have

\begin{prop}
  \label{prop:alg<->nerve}
  The two operations defined above form a natural isomorphism
  \[
  \widehat{\G}(T_G\GD,T_F\GG)_{\alg} \isoto \widehat{\COH_\infty}(N_G\GD,N_F\GG)
  \]
\end{prop}
\begin{proof}
  We have already proved in \cref{lemma:nt-alg-nerve,lemma:nt-nerve-alg} that
  these operations are well-defined, and moreover Lemma~\ref{lemma:restrict-nt}
  shows that the induced transformation of a restriction is the transformation
  itself. So it suffices to show that restricting an induced algebraic natural
  transformation also yields the identity. Consider an algebraic natural
  transformation $\eta \in\widehat{\G}(T_G\GD,T_F\GG)$. By definition, for every
  object $X$ in $\COH_0$ and for all maps $\Gd\in\D(\GD,GX)$ and
  $\Gc \in\COH_0(D^n,X)$, we have the equality
  $\eta((G\Gc)\Gd) = (F\Gc) \eta^\star(\Gd)$. In particular, taking
  $X=D^n$ and $\Gc$ to be the identity yields $\eta(\Gd) =
  \eta^\star(\Gd)$. Hence $\eta^\star$ coincides with $\eta$ on the presheaf
  $T_G\GD$, and thus the induction and restriction operation are inverse
  operations.
\end{proof}

\paragraph{Algebraic natural transformations that agree on the variables.}
An important property of algebraic natural transformations, is that their value
is entirely determined by their values on the variables of the theory. This is
similar to substitutions, and indeed, we show later that this notion captures
exactly the computation of the substitutions.

\begin{lemma}
  \label{lemma:alg-determined-vars}
  Consider a $\CaTT$-category with families $F:\op{\COH_\infty}\to\C$ along with a
  context $\GG$ in $\Syn{\CaTT}$ and an object $\GD$ in $\C$. Two algebraic
  natural transformations
  $\eta,\eta'\in\widehat{\G}(T\GG,T_F\GD)_{\alg}$ are equal if and
  only if we have $\eta(\Gc_x) = \eta'(\Gc_x)$ for every variable $x$ in $\GG$.
\end{lemma}
\begin{proof}
  If two algebraic natural transformations are equal, then they necessarily
  agree on the characteristic maps of the variables, so it suffices to check the
  converse. Consider two algebraic natural transformations
  $\eta,\eta'\in\widehat\G(T\GG,T_F\GD)_{\alg}$ that coincide on all
  the variables. We want to prove that they are equal. For this we show by
  induction on the depth of the term~$t$ that for any derivable term $t$ in
  $\GG$, we have $\eta(\Gc_t) = \eta'(\Gc_t)$.
  \begin{itemize}
  \item Terms of depth $0$ are simply variables, and for those, we have the equality by hypothesis.
  \item A term $t$ of depth $d+1$ is of the form $\cohop_{\GTH,B}[\Gth]$ or of
    the form $\coh_{\GTH,B}[\Gth]$, with $\Gth$ a substitution of depth at most
    $d$. In this case we consider the term $t'$ to be
    $t'=\cohop_{\GTH,B}[\id{\GTH}]$ or $t' = \coh_{\GTH,B}[\id\GTH]$
    respectively. This provides a factorization of the form
    $\Gc_t = \Gc_{t'} \Gth$. Since $\eta$ and $\eta'$ are algebraic, we
    therefore have
    \begin{align*}
      \eta(\Gc_t) &= (F(\Gc_{t'}))\eta^\star(\Gth)
      &
      \eta'(\Gc_t) &= (F(\Gc_{t'}))\eta'^\star(\Gth)
    \end{align*}
    Moreover, for every variable $y$ of $\GTH$, since $\Gth$ is of depth at most
    $d$, so is $y[\Gth]$, and therefore, by induction, we have the following
    equalities:
    \begin{align*}
      (F(\chi_{y}))\eta'^\star(\Gth) &= \eta'(\Gc_{y[\Gth]}) \\
      &= \eta(\Gc_{y[\Gth]}) \\
      &= F(\chi_{y})\eta^\star(\Gth)
    \end{align*}
    Thus $\eta'^\star(\Gth)$ satisfies the defining property of
    $\eta^\star(\Gth)$, and hence $\eta'^\star(\Gth) = \eta^\star(\Gth)$, which
    proves that $\eta(\Gc_t) = \eta'(\Gc_t)$.\qedhere
  \end{itemize}
\end{proof}

\paragraph{Algebraic transformations and pullbacks along display maps.}
We can compute the algebraic natural transformations mapping out of the nerve of
a context extension. This is the main argument making algebraic natural
transformation easy to compute with, is enough to compensate for the functor
\(N\) not sending limits to colimits. We first present a construction needed to
state the result. Consider a context $\GG$ together with a derivable type
$\GG\vdash A$ in $\CaTT$, and a \(\CaTT\)-category with families
\(F : \op{\COH_{\infty}} \to \C\). By Lemma~\ref{lemma:up-disk-catt}, $A$ is
classified by a substitution $\Gc_A:\GG\to S^{n-1}$ in the category
$\Syn{\CaTT}$. By \cref{lemma:ran-V}, this substitution gives rise to a natural
transformation in $\widehat\G(VS^{n-1},T\GG)$, and by precomposition, it induces
a map $\widehat{\G}(T\GG,T_F\GD)\to \widehat{\G}(VS^{n-1},T_F\GD)$. Applying
\cref{lemma:ran-V} again allows us to rewrite this map as
$f_A : \widehat{\G}(T\GG,T_F\GD)\to \C(\GD,FS^{n-1})$. Following the
construction explicitly shows that for a natural transformation
$\eta\in\widehat{\G}(T\GG,T_F\GD)$ the map $f_A(\eta)$ is defined by the
property following property
\[
\text{For any map $\Gc\in\Syn{\Glob}(S^{n-1},D^k)$ (\ie
any variable in $S^{n-1}$), } (F\Gc)f_A(\eta) = \eta(\Gc\Gc_A)
\]
In particular, considering a context of the form $(\GG,x:A)$, we have
the type $\GG\vdash A$. The projection substitution
$\pi : (\GG,x:A)\to\GG$ induces a morphism $T\pi : T\GG\to T(\GG,x:A)$
which can be thought of as a weakening.
Using the commutation of the following pullback square
\[
\begin{tikzcd}
  (\GG,x:A) \ar[r,"\Gc_x"] \ar[d,"\pi"']\ar[dr,phantom,"\lrcorner"very near start] & D^n\ar[d,"\partial_{n}"] \\
  \GG\ar[r,"\Gc_A"'] & S^{n-1}
\end{tikzcd}
\]
The defining property of $f_A(\eta T\pi)$ rewrites as
\[
\text{For every map $\Gc\in\Syn{\Glob}(S^{n-1},D^k)$, }(F\Gc)
f_A(\eta (T\pi)) = \eta(\Gc\partial_n\Gc_x)
\]

\begin{lemma}
  \label{lemma:extends-alg}
  $\widehat{\G}(T(\GG,x:A),T_F\GD)_{\alg}$ is obtained
  as the following pullback
  \[
  \begin{tikzcd}
    \widehat{\G}(T(\GG,x:A),T_F\GD)_{\alg} \ar[r]\ar[d]\ar[dr,phantom,"\lrcorner"very near start] & (T_F\GD)_n \ar[d]\\
    \widehat{\G}(T\GG,T_F\GD)_{\alg} \ar[r] & \C(\GD,FS^{n-1})
  \end{tikzcd}
  \]
  More, explicitly, there is an isomorphism as follows
  \[
    \widehat{\G}(T(\GG,x:A),T_F\GD)_{\alg}
    \isoto
    \setof{(\eta,\Gc_t)\in\widehat{\G}(T\GG,T_F\GD)_{\alg}\times \C(\GD,FD^n)}{\partial_n\Gc_t = f_A(\eta)}
  \]
\end{lemma}
\begin{proof}
  Consider a context $(\GG,x:A)$ in the theory $\CaTT$ and, in order to simplify
  notations, define the set
  \[
    X = \setof{(\eta,\Gc)\in\widehat{\G}(T\GG,T_F\GD)_{\alg}\times\C(\GD,FD^n)}{\partial_n\Gc = f_A(\eta)}
  \]
  We consider the following map:
  \begin{align*}
    \widehat{\G}(T(\GG,x:A),T_F\GD)_{\alg} & \to X \\
    \eta &\mapsto (\eta (T\pi), \eta(\Gc_x))
  \end{align*}
  We show that this map is well defined, \ie we show that for every
  $\eta\in\widehat\G(TD^n,T_F\GD)_{\alg}$, we have
  $(\eta (T\pi),\eta(\Gc_x))\in X$. First note that $\eta$ and
  $\eta (T\pi)$ act in the same way on every term on which they are both
  defined, but $\eta (T\pi)$ is defined on strictly fewer terms than
  $\eta$. Hence, since $\eta$ is algebraic, $\eta (T\pi)$ is also necessarily
  so. Since, by definition, $F$ is a $\CaTT$-category with families, it provides
  $\C$ with a structure of globular category with families, and we have that
  $F\partial_n = \partial_n$. Hence, for every map
  $\Gc\in\Syn{\Glob}(S^{n-1},D^k)$, we have, by naturality of \(\eta\):
  \begin{align*}
    F\Gc \partial_n  \eta(\Gc_x) = F(\Gc\partial_n)\eta(\Gc_x)
    = \eta(\Gc\partial_n\Gc_x)
  \end{align*}
  Hence $\partial_n\eta(\Gc_x)$ satisfies the defining property of
  $f_A(\eta (T\pi))$. Hence $(\eta (T\pi),\eta(\Gc_x))\in X$.

  We prove that this mapping is a bijective. Consider a pair $(\eta,\Gc)\in X$,
  an algebraic natural transformation $\eta'$ mapping onto this pair has its
  action on the variables determined. Indeed, a variable in $(\GG,x:A)$ is
  either the variable $x$, or it is a variable of $\GG$. For the variable $x$,
  we have, by definition of $\eta'$, that $\eta'(\Gc_x) = \Gc$, and for a
  variable $y$ in $\GG$, then we have the factorisation $\Gc_y = \Gc_y\pi$ and
  so $\eta'(\Gc_y) = \eta(\Gc_y)$. By Lemma~\ref{lemma:alg-determined-vars} this
  proves that the mapping is injective.

  Conversely, we show that this mapping is surjective. We construct a natural
  transformation $\eta' \in\widehat{\G}(T(\GG,x:A),T_F\GD)_{\alg}$ which extends
  the algebraic natural transformation $\eta$. First for any term $t$ in
  $(\GG,x:A)$ which does not use the variable $x$, the term $t$ is also
  definable in $\GG$, and we define $\eta'(\Gc_t) = \eta(\Gc_t)$. So it suffices
  to define the natural transformation $\eta'$ on the terms in $(\GG,x:A)$ that
  contain the variable $x$, and to verify the naturality and algebraicity of
  $\eta'$ on those terms. We proceed by induction on the coherence depth of the
  term.

  \begin{itemize}
  \item The term containing $x$ of minimal coherence depth is necessarily the
    variable $x$ itself, and in this case we define $\eta'(\Gc_x) = \Gc$. This
    assignment is natural on the variable $x$ by definition of the set $X$.
  \item Suppose $\eta'\in \widehat{\G}(T_d\GG,T_F\GD)$ to be defined and natural
    on all terms containing the variable $x$ of coherence depth at most $d$, and
    consider a term $t$ of depth $d+1$. Then $t$ is necessarily of the form
    $t=\cohop_{\GTH,B}[\Gth]$ (\resp $t=\coh_{\GTH,B}[\Gth]$), and we define
    $t' = \cohop_{\GTH,B}[\id\GTH]$ (\resp $t'=\coh_{\GTH,B}[\id\GTH]$), in such
    a way that $\Gc_t=\Gc_t'\Gth$. Then note that $\Gth$ is of coherence
    depth at most $d$, and hence defines a natural transformation in
    $\op{\G}(V\GTH,T_d\GG)$, hence by composition with $\eta'$, this provides a
    natural transformation in $\widehat{\G}(V\GTH,T_F\GD)$ which gives a
    morphism $\eta'^\star(\Gth) : \GD \to F\GTH$. We then define
    $\eta'(\Gc_t) = F(\Gc_{t'})(\eta'^\star(\Gth))$. We check that the
    transformation defined this way is natural. Consider a variable $y$ in the
    context $D^n$, corresponding to a morphism $\Gc_y: D^k\to D^n$ in the
    category $\G$, and a term $t$ of coherence depth $d+1$ in the context
    $(\GG,x:A)$ that uses the variable $x$, and denote $t'$ and $\Gth$ as
    above. We have the equalities
    \begin{align*}
      F(\Gc_y) \eta'(\Gc_t) &= F(\Gc_y) F(\Gc_{t'}) \eta'^\star(\Gth) \\
      &= F(\Gc_y\Gc_{t'}) \eta'^\star(\Gth)
    \end{align*}
    If $\Gc_y\Gc_{t'} =\Gc_z$, where $z$ is a variable, then we
    have $\Gc_y\Gc_t = \Gc_z\Gth$, and by definition of
    $\eta'^\star(\Gth)$, we have
    $F(\Gc_z)\eta'^\star(\Gth) = \eta'(\Gc_z\Gth)$. If
    $\Gc_y\Gc_{t'} = \Gc_u$ where $u$ is not a variable, it is
    again of the form $\cohop_{\GX,C}[\Gx]$ (\resp
    $\coh_{\GX,C}[\Gx]$), and we denote $u' = \cohop_{\GTH',C}[\id{}]$
    (\resp $u'=\coh_{\GTH',C}[\id{}]$) in such a way that
    $\chi_{u} = \chi_{'u} \xi$. In this case, we have
    $\Gc_y\Gc_t = \Gc_{u'}\Gx\Gth$, and thus we have
    \begin{align*}
      \eta'(\Gc_y\Gc_t) &= (F(\Gc_{u'}))  \eta'^\star(\Gx\Gth)\\
      &= F(\Gc_{u'}) F(\Gx)  \eta'^\star(\Gth) && \text{by naturality of $\eta'^\star$} \\
      &= F(\Gc_y\Gc_{t'})  \eta'^\star(\Gth)
    \end{align*}
    In both cases, we hwave $\eta'(\Gc_y\Gc_t) = F(\Gc_y) \eta'(\Gc_t)$ which proves that $\eta'$ is natural on $\Gc_t$.
  \end{itemize}
  We now prove that the natural transformation we have just defined is a
  preimage of the couple $(\eta,\Gc)$, and note that by definition, we have
  $\eta' T\pi = \eta$ and $\eta'(\Gc_x) = \Gc$, so it suffices to show that
  $\eta'$ is algebraic. Consider a ps-context $\GTH$ together with a map
  $\Gth : (\GG,x:A)\to \GTH$, and a map $\Gx \in \COH_{\infty}(D^n,\GTH)$. The
  map $\Gx$ corresponds to a term in the ps-context $\GTH$ which is either a
  variable or of the form $\cohop_{\GTH',B}[\Gx']$ (\resp
  $\coh_{\GTH',B}[\Gx']$). If $\Gx$ defines a variable, the equality required
  for the algebraicity is implied by the naturality, so it suffices to verify it
  for the term constructors. We define $t' = \cohop_{\GTH',B}[\id{\GTH'}]$
  (\resp $\coh_{\GTH',B}[\id{\GTH'}']$), in such a way that we have $\Gx = \Gc_{t'} \Gx'$.
  We then have the following equalities
  \begin{align*}
    \eta'(\Gx\Gth) &= F(t')  \eta'^\star(\Gx'\Gth) \\
    &= F(t') F(\Gx') \eta'^\star(\Gth) && \text{by naturality of $\eta'^\star$}\\
    &= F(\Gx) \eta'^\star(\Gth) \tag*{\qedhere}
  \end{align*}
\end{proof}

\subsection{Kan extension of a $\CaTT$-category with families.}
Using algebraic natural transformations, we define and characterize
the right Kan extension of a $\CaTT$-category with family
$F : \op{\COH_\infty}\to\C$ along the functor
$P_\infty : \op{\COH_\infty}\to\Syn{\CaTT}$. Like in \secr{glob-cwf},
this right Kan extension is the key construction to prove the
initiality of the syntactic category. In the case of $\CaTT$, the
presence of term constructors makes the existence harder to prove, as
witnessed by the introduction of algebraic natural transformations.

\paragraph{Existence of the Kan extension.}
The previous results show that algebraic natural transformations can be built
inductively following the structure of contexts, starting with the empty
contexts and computed with a sequence of context comprehension operations. This
lets us define and characterize the right Kan extension of any $\CaTT$-category with
families $\C$ along the functor~$P_\infty$, by proving that all the canonical
diagrams of objects in $\Syn{\CaTT}$ necessarily have a limit in $\C$.

\begin{lemma}
  \label{lemma:ran-exists-catt}
  Given a $\CaTT$-category with families $F: \op{\COH_{\infty}}\to\C$,
  there exists a pointwise right Kan extension
  $\Ran_{P_{\infty}}F : \Syn{\CaTT} \to \C$. Moreover,
  $(\Ran_{P_{\infty}}F) \emptycontext$ is the terminal object of $\C$,
  and $\Ran_{P_{\infty}}F$ satisfies the equation
  $ (\Ran_{P_\infty}F)(\GG,A)\isoto
  ((\Ran_{P_\infty}F)\GG,\partial_n((\Ran_{P_\infty}F) \Gc_A))$.
\end{lemma}
\begin{proof}
  By Lemma~\ref{lemma:ran-V} The existence of the pointwise Kan extension is
  equivalent to showing that for every context $\GG$ in $\Syn{\CaTT}$ there is a
  natural isomorphism \(\C(\GD,X)\isoto \widehat{\COH_\infty}(N\GG,N_F\GD)\), so
  it suffices to construct an object $X$ which satisfies this property.
  Proposition~\ref{prop:alg<->nerve} lets us rewrite as the above isomorphism
  \[
  \C(\GD,X)\isoto \widehat{\G}(T\GG,T_F\GD)_{\alg}
  \]
  We proceed by induction on $\GG$ to show that there exists an object $X$
  satisfying this property.
  \begin{itemize}
  \item For the context $\emptycontext$, an element of the presheaf
    $T\emptycontext$ is a substitution $\emptycontext\vdash \Gg: D^n$, so by
    Lemma~\ref{lemma:up-disk-catt} it is necessarily of the form $\Gc_t$ where
    $t$ is a term in the context $\emptycontext$. Since by
    Lemma~\ref{lemma:empty-ctx-catt} there is no such term, this implies that
    there is no element in $T\emptycontext$, and thus it is the empty globular
    set, that is initial. Moreover the only natural transformation
    $! : T\emptycontext \to T_F\GD$ is vacuously algebraic. Hence
    $\widehat{\G}(T\GG,T_F\GD)_{\alg}=\set{\bullet}$ is a singleton. So the limit
    of the diagram is the terminal object in $\C$, which exists by definition of
    a category with families.
  \item For a context of the form $(\GG,x:A)$, assume that there is an object
    $Y$ in $\C$, together with a natural isomorphism
    $\C(\GD,Y)\isoto\widehat{\G}(T\GG,T_F\GD)_{\alg}$.  We can apply
    Lemma~\ref{lemma:extends-alg}, which provides the following equalities
    \begin{align*}
      \widehat{\G}(T(\GG,A),T_F\GD)_{\alg} & \isoto \lim \pa{\widehat{\G}(T\GG,T_F\GD)_{\alg}\xrightarrow{\phantom{\C(\GD,\partial_n)}}\C(\GD,FS^{n-1})\xleftarrow{\C(\GD,\partial_n)}\C(\GD,F n)} \\
      &\isoto \lim\pa{\C(\GD,Y)\xrightarrow{\phantom{\C(\GD,\partial_n)}}\C(\GD,FS^{n-1})\xleftarrow{\C(\GD,\partial_n)}\C(\GD,F n)}\\
      &\isoto \C\pa{\GD,\lim\pa{Y\overset{f}{\to} FS^{n-1} \overset{\partial_n}{\leftarrow}F n}}
    \end{align*}
    By definition of the structure of a globular category with families, the
    above limit in $\C$ exists and can be computed as $X = (Y,\partial_n(f))$. This
    choice of $X$ by definition is an object such that
    $\C(\GD,X)\isoto \widehat{\G}(T\GG,T_F\GD)_{\alg}$, hence it
    defines a limit for the canonical diagram associated to $\GG$.
  \end{itemize}
  The construction of the object $X = \Ran_{P_{\infty}}(\Gamma)$ shows that the
  right Kan extension sends the terminal to the terminal and satisfies the
  required equation.
\end{proof}

\noindent Considering a \(\CaTT\)-category with families
\(F : \op{\COH_{\infty}} \to \C\), the right Kan extension
\(\Ran_{P_{\infty}}F\) exists, and since \(P_{\infty}\) is fully faithful,
Lemma~\ref{lemma:ran-extension} shows that the universal natural transformation
\(\epsilon(F) : (\Ran_{P_{\infty}}F)P_{\infty} \Rightarrow F\) is a natural
isomorphism.

\paragraph{Functors preserving pullbacks along display maps.}
We have proven that restricting the right Kan extension yields a functor
equivalent to the one from which we started. Conversely, we now show that every
functor preserving pullbacks along display maps is the Kan extension of its
restriction.
\begin{lemma}\label{lemma:ran-extension-pullbacks-catt}
  Given a cartegory \(\C\) and a functor \(F : \Syn{\CaTT} \to \C\)
  that preserves the terminal object and sends pullbacks along the
  projection maps \(\set{\partial_{n} : D^{n} \to S^{n-1}}\) onto
  pullbacks, then \(F\) is the pointwise right Kan extension
  \(F \isoto \Ran_{P_{\infty}}FP_{\infty}\), and in particular this
  Kan extension exists.
\end{lemma}
\begin{proof}
  The proof is the same as the one of \cref{lemma:ran-extension-pullbacks}, but
  with \(\widehat{\G}(T\_{},T_{FP_{\infty}}\_{})_{\alg}\) playing the role of
  \(\widehat{\G}(V\_{},T_{F}\_{})\). If it exists, the right Kan extension
  \((\Ran_{P_{\infty}}FP_{\infty})\Gamma\) is characterized by the fact that for
  all \(\Delta\), we have
  \(\C(\Delta,(\Ran_{P_{\infty}}FP_{\infty})\Gamma) \isoto
  \widehat{\COH_{\infty}}(N\Gamma,N_{FP_{\infty}}\Delta)\). We prove that
  \(F(\Gamma)\) satisfies this equation. First, it holds for the disk contexts
  \(D^{n}\), since \(P_{\infty}\) is fully faithful, \(ND^{n}\) is the
  representable presheaf associated to \(D^{n} \in \op{\COH_{\infty}}\), and by
  the Yoneda lemma,
  \( \widehat{\COH_{\infty}}(ND^{n},N_{FP_{\infty}}(\Delta)) \isoto
  N_{FP_{\infty}}(\Delta)_{D^{n}} = \C(\Delta,FD^{n})\). We now prove this
  equation for the sphere contexts by induction
  \begin{itemize}
  \item The \(-1\) sphere is the empty context and the property
    follows immediately from the preservation of terminal object.
  \item Assume that we have proved the property for \(S^{n}\), then we have,
    using \cref{lemma:extends-alg}
    \begin{align*}
      \C(\Delta,S^{n+1})
      &= \C(\Delta,\lim(D^{n+1}\to S^{n} \leftarrow D^{n+1})) \\
      & \isoto \lim (\C(\Delta,D^{n+1}) \to
        \C(\Delta,S^{n})\leftarrow\C(\Delta,D^{n+1})) \\
      & \isoto \lim(\widehat{\COH_{\infty}}(ND^{n+1}, N_{FP_{\infty}}\Delta) \to
        \widehat{\COH_{\infty}}(NS^{n},N_{FP_{\infty}}\Delta)
        \leftarrow \widehat{\COH_{\infty}}(ND^{n+1},N_{FP_{\infty}}\Delta)) \\
      & \isoto \lim(\widehat{\G}(TD^{n+1}, T_{FP_{\infty}}\Delta)_{\alg} \to
        \widehat{\G}(TS^{n},T_{FP_{\infty}}\Delta)_{\alg}
        \leftarrow \widehat{\G}(TD^{n+1},T_{FP_{\infty}}\Delta)_{\alg}) \\
      & \isoto \widehat{\G}(TS^{n+1},T_{FP_{\infty}}\Delta)
    \end{align*}
  \end{itemize}
  We now prove this same equation for any context by induction on the length.
  \begin{itemize}
  \item For the empty context, this is by preservation of the terminal object.
  \item For an extended context of the form \(\Gamma,A\), a similar computation
    to the sphere case shows that
    \begin{align*}
      \C(\Delta,(\Gamma,A))
      &= \C(\Delta,\lim(\Gamma\to S^{n-1} \leftarrow D^{n})) \\
      & \isoto \lim(\widehat{\G}(T\Gamma, T_{FP_{\infty}}\Delta)_{\alg} \to
        \widehat{\G}(TS^{n-1},T_{FP_{\infty}}\Delta)_{\alg}
        \leftarrow \widehat{\G}(TD^{n},T_{FP_{\infty}}\Delta)_{\alg}) \\
      & \isoto \widehat{\G}(T(\Gamma,A),T_{FP_{\infty}}\Delta) \tag*{\qedhere}
    \end{align*}
  \end{itemize}
\end{proof}

\subsection{Initiality and models of the theory $\CaTT$.}
From now on, the arguments that we give follow closely the structure
of our study of the models of $\Glob$ given in \Cref{sec:glob-cwf}.

\paragraph{Initiality of $\Syn{\CaTT}$.}
We start by proving that the
category $\Syn{\CaTT}$ is initial among $\CaTT$-categories with families,
in the same sense that of Theorem~\ref{thm:weak-initiality-glob}.
\begin{lemma}\label{lemma:ran-definition-catt}
  For a $\CaTT$-category with families $F : \op{\COH_{\infty}}\to\C$,
  there is a morphism of categories with families
  \[
    (\Ran_{P_{\infty}}F, \epsilon(F)) : \Syn{\CaTT} \to \C
  \]
  where the transformation
  $\epsilon(F) : (\Ran_{P_{\infty}}F)P_{\infty} \Rightarrow F$
  is a natural isomorphism.
\end{lemma}
\begin{proof}
  Lemma~\ref{lemma:ran-exists-catt} shows that the right Kan extension exists
  and preserves the terminal object and the pullbacks along display maps
  (provided by the equation concerning the context comprehension operation).
  Hence $\Ran_{P_{\infty}}F$ can be chosen uniquely to be a morphism of
  categories with families. Denote
  $\epsilon(F) : (\Ran_{P_{\infty}}F)P_{\infty}\Rightarrow F$ the universal
  natural transformation obtained as part of the Kan extension. Then, by
  definition, $(\Ran_{P_{\infty}}F,\epsilon(F))$ defines a morphism of
  $\CaTT$-categories with families, and Lemma~\ref{lemma:ran-extension} shows
  that $\epsilon(F)$ is a natural isomorphism.
\end{proof}

\begin{thm}[local initiality of the syntactic category]
  \label{thm:weak-initiality-catt}
  The morphism of $\CaTT$-categories with families
  $(\Ran_{p_{\infty}}F, \epsilon(F)) : \Syn{\CaTT} \to \C$ is a
  terminal object in the category $\cCwFh(\Syn{\CaTT},\C)$.
\end{thm}
\begin{proof}
  This is exactly the universal property of the right Kan extension.
  Consider a morphism of $\CaTT$-category with families
  $(G, \alpha) : \Syn{\CaTT} \to \C$. The universal property of the
  right Kan extension lets us construct a natural transformation
  $\gamma : G \Rightarrow \Ran_{P_{\infty}}F$ such that we have the
  following equality:
 \[
    \begin{tikzcd}
      \Syn{\CaTT}
      \ar[rr, "G"]
      \ar[rrd, phantom, "\overset{\alpha}{\Rightarrow}"{sloped,
      very near start}]
      && \C \\
      \op{\COH_{\infty}} \ar[u,"P_{\infty}"]\ar[urr, "F"'] && \phantom{A}
    \end{tikzcd}
    \quad=\quad
    \begin{tikzcd}
      \Syn{\CaTT}
      \ar[rr,"\Ran_{P_{\infty}}F"]
      \ar[rr, bend left = 60, "G"]
      \ar[rr, bend left = 40, "\Downarrow_{\gamma}", phantom]
      \ar[rrd, phantom, "\overset{\epsilon(F)}{\Rightarrow}"{sloped,
      very near start}]
      && \C \\
      \op{\COH_{\infty}} \ar[u,"P_{\infty}"]\ar[urr, "F"'] && \phantom{A}
    \end{tikzcd}
  \]
  Thus, $\gamma$ is a natural transformation satisfying
  $\gamma : (G,\alpha) \to (\Ran_{P_{\infty}}F,\epsilon(F))$.
\end{proof}

\paragraph{Models of the theory $\CaTT$.}
We use initiality (Theorem~\ref{thm:weak-initiality-catt}) to
characterize the models of the theory $\CaTT$. Denote
$\forget : \gCwF \to \CwF$ the forgetful functor, and consider, for a
given category with families $\C$, the category $\fiber \C$ whose
objects are the $\CaTT$-categories with families $\D$ such that
$\mathcal{U}(\D) = \C$, morphisms are the morphisms of $\CaTT$-categories
with families which project onto $\id\C$ by $\mathcal{U}$.

\begin{prop}\label{prop:models-ccwf}
  Consider a category with families $\C$, there is an equivalence of
  categories $\CwFh(\Syn\CaTT,\C) \simeq \fiber \C$.
\end{prop}
\begin{proof}
  We build a pair of functors
  \[
    \begin{tikzcd}[sep=8ex]
      \CwFh(\Syn{\CaTT},\C) \ar[r, shift left, "P_{\infty}^{*}"]
      & \ar[l,shift left, "\Ran_{P_{\infty}}"]\fiber \C
    \end{tikzcd}
  \]
  and show that they define an equivalence of categories.

  \emph{Definition of the functor $P_{\infty}^{*}$.} This functor is given by
  precomposition. Given a morphism of categories with families
  $F : \Syn{\CaTT}\to\C$, we define $P_{\infty}^{*}(F) = FP_{\infty}$, with the
  structure of globular category with families on $\C$ given by $\ind(FI)$
  (where $I : \Syn{\Glob}\to\Syn{\CaTT}$ is the embedding). The pair $(F,\id{})$
  defines a morphism of $\CaTT$-categories with families
  $\CaTT \to P_{\infty}^{*}(F)$.

  \emph{Definition of the functor $\Ran_{P_{\infty}}$.} This functor is given by
  the right Kan extension. Given a $\CaTT$-category with families
  $G : \op{\COH_{\infty}} \to \C$, it associates the morphism of categories with
  families $\Ran_{P_{\infty}}F$ which exists and is a morphism of categories
  with families by Lemma~\ref{lemma:ran-definition-catt}. Given a morphism
  $\alpha : G \Rightarrow G'$ in $\fiber \C$, we have
  $\alpha\epsilon(G) : (\Ran_{P_{\infty}}G)P_{\infty} \Rightarrow G'$. By
  universal property of the Kan extension, there is a unique
  $\Ran_{P_{\infty}}(\alpha) : \Ran_{P_{\infty}}G \Rightarrow
  \Ran_{P_{\infty}}G'$ such that
  $\Ran_{P_{\infty}}(\alpha) \epsilon(G)= \epsilon(G')\alpha$.

  \emph{Equivalence $P_{\infty}^{*}\Ran_{P_{\infty}} \simeq \id{}$.}
  Consider a $\CaTT$-category with families $G : \op{\COH_{\infty}} \to\C$, then
  we have the map $\epsilon(G) : (\Ran_{P_{\infty}}G)P_{\infty}\Rightarrow G$,
  which is an isomorphism in $\C$. Given a map $\alpha$ in $\fiber\C(G,G')$, by
  definition, $\Ran_{P_{\infty}}(\alpha)\epsilon(G)= \epsilon(G')\alpha$. This
  is exactly the naturality of $\epsilon$.

  \emph{Equivalence $\id{}\simeq \Ran_{P_{\infty}}P_{\infty}^{*}$.} A morphism
  of category with families $F : \Syn{\CaTT}\to\C$, defines a morphism of
  $\CaTT$-categories with families $(F, \id{}) : \Syn{\Glob} \to \ind(F)$. By
  \cref{thm:weak-initiality-catt}, we have a natural transformation
  $\alpha(F) : F \to \Ran_{P_{\infty}}(FP_{\infty})$, obtained by universal
  property of the right Kan extension. It is thus uniquely characterized by
  $\epsilon(F)\alpha(F)_{P_{\infty}} = \id{}$.Since \(\epsilon(\ind(F))\) is an
  isomorphism, so is \(\alpha(F)_{P_\infty}\). Consider the isomorphism
  \(\gamma: F \isoto \Ran_{P_{\infty}}(FP_{\infty})\) obtained by
  Lemma~\ref{lemma:ran-extension-pullbacks-catt}. Then
  \((\alpha(F)\gamma^{-1})_{P_{\infty}}\) is an isomorphism, so by
  Lemma~\ref{lemma:extend-natural-isos}, so is \(\alpha(F)\gamma^{-1}\), and
  thus so is \(\alpha(F)\). We now show that the family \(\alpha(F)\) is natural
  in \(F\): Given two morphisms of categories with families
  \(F,G : \Syn{\Glob} \to \C\), we consider the two following diagram, whose
  compositions are both equal to \(\beta_{P_{\infty}}\), using the equations
  that characterize \(\epsilon\) and \(\Ran_{P_\infty}(\beta_{P_\infty})\).
    \[
     \begin{tikzcd}[column sep=large]
        \Syn{\Glob} \ar[rr] \ar[rr, "F", bend left =
        70] \ar[rr, bend left = 30] \ar[rr, bend left = 15, phantom,
        "{\scriptscriptstyle\alpha(G)}"] \ar[rr, bend left = 50,
        phantom, "{\scriptscriptstyle\beta}"] \ar[rrd, phantom,
        "{\scriptscriptstyle\epsilon(GP_\infty)}"{very near start}]
        && \C \\
        \op\G\ar[u,"P_\infty"]\ar[urr,"FP_\infty"'] && \phantom{(A)}
      \end{tikzcd}=
     \begin{tikzcd}[column sep=large]
       \Syn{\Glob} \ar[rr] \ar[rr, "F", bend left = 70] \ar[rr, bend left = 30]
       \ar[rr, bend left = 50, phantom, "{\scriptscriptstyle\alpha(F)}"] \ar[rr,
       bend left = 15, phantom,
       "{\scriptscriptstyle\Ran_{P_\infty}(\beta_{P_\infty})}"] \ar[rrd, phantom,
       "{\scriptscriptstyle\epsilon(GP_\infty)}"{very near start}]
       && \C \\
       \op\G\ar[u,"P_\infty"]\ar[urr,"FP_\infty"'] && \phantom{(A)}
      \end{tikzcd}
    \]
    By universality of the Kan extension, this shows the equation
    \(\alpha(G)\beta = \Ran_{P_\infty}(\beta_{P_\infty})\alpha(F)\), which is
    exactly the naturality of \(\alpha\). \qedhere

\end{proof}

\paragraph{Set-theoretic models of $\CaTT$.} In the case where we
consider the models in $\Set$, this theory lets us obtain an
equivalence with the Grothendieck-Maltsiniotis weak
$\omega$-categories.

\begin{prop}\label{prop:set-fiber-catt}
  There is an equivalence of categories $\fiber \Set \simeq \wCat$.
\end{prop}
\begin{proof}
  In a $\CaTT$-structure on $\Set$ given by
  $F : \op{\COH_{\infty}}\to \Set$, the functor $F$ preserves the
  globular products, and thus defines a Grothendieck-Maltsiniotis weak
  $\omega$-category. A morphism of $\CaTT$-category with families is
  exactly a morphism of weak $\omega$-categories between them, thus
  this defines a fully faithful functor $\fiber\Set \to \wCat$.
  Moreover, every weak $\omega$-category is obtained this way, the
  $\CaTT$-category with families structure is obtained by considering the
  globular structure induced by $FD_{P}$. Hence this functor is
  essentially surjective so it is an equivalence.
\end{proof}

\begin{thm}\label{thm:models-catt}
  There is an equivalence of categories
  $\Mod{\Syn{\CaTT}}\simeq \wCat$.
\end{thm}
\begin{proof} \belowdisplayskip=-12pt
  By \cref{prop:models-ccwf}
  and \cref{prop:set-fiber-catt} we have the following equivalences of
  categories
  \begin{align*}
    \Mod{\Syn\Glob}=\CwFh(\Syn\CaTT, \Set) \simeq \fiber\Set \simeq
    \wCat
  \end{align*} \qedhere
\end{proof}

\subsection{Interpretation of the proof.}
\label{sec:algebraic-transformations-and-substitutions}
We have now proven the main result of this article. Yet the proof
involves some constructions that may seem \emph{ad-hoc} or unnatural.
the aim of this section is to show that these construction can be
interpreted syntactically and correspond meaningful properties about
the behavior of the dependent type theory $\CaTT$.

\paragraph{Substitutions to a globular context.}
We have proved with Theorem~\ref{lemma:funext-c-system} that a
substitution is completely determined by its action on variables of a
context, we can extract from \cref{thm:weak-initiality-glob} a partial
answer to the converse question: considering a given action on
variables, is there a substitution acting this way? We already know
that the action cannot be completely free, since the substitution must
respect the typing, and hence the source and target. In fact, in the
case where the target context is a context in $\Syn{\Glob}$, this
theorem shows that this is the only obstruction. Consider
$\Syn{\CaTT}$ as a globular category with families, and consider, in
the category~$\Syn{\CaTT}$, a context $\GG$ which comes from the
theory $\Glob$: $\GG$ is constructed only from variables and term
constructors do not appear in it. Then for an arbitrary context $\GD$,
one can apply \cref{thm:weak-initiality-glob} to characterize the
substitutions $\GD\to\GG$ as being equivalent to the natural
transformations $\widehat{\G}(V\GG,T\GD)$. In other words, in this
case, a substitution $\Gg$ is nothing else than the data of, for every
variable $x$ in $\GG$, a term $t$ in $\GD$ with the intent that
$t = x[\Gg]$ in a way that is compatible with the source and target
relations.

\paragraph{Substitutions to an arbitrary context.}
We can interpret \cref{thm:weak-initiality-catt} as a generalization
of the previous discussion, where we characterize substitutions with
an arbitrary context~$\GG$ as target (as opposed to one built from
variables only). We cannot generalize it naively, by requiring that we
associate a term to any variable of $\GG$. Indeed, if we consider the
context
\[
  \GG = \verb?(x : *) (f : id x -> id x)?
\]
then the source of the variable \verb?f? is the term \verb?id x?,
which is not itself a variable, hence the compatibility of the source
and target cannot be expressed as a naturality condition.
Categorically, this means that the set of variables $V\GG$ is not
equipped with a structure of a globular set given by the source and
target: in our example, we would indeed have
$(V\GG)_2 = \left\{\verb?f?\right\}$, but the source of this term is
\verb?id x?, which is not an element of $V(\GG)_1$. The solution we
have chosen is to associate not only a term to any variable of $\GG$,
but also to any term of $\GG$, in a way that respects the source and
target, and thus we now represent substitutions $\Gg : \GD\to\GG$ as
natural transformations in~$\op\G(T\GG,T\GD)$. However, this gives too
much freedom, and there are such natural transformations that are
ill-defined transformations. For instance, consider the contexts
\[
  \GG = \verb?(x : *) (f : x -> x)? \qquad\qquad \GD = \verb?(x : *)?
\]
together with a natural transformation $\eta:T\GD\Rightarrow T\GG$
such that $\eta(\verb?id x?) = \verb?f?$. This can never be the action
of a substitution, since
$(\verb?id?\ x)[\gamma] = \verb?id?\ (x[\Gg])$. The problem is that,
in this representation, we do not account for the fact that a
substitution must respect the term constructors. The notion of
algebraic natural transformation achieves exactly this: an algebraic
natural transformation is a transformation that respects term
constructors, and \cref{thm:weak-initiality-catt} ensures that, by
considering only the algebraic natural transformations, we recover
exactly the data of the substitutions.

\paragraph{Codensity of the functor $P_\infty$.}
\cref{thm:weak-initiality-catt} states that for any $\CaTT$-category
with families $\C$, the right Kan extension along $P_\infty$ gives the
unique morphism of $\CaTT$-category with families from the syntactic
category to $\C$. In particular, applying this theorem to the
$\CaTT$-category $\Syn{\CaTT}$ with the structure given by $P_\infty$
shows that $\Ran_{P\infty}P_\infty$ is this unique morphism. Since the
identity functor $\id{\Syn{\CaTT}}$ is also a morphism, this shows in
particular that $\id{\Syn{\CaTT}} = \Ran_{P_\infty}P_\infty$: in other
words, the functor $P_\infty$ is codense. More concretely, this proves
that every context in the category $\Syn{\CaTT}$ is canonically
obtained as a limit of ps-contexts.

\paragraph{Developing the limits.}
There is also an interesting interpretation of
Proposition~\ref{prop:alg<->nerve}, which establishes the equivalence between
algebraic natural transformations $\widehat\G(T\GG,T\GD)_{\alg}$ and
natural transformations $\widehat{\COH_\infty}(N\GG,N\GD)$. Indeed, consider
that the natural transformations $\widehat{\COH_\infty}(N\GG,N\GD)$ are maps
of cones, between a cone of apex $\GG$ and a cone of apex $\GD$, over the
canonical diagram of $\GG$. Recall that the only objects that appear in the
canonical diagram of $\GG$ are ps-contexts, which are themselves globular
products of disks. Hence one can ``develop'' this diagram, and obtain from the
above map of cones, a new map of cones, with same apex, but over a diagram only
made out of disks. Proposition~\ref{prop:alg<->nerve} shows that algebraic
natural transformations are exactly those maps of cones between diagrams over
disks that can be obtained by such an operation. This theorem can thus be seen
as a way to develop a canonical limit of ps-contexts into a non-canonical limit
of disks. This matches the syntactic construction of context as a succession of
context comprehension, which exhibits each context as a succession of pullback
of disks. In that respect, the contexts in the theory are analogous to the
CW-complexes in topology.

\paragraph{Finitely generated polygraphs.}
Recall that the nerve functor is defined by
\begin{align*}
  N : \Syn{\CaTT}&\to\widehat{\COH_\infty}\\
  \GG&\mapsto\Syn{\CaTT}(\GG,\_)
\end{align*}
A colimit in $\COH_\infty$, which is thus a limit in
$\Syn{\ps,\infty}$, is preserved by the functor $N\GG$, by continuity
of the hom-functor. Hence, $N\GG$ defines a weak $\omega$-category in
the sense of Grothendieck-Maltsiniotis. By \cref{thm:models-catt},
$N\GG$ thus defines a model of the theory $\CaTT$. In fact, one can
describe the corresponding model, given by
$\Syn{\CaTT}(\GG,\_) : \Syn{\CaTT}\to\Set$, which by continuity of the
hom-functor preserves all the limits, and hence is a model. This shows
that we have a functor $\Syn{\CaTT}\to\Mod{\CaTT}$, given by the
coYoneda embedding, which is fully faithful and thus exhibits
$\Syn{\CaTT}$ as a full subcategory of the weak $\omega$-categories.
We call \emph{finitely generated polygraphs} (or \emph{computads}) the
weak $\omega$-categories that come from an object of $\Syn{\CaTT}$,
which generalize similar notions studied in higher category theory
\cite{burroni1993higher, street1976limits}, in particular the
polygraphs play an important role in the theory of strict
$\omega$-categories as they are the cofibrant objects for the folk
model structure~\cite{lafont2010folk}. This remark draws an analogy
between our presentation of weak $\omega$-categories and the
Gabriel-Ülmer duality~\cite{gabriel2006lokal} in which the syntactic
category sits inside the models as the opposite of the free finitely
generated objects.


\section*{Further Work}
In this article, we have presented a construction allowing us to
characterize the semantics of the theory $\CaTT$. The method that we
use roughly breaks down in two steps. First we identified a class of
contexts playing a particular role in the theory, that we called
ps-contexts and showed they formed a subcategory of the syntactic
category that is equivalent to the coherator of the
Grothendieck-Maltsiniotis definition of weak $\omega$-category.
Second, we characterized the models of $\CaTT$ as presheaves over this
coherator satisfying additional conditions. While in the first step
the characterization of the coherator is inherently very specific to
the theory $\CaTT$, the second step is fairly generic, and we believe
that it points towards a more general framework connecting dependent
type theory with notions algebraic theories. This idea has been
present since the early days of dependent type theory and was explored
by Cartmell~\cite{cartmell1986generalised} through the notion of
generalized algebraic theory. However, we believe that our work brings
the formulation closer to that of monads with
arities~\cite{berger2012monads}, and we believe that there is a strong
connection between the method we use and the nerve theorem in this
framework, as investigated in~\cite{subramaniam2021dependent}.

We believe that our work gives a promising approach to tackling the initiality
conjecture, which could be solved in the aforementioned particular case of
dependent type theories which entertain a close enough connection with the
category $\CaTT$. We however point out this theory is much simpler than
Martin-Löf type theory and its variations, particularly since it does not have
any computation rules, and we believe that the presence of those rules strongly
increases the difficulty for proving this conjecture for such theories.

We believe that it would be valuable to establish a connection between our
interpretation of the contexts as finitely generated polygraphs and the notion
of polygraphs usually defined for strict
$\omega$-categories~\cite{lafont2010folk, street1987algebra}. In particular, in
the strict case, polygraphs are used to characterized cofibrant objects for a
model structure, and we would like to investigate whether such a model
structure, or a weaker version of it in the form of a weak factorization system,
would make sense for weak $\omega$-categories~\cite{henry-hott}.

The approach that we have presented in this article generalizes to other related
higher structures, allowing for a type theoretic presentation of these
structures. In particular, similar methods have been conducted in order to study
monoidal weak $\omega$-categories~\cite{benjamin22monoidal}, cubical weak
$\omega$-categories~\cite{benjamin2020type} and strictly unital weak
$\omega$-categories~\cite{finster2020type}. Further work along the lines of this
article includes generalizing the methods that we have presented in order to
study the semantics of such theories. In the case of monoidal weak
$\omega$-categories, the theories are close enough that a transfer of the
semantics can be done~\cite{benjamin22monoidal}, and understanding the
semantics of $\CaTT$ is enough to settle the semantics of the other theory. The
theory for cubical $\omega$-categories, can be presented in a very similar
fashion to the theory $\CaTT$ and we believe that similar methods can be used to
study its semantics. The theory for strictly unital $\omega$-categories is more
complicated since it contains rewriting rules. Understanding the semantics is a
relevant challenge left for future work and we believe that either one could
achieve it by relating this theory to $\CaTT$ or by adapting the methods that we
have presented to account for the rewriting rules.



\bibliographystyle{plain}
\bibliography{papers}

\end{document}